\documentclass[nofootinbib,prc,superscriptaddress,showpacs,tightenlines,preprint,floatfix]{revtex4-1}

\usepackage[utf8]{inputenc}

\usepackage{grffile}
\usepackage{graphicx,amsmath,amssymb,bm,mathtools}
\usepackage{amsfonts}
\usepackage{verbatim}
\usepackage{float}
\usepackage{amsbsy}

\usepackage{fancyhdr,hyperref}

\usepackage{color}
\usepackage[labelfont={normalsize},subrefformat=parens,caption=false]{subfig}
\graphicspath{{./figures/}}

\newcommand{\beq}{\begin{equation}}
\newcommand{\eeq}{\end{equation}}
\newcommand{\bml}{\begin{multline}}
\newcommand{\eml}{\end{multline}}

\newcommand{\bseq}{\begin{subequations}}
\newcommand{\eseq}{\end{subequations}}

\newcommand{\RNN}{R_{NN}}
\newcommand{\RNNN}{R_{3N}}
\newcommand{\RTN}{R_{3N}}

\newcommand{\fm}{\, \text{fm}}

\newcommand{\eft}{$\chi$EFT}
\newcommand{\kf}{k_{\textrm{F}}}

\newcommand{\mpi}{m_{\pi}}
\newcommand{\delm}{M_{\Delta-N}}
\newcommand{\fpi}{f_{\pi}}
\newcommand{\Fpi}{F_{\pi}}
\newcommand{\ga}{g_A}
\newcommand{\ha}{h_A}

\newcommand{\la}{\langle}
\newcommand{\ra}{\rangle}

\newcommand{\F}{\mathcal{F}}
\newcommand{\De}{\mathcal{Z}}
\newcommand{\Se}{\mathcal{W}}
\newcommand{\Fp}{\mathcal{F'}}
\newcommand{\W}{\mathcal{W}}
\newcommand{\Wp}{\mathcal{W'}}
\newcommand{\T}{\mathcal{T}}
\newcommand{\Y}{\mathcal{Y}}
\newcommand{\Z}{\mathcal{Z}}
\newcommand{\Zp}{\mathcal{Z'}}
\newcommand{\B}{\mathcal{B}}
\newcommand{\R}{\mathcal{R}}
\newcommand{\ct}{\cos\theta}

\newcommand{\Sub}{\widetilde{\Pi}}
\newcommand{\Dub}{\widetilde{\Pi}}

\newcommand{\NNLO}{\ensuremath{{\rm N}{}^2{\rm LO}}}
\newcommand{\NNNLO}{\ensuremath{{\rm N}{}^3{\rm LO}}}
\newcommand{\NNNNLO}{\ensuremath{{\rm N}{}^4{\rm LO}}}

\newcommand{\avec}{\mathbf{a}}

\newcommand{\kvec}{\mathbf{k}}
\newcommand{\lvec}{\mathbf{l}}

\newcommand{\jvec}{\mathbf{j}}
\newcommand{\Jvec}{\mathbf{J}}

\newcommand{\Tvec}{\mathbf{T}}

\newcommand{\svec}{\mathbf{s}}

\newcommand{\xvec}{\mathbf{x}}

\newcommand{\rvec}{\mathbf{r}}
\newcommand{\yvec}{\mathbf{y}}

\newcommand{\rhat}{\mathbf{\hat{r}}}

\newcommand{\xhat}{\mathbf{\hat{x}}}

\newcommand{\Rvec}{\mathbf{R}}
\newcommand{\Nvec}{\mathbf{N}}

\newcommand{\rone}{\mathbf{r}_1 \sigma_1 \tau_1}
\newcommand{\rtwo}{\mathbf{r}_2 \sigma_2 \tau_2}
\newcommand{\rthree}{\mathbf{r}_3 \sigma_3 \tau_3}
\newcommand{\rfour}{\mathbf{r}_4 \sigma_4 \tau_4}

\newcommand{\vnn}{V^{\text{NN}}}
\newcommand{\vnnn}{V^{\text{3N}}}
\newcommand{\vnna}{\mathcal{V}^{\text{NN}}}

\newcommand{\sione}{\sigma_1 \tau_1}
\newcommand{\sitwo}{\sigma_2 \tau_2}
\newcommand{\sithree}{\sigma_3 \tau_3}
\newcommand{\sifour}{\sigma_4 \tau_4}

\newcommand{\sigvec}{\boldsymbol{\sigma}}
\newcommand{\tauvec}{\boldsymbol{\tau}}

\newcommand{\xiav}{\xvec_{2}}
\newcommand{\xibv}{\xvec_{3}}

\newcommand{\xia}{\xhat_{2}}
\newcommand{\xib}{\xhat_{3}}

\newcommand{\af}{\alpha_1}
\newcommand{\as}{\alpha_2}
\newcommand{\at}{\alpha_3}

\newcommand{\bi}{\begin{itemize}}
\newcommand{\ei}{\end{itemize}}

\newcommand{\be}{\begin{enumerate}}
\newcommand{\ee}{\end{enumerate}}
\newcommand{\bc}{\begin{center}}
\newcommand{\ec}{\end{center}}

\newcommand{\appropto}{\mathrel{\vcenter{
  \offinterlineskip\halign{\hfil$##$\cr
    \propto\cr\noalign{\kern2pt}\sim\cr\noalign{\kern-2pt}}}}}

\newcommand{\HF}{\text{HF}}
\newcommand{\tr}{\text{Tr}}

\newcommand\numberthis{\addtocounter{equation}{1}\tag{\theequation}}

\begin{document}

\title{Applying the Density Matrix Expansion with \texorpdfstring{\\}{}
Coordinate-Space Chiral Interactions}

\author{A. Dyhdalo}
\email{dyhdalo.2@osu.edu}
\affiliation{Department of Physics, The Ohio State University, Columbus, OH 43210, USA}

\author{S.K. Bogner}
\email{bogner@nscl.msu.edu}
\affiliation{National Superconducting Cyclotron Laboratory and Department of Physics and Astronomy,
Michigan State University, East Lansing, MI 48824, USA}

\author{R.J. Furnstahl}
\email{furnstahl.1@osu.edu}
\affiliation{Department of Physics, The Ohio State University, Columbus, OH 43210, USA}

\date{\today}

\begin{abstract}
  We apply the density matrix expansion (DME) at Hartree-Fock level with long-range
  chiral effective field theory interactions defined in coordinate space up to 
  next-to-next-to-leading order.
  We consider chiral potentials both with and without explicit Delta isobars.
  The challenging algebra associated with applying the DME to three-nucleon forces is tamed 
  using a new organization scheme, which will also facilitate generalizations.
  We include local regulators on the interactions to mitigate the effects of
  singular potentials on the DME couplings and simplify the optimization of generalized Skyrme-like functionals.
\end{abstract}

\maketitle

\tableofcontents





\section{Introduction}
\label{sec:intro}

Despite great progress in recent years in ab initio methods, 
solving the quantum many-body problem
for the entire range of nuclei is currently only feasible with
phenomenological energy density functionals (EDFs)~\cite{Bender:2003jk}.
These methods, which are often justified by theorems from
density functional theory,
have favorable computational scaling to large systems 
and can be used to compute observables such as binding energies, 
radii, electromagnetic transitions, beta-decay rates, and 
fission cross sections across the nuclear chart~\cite{Bender:2003jk,Goriely:2016uhb}.
Skyrme functionals~\cite{SKYRME1958615,PhysRevC.5.626} are a type of
EDF utilizing \emph{local} nuclear densities and their gradients, such as would result
from zero-range interactions treated at the Hartree-Fock level.
These functionals are usually supplemented with a
pairing interaction to account for nuclear superfluidity
\cite{Dobaczewski:1983zc,Dobaczewski:2001ed,Bender:2003jk,Dobaczewski:2012xu,Bogner:2013pxa}. 
The functional is specified by of order ten parameters,
which are fitted to a subset of nuclear data.
An extensive infrastructure for optimizing and applying Skyrme EDFs
has been developed \cite{Bennaceur200596,Stoitsov200543,
Stoitsov20131592,DOBACZEWSKI1997166,Schunck2012166}.

The best Skyrme parameterizations have had many phenomenological successes, 
but there are clear limitations.
A systematic way to improve these functionals within the Skyrme framework
has not been demonstrated
and extrapolations to nuclei for which no
data exists is often model-dependent, which hinders uncertainty quantification.
Sophisticated analyses have concluded that with the standard form,
the accuracy of masses and other observables has reached a 
limit~\cite{Schunck:2015a,PhysRevLett.114.122501,Schunck:2015b,Haverinen:2016eop}.
Yet improving the global performance of mass models is desirable to constrain 
nuclear reactions, e.g., for r-process nucleosynthesis~\cite{Mumpower:2015,Mumpower201686}.
In previous work, a program was initiated to address these issues by incorporating
microscopic physics in Skyrme EDFs using an improved version of the
density matrix expansion (DME)~\cite{PhysRevC.82.014305,Gebremariam201117,Stoitsov:2010ha, Bogner:2011kp}.
The idea is that existing functionals may have too simplistic density dependencies 
to account for long-range pion physics, but it can be incorporated with the DME while
still taking advantage of the Skyrme infrastructure.
Here we present a new implementation of the DME in coordinate space using
pion-exchange interactions with local regulators. 

We adopt the organization of pion-range physics given by chiral effective field
theory, \eft, including both nucleon-nucleon ($NN$) and three-nucleon ($3N$)
forces~\cite{Weinberg:1990rz, Weinberg:1991um}. 
\eft~provides a model-independent low-energy expansion of the long-range forces
between nucleons, see Refs.~\cite{Epelbaum:2008ga,
Machleidt:20111,Machleidt:2016} for reviews.
The relevant degrees of freedom are asymptotic nucleon 
states and pions, with delta isobars ($\Delta$s) sometimes added to improve the convergence 
of the expansion~\cite{JENKINS1991353,HEMMERT199789}.  
The resulting chiral potentials have had many successes in recent
calculations of nuclear phenomena~\cite{RevModPhys.87.1067,
Roth:2011ar,Lahde:2013uqa,
Tsukiyama:2012sm,PhysRevLett.113.142502,
PhysRevC.93.051301,Soma:2011aj,
PhysRevC.89.061301,
Hergert:2013uja,PhysRevC.90.041302,
PhysRevC.90.061306,
Lee:2008fa,Hagen:2013nca,
Hergert:2015awm,Barrett:2013nh}.
By including these chiral potentials, we set the stage for
connecting the physics of QCD to calculations of the
full table of nuclides.
In the present work, the physics of the chiral potentials is manifested in a generalized Skyrme EDF
as density-dependent couplings multiplying bilinears and trilinears of the local
densities. These chiral couplings are derived starting from Hartree-Fock in many-body
perturbation theory (MBPT).
The chiral physics will not be explicitly built into pairing terms 
as we expect the pairing channel to be well represented
by contact interactions.

We are not yet ready to ``replace phenomenological models of nuclear structure and
reactions with a well-founded microscopic theory that delivers maximum predictive
power with well-quantified uncertainties''~\cite{Bogner:2013pxa} via the construction
of a fully \textit{ab initio} functional based on model-independent chiral interactions.
Even working with renormalization group softened interactions~\cite{Bogner:2009bt}, 
it is necessary to go to at least second order in MBPT for convergence in infinite matter~\cite{PhysRevC.83.031301}.
Instead, we follow the semi-phenomenological philosophy outlined in 
Refs.~\cite{PhysRevC.82.014305,Gebremariam201117} and implemented in 
Refs.~\cite{Stoitsov:2010ha, Bogner:2011kp}.
The idea is to add in the microscopic pion 
and delta physics from the chiral potentials using MBPT 
\textit{without} including the systematic short-range terms from free-space power counting;%
\footnote{Note that we also do not include the contribution from the intermediate-range \NNLO~$3N$ interaction (see Sec.~\ref{sec:chiral_potentials}).}
instead afterwards a global refit of the Skyrme parameters is performed.
The chiral couplings are parameter-free in the sense that they are frozen while 
the Skyrme contacts are adjusted to data.
Thus this is an intermediate approach between \textit{ab initio} and phenomenology
that seeks to constrain the form and couplings of the functional via the underlying
vacuum $NN$ and $3N$ interactions \cite{Bogner:2008kj,Drut:2009ce}. 
This enables a comparison with conventional Skyrme EDFs to assess the role
of explicit pion and delta physics in nuclear structure.
  
When working with finite-range potentials, Fock energy terms in MBPT consist of 
one-body density matrices (OBDMs),
which are inherently nonlocal objects, including correlations between spatially
separated points.
This nonlocality in the OBDM significantly complicates the iteration and computational
cost of solving the Skyrme Hartree-Fock-Bogoliubov equations.
The DME, first formulated by Negele and Vautherin in Refs.~\cite{Negele:1972zp, Negele:1975zz}, 
provides a general way to map nonlocal functionals into local ones by converting OBDMs
into local densities.
In particular, the nonlocality in the OBDMs arising from the finite-range potentials
is factorized into products of local densities multiplied by density-dependent
couplings.
The expansion is \textit{not} a naive Taylor expansion about the OBDM diagonal but
instead resums certain contributions such that it is an expansion about the
homogeneous nuclear matter limit.
The DME is not uniquely defined; in this paper we adopt the 
phase-space-averaging formulation of 
Ref.~\cite{PhysRevC.82.014305}, which showed dramatic improvements in reproducing the 
vector parts of the OBDM over the original Negele-Vautherin prescription.

The algebraic complexity of applying the DME rapidly increases as one goes from 
$NN$ to $3N$ forces, 
which necessitates a robust yet practical organization scheme.
As part of a previous DME implementation using chiral forces, 
the spin-isospin traces over the \NNLO~three-body diagrams were carried out
in Ref.~\cite{Gebremariam20101167} using symbolic software, with an emphasis on 
analytic derivation of the couplings.
The DME for the chiral potentials was implemented in momentum space and did
not include ultraviolet regulators. 
The resulting functional was then used in pre-optimization tests in
Ref.~\cite{Stoitsov:2010ha}. 
While it showed indications of slight improvements in the reproduction of data, 
there were notable complications with stability and optimization.
An alternative implementation of the DME with chiral potentials is given
in Ref.~\cite{Holt201335}.

We instead derive the couplings in coordinate space, using a new organization
scheme, and numerically perform the final integrals.
The resulting DME algebra is much simpler for the $3N$ potentials. 
Coordinate-space chiral potentials have been recently applied in quantum Monte Carlo
calculations in Refs.~\cite{Gezerlis:2013ipa,
Gezerlis:2014zia,
PhysRevLett.113.192501,
PhysRevLett.116.062501}, and allow the natural
use of coordinate-space regulators.
There are several reasons to take this approach:
\be  
  \item The DME is most naturally formulated as an expansion in coordinate space. 
  Working with a coordinate space interaction simplifies the DME coupling
  calculations, as no Fourier integrals need to be done. 
  \item The use of regulators has been shown to have a significant
  influence on many-body calculations even at the Hartree-Fock 
  level~\cite{PhysRevC.93.024305,
PhysRevC.94.034001}. 
  Using regulators allows for the DME couplings to be less affected 
  by the singular short-distance parts of the chiral potentials.
  \item Coordinate-space interactions allow for the natural implementation of
   local coordinate-space regulators.
   These regulators are thought to suppress more effectively the singular parts of 
   the one-pion-exchange (OPE) tensor and two-pion-exchange (TPE) 
   potentials and have smaller artifacts
   than alternative non-local, momentum-space regulators. 
   Recent calculations imply that the convergence of the chiral expansion for
   certain $NN$ cross sections is more systematic with such
   regulators~\cite{Epelbaum:2014efa}.
   \item 
  By including a regulator and varying the cutoff, we allow for an 
  adiabatically turning on of the finite-range physics.
  This enables a more controlled and stable implementation of the EDF
  optimization, as one can ``boot-strap'' the finite-range forces starting
  from the well-studied parameter space of conventional Skyrme functionals.
\ee

The plan of the paper is as follows.
In Sec.~\ref{sec:Hartree-Fock}, we set up the framework starting from  
Hartree-Fock for two- and three-body forces.
The $NN$ and $3N$ coordinate-space chiral potentials used in this paper, 
along with the chosen regularization scheme, are described in Sec.~\ref{sec:chiral_potentials}. 
The relevant quantities in our EDF, the form of the EDF, and the DME parametrization
are enumerated in Sec.~\ref{sec:edf}.
In Secs.~\ref{sec:DME_NN} and \ref{sec:DME_3N} we perform the DME for $NN$ and 
$3N$ forces, respectively. 
Sample results for the $NN$ and $3N$ DME-derived couplings are shown in Sec.~\ref{sec:results} and 
Sec.~\ref{sec:conclusions} has a summary and outlook for the next steps in
this program.
Detailed derivations, formulas, and mathematica notebooks are collected in appendices.


\section{Hartree-Fock for \texorpdfstring{$NN$}{NN} and  \texorpdfstring{$3N$}{3N} Forces} \label{sec:Hartree-Fock}

The Hartree-Fock energy for an antisymmetrized two-body potential is given by
\beq
  V^{\text{NN}}_{\text{HF}} = \frac{1}{2} \sum_{i j} \la i j | \vnna | i j \ra
  \;,
  \label{eq:bare_hf}
\eeq  
with the sums over occupied orbitals and the antisymmetrized $NN$ interaction and exchange operators defined as
\beq
  \vnna \equiv \vnn \left(1 - P_{\text{12}}\right) \;,
  \qquad
  P_{\text{12}} \equiv P^{\sigma} P^{\tau} P^r \;,
\eeq
where
\beq
  P_{\text{12}}^{\sigma} = \frac{1}{2}
  \left(1 + \sigvec_1 \cdot \sigvec_2 \right) \;,
  \qquad
  P_{\text{12}}^{\tau} = \frac{1}{2}
  \left(1 + \tauvec_1 \cdot \tauvec_2 \right) \;.
\eeq 
By inserting resolutions of the identity,
\beq
  1 = \sum_{\sigma \tau} \int
  d\rvec \, |\rvec \sigma \tau \ra
  \la \rvec \sigma \tau |
  \;,
\eeq
Eq.~\eqref{eq:bare_hf} can be rendered in coordinate space: 
\begin{align*}
  V^{\text{NN}}_{\HF} = \frac{1}{2} \sum_{\sigma\tau} \int \prod_{i=1}^4 
  d\mathbf{r}_i \; 
  \la \rone \; \rtwo | \mathcal{\vnna} | \rthree \; \rfour \ra
  \\  \null \times
  \rho_1 \left(\rthree, \rone \right)
  \, 
  \rho_2 \left(\rfour, \rtwo \right)
  \;.
  \numberthis
  \label{eq:coor_nn_hf}
\end{align*}
The $\rho$ terms are the OBDMs that encapsulate all the information about the Hartree-Fock orbitals,
\beq
  \label{eq:obdm_sum}
  \rho(\rthree, \rone) 
  \equiv
  \sum_i 
  \phi^*_i(\rone)
  \phi_i(\rthree) \;,
\eeq
where the sum runs over the occupied orbitals in the system.
The OBDM subscript 1 (2) in Eq.~\eqref{eq:coor_nn_hf} defines the term to act 
on the first (second) part of the two-body product space.
Extending the formalism to include pairing via Hartree-Fock-Bogoliubov,
the orbital occupation in Eq.~\eqref{eq:obdm_sum} becomes fractional \cite{Bender:2003jk}. 
Note again that we are only including chiral physics in the particle-hole
channel as the pairing channel will be represented by contact interactions.

Switching to relative ($\rvec$) and center-of-mass ($\Rvec$) coordinates  and assuming translational invariance
and locality of the potential, two of the coordinate space integrals can be done.
Omitting the spin-isospin arguments in the OBDMs, Eq.~\eqref{eq:coor_nn_hf} then becomes,
\begin{align*}
  V^{\text{NN}}_{\HF} &=  \frac{1}{2} \tr_\text{1} \tr_\text{2} 
  \int d\Rvec d\rvec \; 
  \la \rvec \sione \sitwo 
  | \vnn | 
  \rvec  \sithree \sifour \ra
\\
  & \quad \null \times
  \bigg[
  \rho_1 \left(\Rvec + \frac{\rvec}{2} \right)
  \; 
  \rho_2 \left(\Rvec - \frac{\rvec}{2} \right)
  -
  \rho_1 \left(\Rvec - \frac{\rvec}{2}, 
  \Rvec + \frac{\rvec}{2} \right)
  \; 
  \rho_2 \left(\Rvec + \frac{\rvec}{2}, 
  \Rvec - \frac{\rvec}{2} \right) P_{12}^{\sigma \tau}
  \bigg] \; ,
  \numberthis
  \label{eq:vhf_energy}
\end{align*}
where
\beq
\rho(\xvec, \xvec) \equiv \rho(\xvec) \;,
\eeq
and the traces $\tr_\text{1}$, $\tr_\text{2}$ are over the spin-isospin parts of our product space. 
The first term in brackets is the direct (Hartree) term while the second 
is the exchange (Fock) term.

As the OBDMs for the Hartree term are diagonal, they are written as products of local
densities multiplying a finite-range potential.  Although it is possible to apply the DME to
the Hartree term as well, in practice error propagation in the self-consistent iteration is
reduced when applying the DME only to the Fock term~\cite{Negele:1975zz,SPRUNG19751}. 
Treating the Hartree term exactly also provides a better reproduction of the full 
Hartree-Fock energy and density fluctuations \cite{Negele:1975zz,Bogner:2011kp}. 
This exact treatment does not cause a significant increase in computational complexity 
when solving for self-consistency~\cite{Negele:1975zz,Bogner:2011kp}.  

The Hartree-Fock energy for a three-body potential is given by
\beq
  \vnnn_{\text{HF}}
  =
  \frac{1}{6}
  \sum_{ijk}
  \la i j k |
  \vnnn
  \left(
  1 + P_{13}P_{12}
  + P_{23} P_{12}
  \right)
  \left(
  1 - P_{12}
  \right)
  | ijk \ra
  \;,
  \label{eq:HF_3N_general}
\eeq
where again the sums are over occupied orbitals, and $\vnnn$ refers to the full three-body force (see Eq.~\eqref{eq:full_3n} below) 
with the various exchange operators accounting for three-body antisymmetry.
Using the symmetry of $\vnnn$ under
subscript interchange, Eq.~\eqref{eq:HF_3N_general} can be rewritten using only one of 
the three-body potentials \cite{Bogner:2008kj},
\beq
  \vnnn_{\text{HF}}
  =
  \frac{1}{2}
  \sum_{ijk}
  \la i j k |
  V_{23} 
  \left(
  1 - 2 P_{12} - P_{23}
  + 2 P_{23} P_{12}
  \right)
  | i j k \ra
  \;.
\eeq
  Therefore, there is one direct
  or Hartree term (H) and two exchange pieces. 
  The sum of the single exchange (SE) and the double exchange (DE) terms give us our Fock (F) term,
\bseq
\begin{align*}
  \vnnn_{\text{H}}
  &\equiv \frac{1}{2}
  \sum_{ijk}
  \la i j k |
  V_{23} | i j k \ra
  \;,
  \numberthis
\\
  \vnnn_{\text{SE, F}}
  &\equiv \frac{1}{2}
  \sum_{ijk}
  \la i j k |
  V_{23}
  (-2 P_{12} - P_{23})
  | i j k \ra
  \;,
  \numberthis
\\
  \vnnn_{\text{DE, F}}
  &\equiv \sum_{ijk}
  \la i j k |
  V_{23} P_{23} P_{12}
  | i j k \ra
  \;,
  \numberthis
\\
  \vnnn_{\text{F}} 
  &\equiv 
  \vnnn_{\text{SE, F}}
  +
  \vnnn_{\text{DE, F}}
  \;.
  \numberthis
\end{align*}
\eseq

  Analogously to the $NN$ sector, for the $3N$ sector
  we again insert completeness relations and work in coordinate space, encapsulating
  information about the Hartree-Fock orbitals in the OBDMs.
  The spin-isospin part of the exchange operator is kept with the interaction while 
  the spatial part of the exchange operator acts on the arguments of the OBDMs.
  Enforcing the locality of our potential and suppressing the spin and isospin 
  arguments in the OBDMs and potential, we find \cite{Gebremariam20101167,Gebremariam:2010},
\bseq
\begin{align*}
  \vnnn_{\text{H}}
  &= 
  \frac{1}{2} \tr_1 \tr_2 \tr_3
  \int d\rvec_1 d\rvec_2 d\rvec_3
  \; \rho_1 (\rvec_1)
  \rho_2 (\rvec_2)
  \rho_3 (\rvec_3)
  \;
  V_{23} (\rvec_{21}, \rvec_{31})
  \;,
  \numberthis
  \label{eq:3n_hartree}
\\
  \vnnn_{\text{SE, F}}
  &=
  {-} \tr_1 \tr_2 \tr_3
  \int d\rvec_1 d\rvec_2 d\rvec_3
  \; \rho_1 (\rvec_2, \rvec_1)
  \rho_2 (\rvec_1, \rvec_2)
  \rho_3 (\rvec_3)
  \;
  V_{23} (\rvec_{21}, \rvec_{31})
  \;
  P^{\sigma \tau}_{12}
  \\
  & \quad \null - \frac{1}{2} \tr_1 \tr_2 \tr_3
  \int d\rvec_1 d\rvec_2 d\rvec_3
  \; \rho_1 (\rvec_1)
  \rho_2 (\rvec_3, \rvec_2)
  \rho_3 (\rvec_2, \rvec_3)
  \;  V_{23} (\rvec_{21}, \rvec_{31})
  \;
  P^{\sigma \tau}_{23}
  \;,
  \numberthis
  \label{eq:3n_fock_se}
\\
  \vnnn_{\text{DE, F}}
  &=
  \tr_1 \tr_2 \tr_3
  \int d\rvec_1 d\rvec_2 d\rvec_3
  \; \rho_1 (\rvec_3, \rvec_1)
  \rho_2 (\rvec_1, \rvec_2)
  \rho_3 (\rvec_2, \rvec_3)
  \;
  V_{23} (\rvec_{21}, \rvec_{31})
  \;
  P^{\sigma \tau}_{23}
  P^{\sigma \tau}_{12}
  \;,
  \numberthis
  \label{eq:3n_fock_de}
\end{align*}
\eseq
where the traces go over the spin and isospin of the three particles. 
For the purposes of performing the DME in the $3N$ system, the above equations
must be converted into a more convenient form.
We instead work with the variables,
\beq
  \rvec_{1} \; , \quad
  \xvec_2 = \rvec_{21} \; , \quad 
  \xvec_3 = \rvec_{31} \; ,
  \label{eq:var_trans}
\eeq
where the Jacobian of the transformation is unity. 
Applying this transformation yields, 
\bseq
\label{eq:3n_hf}
\begin{align*}
  \vnnn_{\text{H}}
  &= 
  \frac{1}{2} \tr_1 \tr_2 \tr_3
  \int d\rvec_1 d\xiav d\xibv
  \; \rho_1 (\rvec_1)
  \rho_2 (\rvec_1 + \xiav)
  \rho_3 (\rvec_1 + \xibv)
  \;
  V_{23} (\xiav, \xibv)
  \;,
  \numberthis
\\
  \vnnn_{\text{SE, F}}
  &=
  {-} \tr_1 \tr_2 \tr_3
  \int d\rvec_1 d\xiav d\xibv
  \; \rho_1 (\rvec_1 + \xiav, \rvec_1)
  \rho_2 (\rvec_1, \rvec_1 + \xiav)
  \rho_3 (\rvec_1 + \xibv)
  \\
   & \qquad\qquad\qquad\qquad\qquad \null \times
  V_{23} (\xiav, \xibv)
  \;
  P^{\sigma \tau}_{12}
  \\
  & \quad\null - \frac{1}{2} \tr_1 \tr_2 \tr_3
  \int d\rvec_1 d\xiav d\xibv
  \; \rho_1 (\rvec_1)
  \rho_2 (\rvec_1 + \xibv, \rvec_1 + \xiav)
  \rho_3 (\rvec_1 + \xiav, \rvec_1 + \xibv)
  \\
   & \qquad\qquad\qquad\qquad\qquad \null \times
  V_{23} (\xiav, \xibv)  
  \;
  P^{\sigma \tau}_{23}
  \;,
  \numberthis
\\
  \vnnn_{\text{DE, F}}
  &=
  \tr_1 \tr_2 \tr_3
  \int d\rvec_1 d\xiav d\xibv
  \; \rho_1 (\rvec_1 + \xibv, \rvec_1)
  \rho_2 (\rvec_1, \rvec_1 + \xiav)
  \rho_3 (\rvec_1 + \xiav, \rvec_1 + \xibv)
  \\
   & \qquad\qquad\qquad\qquad\qquad \null \times
  V_{23} (\xiav, \xibv)
  \;
  P^{\sigma \tau}_{23}
  P^{\sigma \tau}_{12}
  \;.
  \numberthis
\end{align*}
\eseq
As in the $NN$ system, the Hartree term is already diagonal in each of the OBDMs. 
As such the Hartree term is evaluated exactly while DME expansions are
performed on the two exchange terms.



\section{Chiral Potentials}
\label{sec:chiral_potentials}

\begin{figure}[tbh]
  \includegraphics[width=0.95\textwidth]{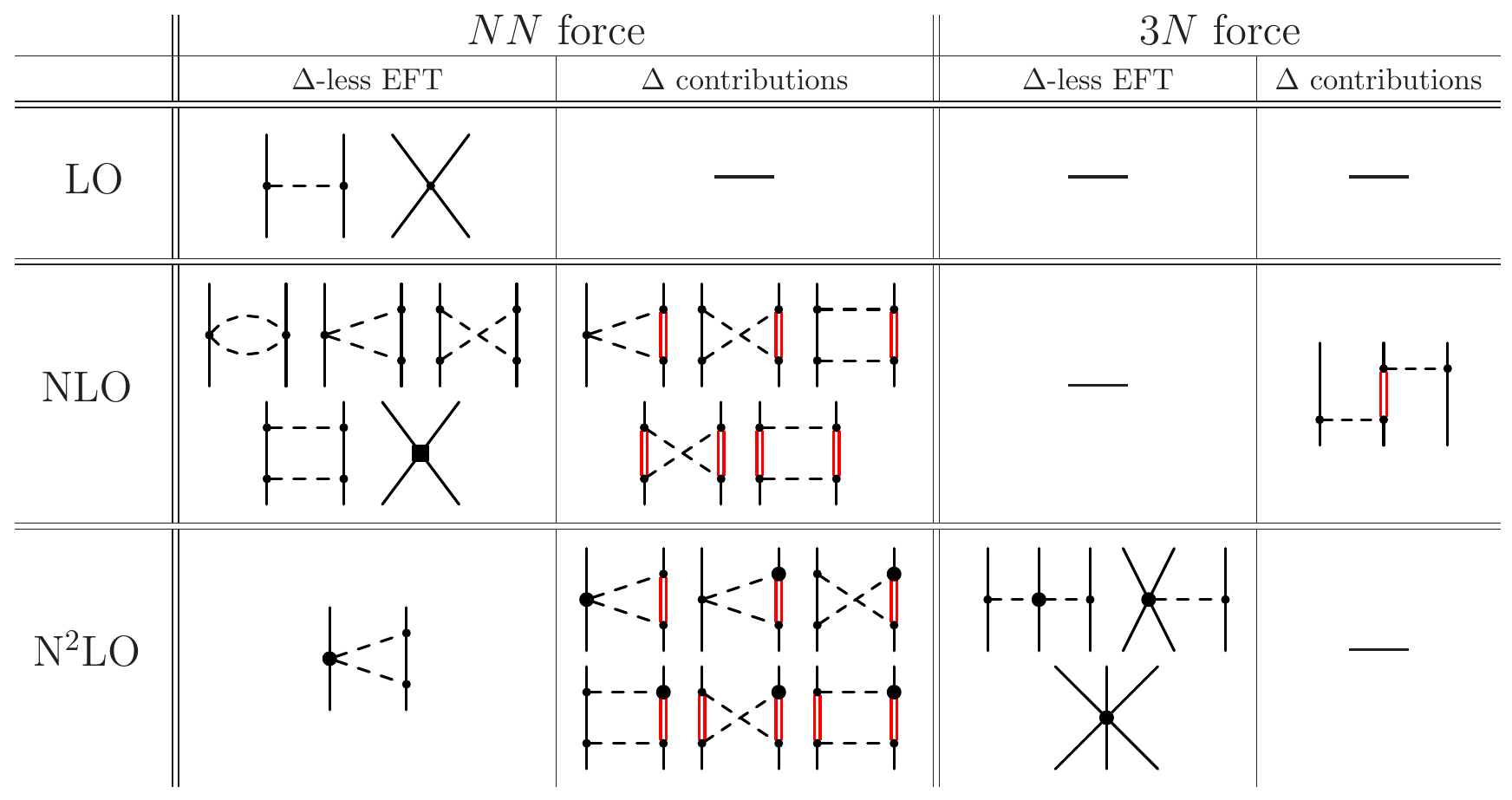}
  \caption{Diagrams in the chiral expansion up to \NNLO~for $NN$ and $3N$ forces. 
  Solid lines are nucleons, dashed lines are pions, and double solid lines (red) are delta isobars.
  Small dots, filled circles, and squares denote vertices of chiral index 0, 1, and 2 respectively \cite{Machleidt:20111}.
  Only one representative example of each type of diagram has been shown.
  In the $\Delta$-less theory, only the diagrams in the two corresponding columns contribute to the potential. 
  In the theory with $\Delta$s, all the columns contribute. 
  }
  \label{fig:delta_chiral}
\end{figure}

We derive DME couplings by working up to \NNLO~in the chiral expansion
including $3N$ forces, both with and without the $\Delta$.
Specially optimized \NNLO~forces~\cite{PhysRevLett.110.192502,PhysRevC.91.051301} 
have recently been used with success in nuclear calculations of medium-mass nuclei
by fine-tuning the
predicted saturation properties through fitting to properties of selected nuclei
up to oxygen. 
This approach is philosophically similar to the global refit we will perform for
our functional after adding in finite-range chiral physics.
The various diagrams that contribute to the chiral expansion
at different orders are displayed in 
Fig.~\ref{fig:delta_chiral}.

Up to NLO, the chiral potentials depend on the pion mass $\mpi$, the $\Delta$-$N$ mass
splitting $\delm$, the pion decay constant $\fpi$, the nucleon axial vector 
coupling $\ga$, the $N$-to-$\Delta$ axial vector coupling $\ha$, and various $NN$ contacts. 
For $\ha$, we use the large $N_C$ value and adopt the convention of Ref.~\cite{PhysRevC.91.024003} 
to define it as $\ha = 3 \ga / \sqrt{2}$. 
Note the factor of 2 difference compared to Ref.~\cite{EPELBAUM200865}.
At \NNLO, the potentials further depend on the subleading LECs $c_1$, $c_2$, $c_3$, and $c_4$
derived from $\mathcal{L}_{\pi N}^{(2)}$, along with the subleading $b_3+b_8$ combination
derived from $\mathcal{L}_{\pi N \Delta}^{(2)}$.
The $c_2$ and $b_3+b_8$ LECs do not contribute to the potential in the $\Delta$-less theory.
The $c_i$ LECs are unnaturally large in the $\Delta$-less theory due to absorbing
the contribution of the $\Delta$ isobar \cite{BERNARD1997483}. 
It is well known that this property weakens the NLO potential and causes the 
\NNLO~potential to be unnaturally strong.
The same effect happens for the $3N$ potential where the \NNNLO~contribution is small but the \NNNNLO~term is large \cite{Machledit:2010}.
Including $\Delta$ isobars in the theory explicitly makes the $c_i$ terms more natural,
shifts more physics to NLO (\NNNLO) for the $NN$ ($3N$) potential, and improves the convergence of the chiral expansion.
At \NNLO, the $3N$ potential additionally depends on an OPE contact term $c_D$ and a $3N$ contact $c_E$.

\begin{figure}[tbh]
  \includegraphics[scale=0.85]{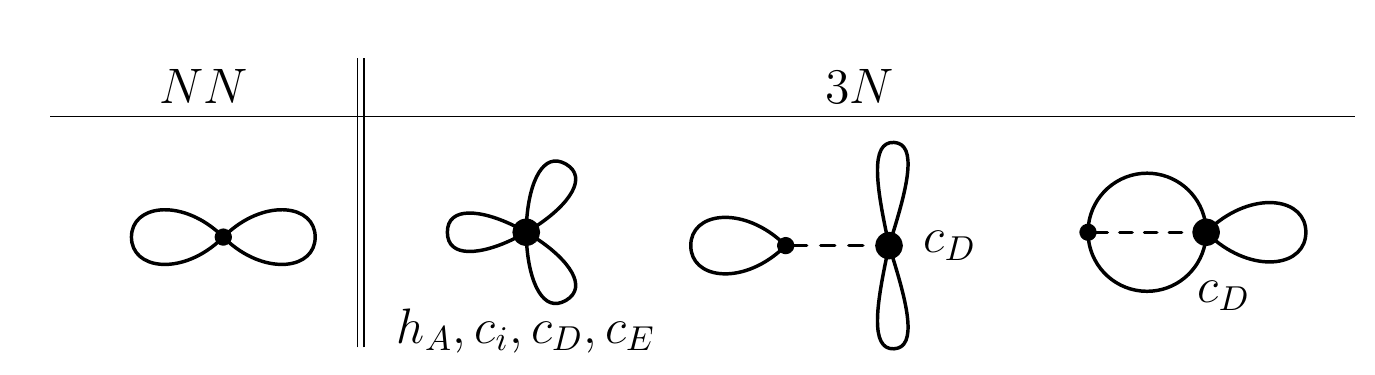}
  \caption{Diagrams given above correspond to the Hartree-Fock topologies absent in our DME couplings with all $3N$ vertices explicitly labelled. Notation is as in Fig.~\ref{fig:delta_chiral}. Diagrams without a dashed line have no finite-range radial function.
  }
  \label{fig:discarded_diagrams}
\end{figure}

In this paper, we \textit{only} consider the finite-range part of the chiral potentials,
with short-range contributions to be absorbed into a refit of Skyrme parameters. 
We omit from the DME couplings all pure contact terms in the coordinate space chiral 
potentials that carry Dirac $\delta$ 
functions in the relative distance variables. 
For the $NN$ case, these contributions have the same structures at Hartree-Fock
as the Skyrme contacts of the EDF.  
Due to the presence of two relative distance variables in the $3N$ case, contributions are classified into long-range (LR) parts with $c_i$ or $\ha$ vertices and no Dirac $\delta$ functions, intermediate-range (IR) parts with $c_i$, $\ha$ or $c_D$ vertices and one Dirac $\delta$ function, and short-range (SR) parts with $c_i$, $\ha$, $c_D$ or $c_E$ vertices and two Dirac $\delta$ functions.
Our DME couplings include contributions from all of the LR terms along with all $c_i$ and $\ha$ IR terms.
For the $3N$ case, there is not an exact correspondence between the omitted IR $c_D$ terms and the EDF Skyrme contacts but for the present work we assume that these contributions can be approximately absorbed in the global refit.
The omitted contributions to the DME couplings are shown in Fig.~\ref{fig:discarded_diagrams} with the $3N$ vertices explicitly labelled.
In Secs.~\ref{sec:nn_forces} and \ref{sec:3N_forces} we give the unregulated forms 
of the $NN$ and $3N$ potentials in coordinate space.
Regularization is then briefly discussed in Sec.~\ref{sec:regularization}.

\subsection{\texorpdfstring{$NN$}{NN} Forces}
\label{sec:nn_forces}

The coordinate space vacuum $NN$ potential can be conveniently decomposed in 
spin-isospin space in a purely local form. 
Up to \NNLO~in the chiral expansion, the $NN$ potential can be written as scalar functions of the relative distance $r$ multiplying various spin-isospin operators:
\begin{align*}
  V^{NN} (\rvec, \{\sigma \tau \}) &= \left[
  V_C(r) + W_C(r) \; \tauvec_1 \cdot \tauvec_2
  \right]
\\
  & \quad\null + \left[
  V_S(r) + W_S(r) \; \tauvec_1 \cdot \tauvec_2
  \right] \sigvec_1 \cdot \sigvec_2
\\
  & \quad\null + \left[
  V_T(r) + W_T(r) \; \tauvec_1 \cdot \tauvec_2
  \right] S_{12} (\rhat) \;,
  \numberthis
  \label{eq:nn_potential}
\end{align*}
where  $r \equiv |\rvec|$, $S_{12} (\rhat)$ is the usual tensor operator,
\beq  
S_{12} (\rhat) = 3 (\sigvec_1 \cdot \rhat) (\sigvec_2 \cdot \rhat) - \sigvec_1 \cdot \sigvec_2 \;,
\eeq
and $\sigvec_i$ ($\tauvec_i$) is the spin (isospin) operator for particle $i$. 
The isoscalar and isovector form factors, $V_i$ and $W_i$ respectively, have 
central ($C$), spin ($S$), and tensor ($T$) components.
Thus, in the operator basis of Eq.~\eqref{eq:nn_potential} for the NN potential,
we only need to consider three spin operator structures,
\bseq
\begin{align*}
  \mathcal{J}_1 &= 1 \; ,
  \numberthis
\\
  \mathcal{J}_2 &= \sigvec_1 \cdot \sigvec_2 \; ,
  \numberthis
\\
  \mathcal{J}_3 &= S_{12} (\rhat) \; ,
  \numberthis
\end{align*}
\label{eq:nn_spin_ops}
\eseq
and two isospin operators,
\bseq
\begin{align*}
  \mathcal{K}_1 &= 1 \; ,
  \numberthis
\\
  \mathcal{K}_2 &= \tauvec_1 \cdot \tauvec_2
  \numberthis 
  \; .
\end{align*}
\label{eq:nn_iso_ops}
\eseq

Deriving $V^{NN}$ from \eft~using Weinberg power counting, the long-range form factors 
at LO are given, with $x \equiv r \mpi$ by~\cite{PhysRevC.91.024003},%
\footnote{Note that the potentials in Ref.~\cite{PhysRevC.91.024003} use $\Fpi = 2 \fpi = 184.80$ MeV.}
\bseq
\begin{align*}
  W_S^{(0)}(r) &= \frac{\mpi^3}{12 \pi} 
  \left(\frac{g_A}{2 \fpi} \right)^2
  \frac{e^{-x}}{x} \;,
  \numberthis
\\
  W_T^{(0)}(r) &= \frac{\mpi^3}{12 \pi} 
  \left(\frac{g_A}{2 \fpi} \right)^2
  \frac{e^{-x}}{x} 
  \bigg(1 + \frac{3}{x} + \frac{3}{x^2}
  \bigg) \; ,
  \numberthis
\end{align*}
\eseq
which is the familiar OPE potential without the Dirac $\delta$ function.

The form factors at NLO including only nucleons and pions are given by \cite{PhysRevC.91.024003},
\bseq
\begin{align*}
  W_C^{(2)}(r) &= \frac{\mpi^5}{8 \pi^3 (2 \fpi)^4} 
  \frac{1}{x^4} 
  \Bigl\{ 
  x \left[ 1 + 10 \ga^2 - \ga^4 (23 + 4 x^2)
  \right]  K_0(2x) 
  \\
  & \quad\null + \left[
  1 + 2 \ga^2(5 + 2x^2) - \ga^4 (23 + 12x^2)
  \right] K_1(2x)
  \Bigr\}
  \;,
  \numberthis
\\
  V_S^{(2)}(r) &= 
  \frac{\mpi^5}{2 \pi^3}  
  \left( \frac{\ga}{2 \fpi}\right)^4
  \frac{1}{x^4}
  \left[
  3x K_0(2x) + (3 + 2 x^2) K_1(2x)  
  \right]
  \;,
  \numberthis
\\ 
  V_T^{(2)}(r) &= 
  {-} \frac{\mpi^5}{8 \pi^3}
  \left( \frac{\ga}{2 \fpi}\right)^4
  \frac{1}{x^4}
  \left[
  12 x K_0 (2x) + (15 + 4 x^2) K_1(2x)
  \right]
  \;,
  \numberthis
\end{align*}
\eseq
where $K_0(x)$ and $K_1(x)$ are modified Bessel functions of the second kind.
The form factors at \NNLO\ including only nucleons and pions are given by \cite{PhysRevC.91.024003},
\bseq
\begin{align*}
  V_C^{(3)}(r) &= 
  \frac{3}{2}
  \frac{\ga^2 \mpi^6}
  {(2 \fpi)^4 \pi^2}
  \frac{e^{-2x}}{x^6}
  \left[
  2 c_1 x^2 (1 + x)^2 
  + 
  c_3 (6 + 12x + 10 x^2 + 4 x^3 + x^4)
  \right]
  \;,
  \numberthis
\\
  W_S^{(3)}(r) &= 
  \frac{1}{3}
  \frac{\ga^2 \mpi^6}{(2\fpi)^4 \pi^2}
  \frac{e^{-2x}}{x^6} 
  c_4
  (1 + x)
  (3 + 3x + 2 x^2)
  \;,
  \numberthis
\\ 
  W_T^{(3)}(r) &= {-}
  \frac{1}{3}
  \frac{\ga^2 \mpi^6}{(2\fpi)^4 \pi^2}
  \frac{e^{-2x}}{x^6}
  c_4
  (1+x)(3 + 3x + x^2)
  \; .
  \numberthis
\end{align*}
\eseq
The potentials given above need to be regulated to tame the short-distance singularities before being inserted in the Schr\"odinger equation 
(see Sec.~\ref{sec:regularization}). 
We emphasize again that the potentials above correspond \textit{only} to the 
long-range chiral potentials (excluding terms with Dirac $\delta$ functions); short-range contact terms in the EFT have not been included.
  
When including explicit delta isobars in the EFT, additional diagrams appear at 
both NLO and \NNLO~consisting of single and double delta excitations.
At both NLO and \NNLO, these additional parts of the potential contribute 
to all 6 form factors given in Eq.~\eqref{eq:nn_potential}.
The explicit form of these potentials with $\Delta$s are given in Ref.~\cite{PhysRevC.91.024003} 
and are written in Appendix~\ref{sec:chiral_appendix} for completeness.

\subsection{\texorpdfstring{$3N$}{3N} Forces}
\label{sec:3N_forces}

As in the $NN$ sector, the $3N$ potentials here will be purely local and 
defined in coordinate space. A general local three-body force will include permutations with respect to three different subsystems,
\beq
  \vnnn = V_{12} + V_{23} + V_{13}
  \;,
  \label{eq:full_3n}
\eeq
where a potential $V_{ij}$ will depend in general on two relative distances related 
to the choice of subscripts and the spin-isospin of the three particles.
That is,
\beq
  V_{ij} \equiv
  V (\rvec_{ik}, \rvec_{jk}, 
  \sigma_1, \tau_1,
  \sigma_2, \tau_2,
  \sigma_3, \tau_3) 
  \;, \quad
  \rvec_{ab} \equiv \rvec_a - \rvec_b \;.
\eeq
It is only necessary to include one of the three terms 
in Eq.~\eqref{eq:full_3n} due to the symmetry of our three-body potentials under 
subscript interchange, see Sec.~\ref{sec:Hartree-Fock}.
As such, in the $3N$ potentials given below, we arbitrarily choose the 
$V_{23} (\rvec_{21}, \rvec_{31}, \{\sigma \tau\})$ piece such that our potentials 
will have no direct dependence on the relative distance coordinate $\rvec_{23}$. 
In anticipation of the $3N$ potentials below, we define the following dimensionless 
functions \cite{Tews:2015}:
\bseq
\begin{align*}
  Y(r) &\equiv \frac{\exp{\left[-\mpi r \right]}}{\mpi r}
  \;, \numberthis
\\
  U(r) &\equiv 1 + \frac{1}{\mpi r}
  \;, \numberthis
\\
  T(r) &\equiv 1 + \frac{3}{\mpi r} + \frac{3}{\left( \mpi r \right)^2}
  \;, \numberthis
\end{align*}
\eseq
denoting the Yukawa function $Y(r)$, scalar function $U(r)$, and singular tensor 
function $T(r)$.

The $3N$ force at NLO is zero in the $\Delta$-less case but has a Fujita-Miyazawa 
term when including explicit $\Delta$s \cite{Fujita01031957,VanKolck:1994yi,EPELBAUM200865},
\beq
  V_{3N}^{(2)} = 
  \alpha_1^{(2)} V_{C,1}
  +
  \alpha_2^{(2)} V_{C,2}
  +
  \alpha_3^{(2)} V_{C,3}
  \;,
\eeq
with both $\alpha_i^{(2)}$ and $V_{C,i}$ specified below.
Conveniently, the potential here has the same structure as the TPE term appearing at \NNLO.
At \NNLO~the $3N$ force has the structure \cite{EPELBAUM200865,VanKolck:1994yi,Epelbaum:2002vt}: 
\beq
  V_{3N}^{(3)} = 
  \alpha_1^{(3)} V_{C,1}
  +  
  \alpha_2^{(3)} V_{C,2}
  +
  \alpha_3^{(3)} V_{C,3}
  +
  V_{D}
  +
  V_{E}
  \;,
\eeq
where the $\alpha_i^{(3)} V_{C,i}$ are TPE terms, $V_D$ is an OPE term, and $V_E$ is a $3N$ contact.
As mentioned already, for the purposes of calculating the DME couplings up 
to \NNLO, only the LR and IR $V_{C,i}$ terms (Fujita-Miyazawa and TPE) are included.
The SR $V_{C,i}$ terms as well as the $V_{D}$ and $V_{E}$ potentials
are assumed to be approximately accounted for by the re-fit Skyrme 
terms, so we do not list here these operator structures. 

  The NLO prefactors for the $V_C$ terms are given by \cite{EPELBAUM200865}:
\beq
  \alpha_1^{(2)} \equiv
  0 \; , 
  \quad
  \alpha_2^{(2)} \equiv
  - \frac{\ha^2 \mpi^6 \ga^2}{2592 \fpi^4 \pi^2 \delm}
  \; , 
  \quad
  \alpha_3^{(2)} \equiv
  \frac{\ha^2 \mpi^6 \ga^2}{10368 \fpi^4 \pi^2 \delm}
  \; ,
\eeq
while the \NNLO~$V_C$ prefactors are~\cite{EPELBAUM200865}
\beq
  \alpha_1^{(3)} \equiv
  \frac{c_1 \mpi^6 \ga^2}{16 \fpi^4 \pi^2}
  \; , 
  \quad
  \alpha_2^{(3)} \equiv
  \frac{c_3 \mpi^6 \ga^2}{288 \fpi^4 \pi^2}
  \; , 
  \quad
  \alpha_3^{(3)} \equiv
  \frac{c_4 \mpi^6 \ga^2}{576 \fpi^4 \pi^2}
  \;.
\eeq

For the purposes of performing the DME, it is convenient to organize the $V_C$ operator structures. 
First we enumerate the spin-isospin structures.
The full $V_{C,i}$ expressions including all contacts are given in Appendix~\ref{sec:chiral_appendix}.
The $V_{C,1}$ potential only has one spin operator, which is a tensor-like term,
\begin{align*}
  \label{eq:nnn_spin_op_1}
  \mathcal{S}_{1} &\equiv (\sigvec_2 \cdot \xia)
  (\sigvec_3 \cdot \xib) \; ,
  \numberthis
\end{align*}
where we have used the variable transformation in Eq.~\ref{eq:var_trans}.
The $V_{C,2}$ potential has one spin-spin term and various other tensor terms,
\bseq
\label{eq:nnn_spin_op_2}
\begin{align*}
  \mathcal{S}_{2} &\equiv \sigvec_2 \cdot \sigvec_3 \; ,
  \numberthis
\\
  \mathcal{S}_{3} &\equiv (\sigvec_2 \cdot \xia)
  (\sigvec_3 \cdot \xia) \; ,
  \numberthis
\\
  \mathcal{S}_{4} &\equiv (\sigvec_2 \cdot \xib)
  (\sigvec_3 \cdot \xib) \; ,
  \numberthis
\\
  \mathcal{S}_{5} &\equiv (\sigvec_2 \cdot \xia)
  (\sigvec_3 \cdot \xib) (\xia \cdot \xib) \; .
  \numberthis
\end{align*}
\eseq
The final $V_{C,3}$ potential has various spin cross products, 
\bseq
\label{eq:nnn_spin_op_3}
\begin{align*}
  \mathcal{S}_{6} &\equiv \sigvec_1 \cdot (\sigvec_2 \times \sigvec_3) \; ,
  \numberthis
\\
  \mathcal{S}_{7} &\equiv (\sigvec_2 \cdot \xia) 
  \; \xia \cdot (\sigvec_3 \times \sigvec_1) \; ,
  \numberthis
\\
  \mathcal{S}_{8} &\equiv(\sigvec_3 \cdot \xib) 
  \; \xib \cdot (\sigvec_1 \times \sigvec_2) \; ,
  \numberthis
\\
  \mathcal{S}_{9} &\equiv (\sigvec_2 \cdot \xia) (\sigvec_3 \cdot \xib) 
  \; 
  \sigvec_1 \cdot (\xia \times \xib) 
  \; .
  \numberthis
\end{align*}
\eseq
There are only two isospin operators, one for the $V_{C,1}$, $V_{C,2}$ potentials 
and one for the $V_{C,3}$ potential,
\bseq
\label{eq:nnn_iso_op_1}
\begin{align*}
  \mathcal{T}_1 &\equiv \tauvec_2 \cdot \tauvec_3 \; ,
  \numberthis
\\
  \mathcal{T}_2 &\equiv \tauvec_1 \cdot (\tauvec_2 \times \tauvec_3) \; .
  \numberthis
\end{align*}
\eseq

The $V_{C,i}$ operator structures can then be written in 
terms of $\mathcal{S}$ and $\mathcal{T}$ \cite{Tews:2015,PhysRevC.85.024003},
\begin{align*}
  V_{C,1} &= 
  \mathcal{T}_1 
  \left\{
  \begin{array}{lr}
  \mathcal{S}_{1}  &
  \mathcal{Y}_1
  \end{array}
  \right\} \; ,
  \numberthis
  \label{eq:vc1}
\end{align*}
\begin{align*}
  V_{C,2}  
  &= \mathcal{T}_1 
  \left\{\begin{array}{lrl}
  & \mathcal{S}_{2}  & 
  \bigl(\mathcal{Y}_2 
  + \Y_{c,2} (x_2) \delta^3(\xvec_3) 
  + \Y_{c,2} (x_3) \delta^3(\xvec_2) \bigl)
\\
  + & \mathcal{S}_{3}  & 
  \bigl(\mathcal{Y}_3
  + \Y_{c,3} (x_2) \delta^3(\xvec_3) \bigl)
\\
  + & \mathcal{S}_{4}  & 
  \bigl(\mathcal{Y}_4
  + \Y_{c,4} (x_3) \delta^3(\xvec_2) \bigl)
\\
  + & \mathcal{S}_{5}   &
  \mathcal{Y}_5
  \end{array}
  \right\} \; ,
  \numberthis
  \label{eq:vc2}
\end{align*}
\begin{align*}
  V_{C,3}  
  &= \mathcal{T}_2 
  \left\{ \begin{array}{lrl}
  & \mathcal{S}_{6}  & 
  \bigl( \mathcal{Y}_6
  + \Y_{c,6} (x_2) \delta^3(\xvec_3)
  + \Y_{c,6} (x_3) \delta^3(\xvec_2) \bigl)
\\
  + & \mathcal{S}_{7}  & 
  \bigl( \mathcal{Y}_7
  + \Y_{c,7} (x_2) \delta^3(\xvec_3) \bigl)
\\
  + & \mathcal{S}_{8}  & 
  \bigl( \mathcal{Y}_8
  + \Y_{c,8} (x_3) \delta^3(\xvec_2) \bigl)
\\
  + & \mathcal{S}_{9}  & 
  \mathcal{Y}_9
  \end{array}
  \right\} \; ,
  \numberthis
  \label{eq:vc3}
\end{align*}
where the $\Y$ functions describe the radial dependence of the long-range physics associated with each spin operator. 
  The $\Y_i (x_2, x_3)$ functions are explicitly given by
\bseq
\label{eq:y_functions}
\begin{align*}
  \mathcal{Y}_1 &=
  U(x_2) Y(x_2)
  U(x_3) Y(x_3) \; ,
  \numberthis  
\\
  \mathcal{Y}_2 =  
  \Y_6 &=
  \left[
  1 - T(x_2)
  \right]
  \left[
  1 - T(x_3)
  \right]
  Y(x_2) Y(x_3) \; ,
  \numberthis
\\
  \mathcal{Y}_3 = 
  \Y_7 &=
  3 T(x_2)
  \left[
  1 - T(x_3)
  \right]
  Y(x_2) Y(x_3) 
  \; ,
  \numberthis
\\
  \mathcal{Y}_4 = 
  \Y_8 &=
  3 T(x_3)
  \left[
  1 - T(x_2)
  \right]
  Y(x_2) Y(x_3)
  \; ,
  \numberthis
\\
  \mathcal{Y}_5 =
  \Y_9 &= 
  9 Y(x_2) Y(x_3)
  T(x_2) T(x_3) \; .
  \numberthis
\end{align*}
\eseq
The $\Y_{c,i} (x)$ functions multiplying the contacts are given by,
\bseq
\label{eq:yc_functions}
\begin{align*}
	\Y_{c,2} 
	= \Y_{c,6}	
	&=  
	- \frac{4 \pi}{\mpi^3}
	(1 - T(x) )Y(x) \; ,
	\numberthis
\\
	\Y_{c,3} = \Y_{c,4} 
	= \Y_{c,7} = \Y_{c,8}	
	&=
	- \frac{12 \pi}{\mpi^3}
	T(x) Y(x) \; .
	\numberthis
\end{align*}
\eseq
Note that in all cases, we have matched the index of the $\Y$ functions to the associated spin operator for clarity.
For later use, we also define $\B$ functions that incorporate the chiral prefactor 
associated with each potential,
\beq
  \B_1 \equiv \Y_1 \af \; ,
  \qquad
  \B_i \equiv \Y_i \as \quad i \in \{2,3,4,5\} \; ,
  \qquad
  \B_i \equiv \Y_i \at \quad i \in \{6,7,8,9\} \; ,
  \label{eq:b_functions}
\eeq
\beq
  \B_{c,i} \equiv \Y_{c,i} \as \quad i \in \{2,3,4\} \; ,
  \qquad
  \B_{c,i} \equiv \Y_{c,i} \at \quad i \in \{6,7,8\} \; .
  \label{eq:bc_functions}
\eeq

\subsection{Regularization}
\label{sec:regularization}

The potentials above are calculated perturbatively from the chiral expansion 
with all divergences usually regulated with dimensional regularization.
However, when they are iterated in the Schr\"odinger or Lippmann-Schwinger
equation, an additional cutoff regularization scheme is required.
For our $NN$ potentials, each form factor $V_i(r)$, $W_i(r)$ is multiplied by a 
regulator function $f(r/R)$, where $R$ is the cutoff~\cite{Epelbaum:2014efa},
\beq
  \label{eq:regulator}
  f\left(\frac{r}{R}\right) = \left[
  1 - \exp \left( 
  - \frac{r^2}{R^2}
  \right)
  \right]^n
  \;.
\eeq
This ensures that the long-range physics ($r \gg R$) is unsuppressed while the 
short-range parts of the potential ($r \ll R$) are cut off. 
This is only one possible form of the regulator in coordinate space, e.g., 
see Refs.~\cite{Gezerlis:2013ipa,Gezerlis:2014zia}.
Typical values used are $R = 0.8\mbox{--}1.2\,\fm$ and $n=4\mbox{--}6$.
For $3N$ potentials, we use the same regulator function,
including it as a part of each Yukawa function $Y(r)$.
So after regularization, the finite-range chiral potentials are modified in the following manner:
\bseq
\begin{align*}
  V_i (r)
  &\xrightarrow{\text{reg.}}
  V_i (r) \;
  f\left(\frac{r}{\RNN}\right) 
  &&\qquad \text{for } NN
  \;,
  \numberthis
\\
  W_i(r)  
  &\xrightarrow{\text{reg.}}
  W_i(r) \;
  f\left(\frac{r}{\RNN}\right) 
  &&\qquad \text{for } NN
  \;,
  \numberthis
\\
  Y(r) 
  &\xrightarrow{\text{reg.}} 
  Y(r) \; 
  f\left(\frac{r}{\RTN}\right)
  &&\qquad \text{for } 3N 
  \;,
  \numberthis
\end{align*}
\eseq
where we allow for the $NN$ and $3N$ potentials to have two different values for 
the cutoff, $\RNN$ and $\RTN$, respectively. 



\section{EDF Technology}
\label{sec:edf}

\subsection{Local Densities}
\label{sec:local_densities}

Here we define the basic variables we will be working with in our Skyrme-like EDF.
The OBDM can be decomposed into scalar-isoscalar, scalar-isovector, vector-isoscalar, 
and vector-isovector channels respectively \cite{PhysRevC.69.014316,PhysRevC.81.014313}:
\begin{align*}
  \rho (\xvec, \yvec)
  &=
  \frac{1}{4}
  \Bigl[
  \rho_0 (\xvec, \yvec)
  +
  \rho_1 (\xvec, \yvec) \tau^z
  +
  \svec_0(\xvec, \yvec) \cdot \sigvec
  +
  \svec_1 (\xvec, \yvec)  \cdot \sigvec
  \tau^z
  \Bigr]
\\
  &= \frac{1}{4}
  \Bigl[
  \rho_t (\xvec, \yvec)
  + 
  \svec_t (\xvec, \yvec)
  \cdot \sigvec
  \Bigr]
  \Bigl[
  \delta^{t,0} 
  + \delta^{t,1} \tau^z
  \Bigr]
  \;,
  \numberthis
  \label{eq:obdm_decomposed}
\end{align*}  
where, by using $\tau_z$ instead of $\tauvec$, we have assumed the OBDM is diagonal in isospin space.
  For time-reversal invariant systems, the scalar and vector OBDMs have the symmetry:
\beq
  \rho_t(\xvec, \yvec) = 
  \rho_t(\yvec, \xvec)
  \; ,
  \qquad
  \svec_t(\xvec, \yvec) =
  - \svec_t(\yvec, \xvec) \; .
  \label{eq:tr_symmetry}
\eeq  
  
	Our functional is built from a set of local densities including up to two derivatives \cite{Bender:2003jk}.
Enumerating the different possibilities with derivatives acting on the scalar 
$\rho (\rvec, \rvec')$ or vector $\svec (\rvec, \rvec')$ part of the OBDM, we get:
\bseq
\label{eq:local_densities}
\begin{align*}
  \rho_t (\rvec) &= \rho_t (\rvec, \rvec') |_{\rvec = \rvec'}
  \; ,
  \numberthis
  \label{eq:particle_density}
\\
  s_{a,t} (\rvec) &= s_{a,t} (\rvec, \rvec') |_{\rvec = \rvec'}
  \; ,
  \numberthis
  \label{eq:spin_density}
\\
  \tau_t (\rvec) &= \nabla \cdot \nabla' \rho_t (\rvec, \rvec') |_{\rvec = \rvec'}
  \; ,
  \numberthis
  \label{eq:kinetic_density}
\\
  T_{a,t} (\rvec) &= \nabla \cdot \nabla' s_{a,t} (\rvec, \rvec') |_{\rvec = \rvec'}
  \; ,
  \numberthis
  \label{eq:spinkinetic_density}
\\
  j_{a,t} (\rvec) &= {-}\frac{i}{2} (\nabla_a - \nabla'_a) \rho_t (\rvec, \rvec')  |_{\rvec = \rvec'}
  \; ,
  \numberthis
  \label{eq:current_density}
\\
  J_{ab,t} (\rvec) &= {-}\frac{i}{2} (\nabla_a - \nabla'_a) s_{b,t} (\rvec, \rvec')  |_{\rvec = \rvec'}
  \; ,
  \numberthis
  \label{eq:spincurrent_density}
\end{align*}
\eseq
where we have defined the matter density $\rho$, spin density $\svec$, 
kinetic density $\tau$, spin-kinetic density $\Tvec$, 
current density $\jvec$, and spin-current density $\Jvec$.
The subscript $t$ denotes either isoscalar ($t=0$) or isovector ($t=1$) densities.
Isoscalar and isovector quantities are defined to be either the sum or difference 
of neutron and proton densities, i.e.,
\beq
  \rho_0 (\rvec) \equiv \rho_n (\rvec) + \rho_p (\rvec) \;,
  \quad
  \rho_1 (\rvec) \equiv \rho_n (\rvec) - \rho_p (\rvec) \;.
\eeq  
Note that $\rho$, $\tau$, and $\Jvec$ are time-even while 
$\svec$, $\Tvec$, and $\jvec$ are time-odd.
Here we restrict ourselves to time-reversal invariant systems such that 
all time-odd densities vanish. 
Therefore, for a first application, our results will only be applicable to even-even nuclei, 
which have time-reversal symmetry. 

\subsection{EDF Form}

The finite-range physics associated with pion exchange and delta excitations are 
encoded as density dependent couplings $g(\Rvec)$ in our EDF.
The couplings $g(\Rvec)$ multiply various products of local densities and are 
added to the standard Skyrme functionals such that, e.g.,
\beq
  U^{\rho \rho}_t \equiv  
  g^{\rho \rho}_t(\Rvec)
  + C^{\rho \rho}_t
  \;,
\eeq
where $U$ is the new coupling term and $C$ is a standard term appearing in the Skyrme functional.
For an $NN$ potential given by Eq.~\eqref{eq:nn_potential}, 
the EDF that results after performing the DME for the Fock term 
will consist of 12 bilinears of local densities with the form:
\begin{align*}
  V_{\text{F}} \approx \sum_{t=0}^1 
  \int d\Rvec \; 
  & g^{\rho \rho}_t \rho_t \rho_t
  + g^{\rho \tau}_t \rho_t \tau_t
  + g^{\rho \Delta \rho}_t \rho_t \Delta \rho_t
\\
  & \null + g^{JJ,1}_t J_{t,aa} J_{t,bb}
  + g^{JJ,2}_t J_{t,ab} \; J_{t, ab}
  + g^{JJ,3}_t J_{t,ab} \; J_{t, ba}
  \;.
  \numberthis
  \label{eq:nn_edf}
\end{align*}
This is the same form as the EDF given in Ref.~\cite{PhysRevC.82.014305}, modulo the spin-orbit contributions arising from the short-range $NN$ contact interaction. For our $3N$ potentials, the resulting EDF for the Fock term will consist of 23
trilinears of local densities with the form:
\begin{align*}
  V_{\text{F}} \approx& \ \int d\Rvec \;
  g^{\rho_0^3} \rho_0^3 
  + 
  g^{\rho_0^2 \tau_0} \rho_0^2 \tau_0 
  +
  g^{\rho_0^2 \Delta \rho_0}\rho_0^2 \Delta \rho_0 
  +
  g^{\rho_0 (\nabla \rho_0)^2}
  \rho_0 \nabla \rho_0 \cdot \nabla \rho_0 
  +
  g^{\rho_0 \rho_1^2}\rho_0 \rho_1^2
\\
  & \quad\null + 
  g^{\rho_1^2 \tau_0} \rho_1^2 \tau_0
  +
  g^{\rho_1^2 \Delta \rho_0}
  \rho_1^2 \Delta \rho_0
  +
  g^{\rho_0 \rho_1 \tau_1}
  \rho_0 \rho_1 \tau_1
  +
  g^{\rho_0 \rho_1 \Delta \rho_1}
  \rho_0 \rho_1 \Delta \rho_1
  +
  g^{\rho_0 (\nabla \rho_1)^2}
  \rho_0 \nabla \rho_1 \cdot \nabla \rho_1
\\
  & \quad\null + 
  \rho_0 \epsilon_{ijk}
  \left[
  g^{\rho_0 \nabla \rho_0 J_0}
  \nabla_i \rho_0 J_{0,jk}
  +
  g^{\rho_0 \nabla \rho_1 J_1}
  \nabla_i \rho_1 J_{1,jk}
  \right]
\\
  & \quad\null + 
  \rho_1 \epsilon_{ijk}
  \left[
  g^{\rho_1 \nabla \rho_1 J_0}
  \nabla_i \rho_1 J_{0,jk}
  +
  g^{\rho_1 \nabla \rho_0 J_1}
  \nabla_i \rho_0 J_{1,jk}
  \right]
\\
  & \quad\null + 
  \rho_0 \left[
  g^{\rho_0 J_0^2, 1}
  J_{0,aa} J_{0,bb}
  +
  g^{\rho_0 J_0^2, 2}
  J_{0,ab} J_{0,ab} 
  +
  g^{\rho_0 J_0^2, 3}
  J_{0,ab} J_{0,ba}
  \right]
\\
  & \quad\null + 
  \rho_0 \left[
  g^{\rho_0 J_1^2, 1}
  J_{1,aa} J_{1,bb}
  +
  g^{\rho_0 J_1^2, 2}
  J_{1,ab} J_{1,ab} 
  +
  g^{\rho_0 J_1^2, 3}
  J_{1,ab} J_{1,ba}
  \right]
\\
  & \quad\null + 
  \rho_1 \left[
  g^{\rho_1 J_0 J_1, 1}
  J_{1,aa} J_{0,bb}
  +
  g^{\rho_1 J_0 J_1, 2}
  J_{1,ab} J_{0,ab} 
  +
  g^{\rho_1 J_0 J_1, 3}
  J_{1,ab} J_{0,ba}
  \right]
  \;.
  \numberthis
  \label{eq:nnn_edf}
\end{align*}
Our EDF above is similar to the one in Ref.~\cite{Stoitsov:2010ha}, though more general 
as we have not assumed spherical symmetry for the self-consistent solutions.
Also the DME parametrization we adopt below in Sec.~\ref{sec:dme_parametrization} 
is not identical to the one used in Ref.~\cite{Stoitsov:2010ha}.

\subsection{DME Parametrization}
\label{sec:dme_parametrization}

There are three steps in applying the DME to derive couplings:
\be
  \item perform spin-isospin traces on the operators present in the potential;
  \item expand the resulting OBDM structures using the DME;
  \item combine as needed the local densities, DME functions, and potentials for 
  each coupling term and numerically perform the relevant integrals.
\ee
  When performing the $NN$ DME, non-diagonal OBDMs are expanded about the diagonal such that the nonlocality is factorized using the following formulas:
\beq
  \rho_t \left(\Rvec + \frac{\rvec}{2},\Rvec - \frac{\rvec}{2}\right) 
  \approx 
  \sum_{n=0}^{n_{\text{max}}} \Pi_{n}^{\rho} (kr) \mathcal{P}_n (\Rvec)
  \;,
  \label{eq:nn_scalar_dme}
\eeq

\beq
  \svec_t \left(\Rvec + \frac{\rvec}{2}, \Rvec - \frac{\rvec}{2}\right) 
  \approx 
  \sum_{m=0}^{m_{\text{max}}} \Pi_{m}^{s} (kr) \mathcal{Q}_m (\Rvec)
  \;,
  \label{eq:nn_vector_dme}
\eeq
where the $\Pi$ functions are specified by the DME variant and 
$\mathcal{P}_n (\Rvec)$, $\mathcal{Q}_m (\Rvec)$ denote various local densities.
The momentum scale $k$ in the $\Pi$ functions sets the scale for fall off in the 
off-diagonal direction of the OBDM; one is free to choose $k$ in such a way that 
the expansion is optimized.  
We define the momentum scale $k$ to be the local Fermi momentum, 
\beq
  k \equiv 
  \kf (\Rvec) = 
  \left(
  \frac{3 \pi^2}{2} 
  \rho_0(\Rvec)
  \right)^{1/3}
  \;,
\eeq
where $\rho_0$ is the isoscalar density.
However alternative choices of $k$, for example involving $\tau(\Rvec)$ and 
$\Delta \rho(\Rvec)$, are also possible \cite{CAMPI1978263}. 

We follow past practice and truncate the DME expansion at 
$n_{\text{max}} = 2$ and $m_{\text{max}} = 1$ such that:
\beq
  \rho \left(\Rvec + \frac{\rvec}{2}, \Rvec - \frac{\rvec}{2}\right) 
  \approx
  \Pi_0^\rho (\kf r) \rho(\Rvec) + 
  \frac{r^2}{6}
  \Pi_2^\rho (\kf r) \left[
  \frac{1}{4} \Delta \rho (\Rvec)
  - \tau (\Rvec) + 
  \frac{3}{5} \kf^2 \rho (\Rvec)
  \right] \;,
\eeq
\beq
  s_b \left(\Rvec + \frac{\rvec}{2}, \Rvec - \frac{\rvec}{2}\right) 
  \approx
  i \Pi_1^s(\kf r) 
  \sum_{a=x}^z r_a J_{ab} (\Rvec) \;.
\eeq 
The DME parameterization we adopt is the simplified phase-space-averaging 
(PSA)~\cite{PhysRevC.82.014305,Gebremariam201117} choice, which was shown
to better reproduce the vector part of the OBDM over the original Negele-Vautherin prescription.
For this choice, the $\Pi$ functions are given by:
\beq
  \label{eq:dme_pi_functs}
  \Pi_0^\rho (\kf x) = \Pi_2^\rho (\kf x)
  = \Pi_1^s (\kf x) = 3 \frac{j_1(\kf x)}{\kf x} \; ,
\eeq
where $j_1$ is a spherical Bessel function of the first kind. 
Applying the symmetry principle in Eq.~\eqref{eq:tr_symmetry}, the DME expansion 
can also be applied to OBDMs with reversed arguments.

For the three-body system, as anticipated by the transformation to Eq.~\eqref{eq:3n_hf},
a different set of coordinates than the center-of-mass choice is needed.
A successful coordinate choice for the DME allows for factorization of the 
variable appearing in the local densities and the variables appearing in the potential and DME $\Pi$ functions.
For the three-body system, these conditions can be satisfied by performing the DME
expansion about the location of one of the particles, i.e., $\rvec_1$ in Eq.~\eqref{eq:3n_hf}.
However, this choice of coordinates leads to OBDMs with two nonlocality coordinates in some cases; this is in contrast to the one nonlocality variable that occurs in two-body systems.
Assessing the ultimate accuracy of the simplified PSA-DME in such cases is an open research topic. 
In the following, we follow the straightforward generalization articulated in Ref.~\cite{Gebremariam:2010}.
For the scalar part of the OBDM, this leads to the following DME expansion equations \cite{Gebremariam:2010}:
\begin{align*}
  \label{eq:dme_sca_1}
  \rho(\rvec_1, \rvec_1 + \xvec) &\approx
  \Pi_0^\rho (\kf x) \rho(\rvec_1) + \frac{x^2}{6} \Pi_2^\rho (\kf x) 
  \Bigl(
  \frac{1}{2} \Delta \rho(\rvec_1)
  - \tau(\rvec_1)
  + \frac{3}{5} \kf^2 \rho (\rvec_1)
  \Bigr)
  \;,
  \numberthis
\end{align*}
\begin{align*}
  \label{eq:dme_sca_2}
  \rho(\rvec_1 + \xvec_2, \rvec_1 + \xvec_3) &\approx
  \Pi_0^\rho (\kf l) \Bigl[
  \rho(\rvec_1) + \Nvec \cdot \nabla \rho (\rvec_1)
  + \frac{1}{2} \left(
  \Nvec \cdot  \nabla \right)^2 \rho(\rvec_1)
  \Bigr]
\\
  & \quad\null + \frac{l^2}{6} \Pi_2^\rho (\kf l) 
  \Bigl[
  \gamma \Delta \rho(\rvec_1)
  - \tau(\rvec_1)
  + \frac{3}{5} \kf^2 \rho (\rvec_1)
  \Bigr]
  \;,
  \numberthis
\end{align*}
where
\beq
  \lvec \equiv \xvec_2 - \xvec_3 \; ,
  \qquad
  \Nvec \equiv (1-a) \xvec_2 + a \xvec_3 \; ,
  \qquad
  \gamma \equiv a^2 - a + 1/2 \; .
\eeq
The variable $a$ reflects our freedom in choosing how to perform the DME expansion 
with respect to the $23$ particle subsystem.
For the choice $a=1/2$, the usual center-of-mass choice is recovered.
For the vector part, the expansion equations are \cite{Gebremariam:2010}:
\beq
  \label{eq:dme_vec_1}
  s_b (\rvec_1 + \xvec, \rvec_1)
  \approx i \Pi_1^s (\kf x) \
  \sum_{a=x}^z
  x_{a} \; J_{ab} (\rvec_1)
  \;,
\eeq
\beq
  \label{eq:dme_vec_2}
  s_b (\rvec_1 + \xvec_2, \rvec_1 + \xvec_3)
  \approx i \Pi_1^s (\kf l) \
  \sum_{a=x}^z
  l_a \; J_{ab} (\rvec_1)
  \;.
\eeq
For the expansions in Eqs.~\eqref{eq:dme_sca_1}, \eqref{eq:dme_sca_2}, 
\eqref{eq:dme_vec_1}, and \eqref{eq:dme_vec_2}, the
$\Pi$ functions are again given by Eq.~\eqref{eq:dme_pi_functs}.
As before, DME expansions for OBDMs with reversed arguments can be found using 
Eq.~\eqref{eq:tr_symmetry}.
In Appendix~\ref{sec:dme_expansion_example} we show an example of how these expansions are performed.
  


\section{DME for \texorpdfstring{$NN$}{NN} Forces}
\label{sec:DME_NN}

Because we do not plan to apply the DME to the Hartree term,
we concentrate only on the exchange term. 
The Fock energy is given by:
\begin{align*}
  V_\text{F} = -\frac{1}{2} \tr_1 \tr_2 
  \int d\Rvec \; d\rvec \; 
  \rho_1 \left(\Rvec - \frac{\rvec}{2}, \Rvec + \frac{\rvec}{2}\right) \;
  \rho_2 \left(\Rvec + \frac{\rvec}{2}, \Rvec - \frac{\rvec}{2}\right) \; 
  \vnn(\rvec, \{\sigma \tau\}) P^{\sigma\tau}_{12}
  \;.
  \numberthis
  \label{eq:nn_fock}
\end{align*}
Note that in applying Eqs.~\eqref{eq:nn_scalar_dme} and \eqref{eq:nn_vector_dme} 
to the Fock energy, the integrations over $\Rvec$ and $\rvec$ will factorize.
The $\rvec$ integral will go over the $\Pi$ functions and $\vnn$ while the $\Rvec$ 
integral will be the remaining integral in the EDF of Eq.~\eqref{eq:nn_edf}.
Eq.~\eqref{eq:nn_fock} can be rendered in a more compact form via
\beq
  V_{\text{F}} =
  -\frac{1}{2} 
  \sum_{ij} \int d\Rvec \; d\rvec
  A_i B_j
  \widetilde{V}(r)_{ij}
  \;,
  \label{eq:fock_compact}
\eeq
where $\widetilde{V}(r)_{ij}$ is an isoscalar or isovector form factor from 
Eq.~\eqref{eq:nn_potential} with pure radial dependence, and the large Latin letters include information 
about the spin-isospin traces:
\bseq
  \label{eq:nn_traces}
\begin{align*}
  A_m &= \frac{1}{4}
  \prod_{i=1}^2   
  \tr_{\sigma_i} 
  \left[
  \bigg(
  \rho_{t,i} (\avec_i)
  + 
  \svec_{t,i} (\avec_i)
  \cdot \sigvec_i
  \bigg)
   \; \mathcal{J}_m
  P_{12}^{\sigma}
  \right]
  \;,
  \numberthis
  \label{eq:nn_traces_1}
\\
  B_m &= \frac{1}{4}
  \prod_{i=1}^2
  \tr_{\tau_i}
  \left[
  \bigg(
  \delta^{t,0}_i
  + \delta^{t,1}_i \tau^z_i
  \bigg)
  \mathcal{K}_m P_{12}^{\tau}
  \right]
  \;.
  \numberthis
  \label{eq:nn_traces_2}
\end{align*}  
\eseq
Note that in Eq.~\eqref{eq:nn_traces}, the OBDMs have been decomposed into scalar and vector 
parts with the isospin part factorized by Eq.~\eqref{eq:obdm_decomposed}.
The $\avec$ variables are schematic stand-ins for the arguments of the OBDMs appearing 
in Eq.~\eqref{eq:nn_fock} and $\mathcal{J}_m$, and $\mathcal{K}_m$ are, respectively, the spin and isospin 
operators in Eq.~\eqref{eq:nn_spin_ops} and Eq.~\eqref{eq:nn_iso_ops}. 
Now, we follow the steps outlined in Sec.~\ref{sec:dme_parametrization}.

\subsection{\texorpdfstring{$NN$}{NN} DME Step 1 - Traces}

Inserting the $\mathcal{J}$ and $\mathcal{K}$ operators into Eqs.~\eqref{eq:nn_traces_1} 
and \eqref{eq:nn_traces_2} and evaluating the traces yields:
\bseq
\begin{align*}
  A_1 &= \frac{1}{2}
  \left(
  \rho_1 \; \rho_2 + \svec_1 \cdot \svec_2
  \right)
  \;,
  \numberthis
\\
  A_2 &= \frac{1}{2}
  \left(
  3 \rho_1 \; \rho_2 - \svec_1 \cdot \svec_2
  \right)
  \;,
  \numberthis
\\
  A_3 &= 3 (\svec_1 \cdot \rhat) (\svec_2 \cdot \rhat) - \svec_1 \cdot \svec_2
  \;,
  \numberthis
\end{align*}
\label{eq:NN_spin_traces}
\eseq
and
\bseq
\begin{align*}
  B_1 &= \frac{1}{2}
  \left(
  \delta_1^{t,0} \; \delta_2^{t,0}
  +
  \delta_1^{t,1} \; \delta_2^{t,1}
  \right)
  \;,
  \numberthis
\\
  B_2 &= \frac{1}{2}
  \left(
  3 \delta_1^{t,0} \; \delta_2^{t,0} 
  -
  \delta_1^{t,1} \; \delta_2^{t,1}
  \right)
  \;,
  \numberthis
\end{align*}
\label{eq:NN_iso_traces}
\eseq
where the arguments of the scalar ($\rho$) and vector ($\svec$) density matrices 
along with the isoscalar or isovector subscript $t$ have been suppressed for brevity.

\subsection{\texorpdfstring{$NN$}{NN} DME Step 2 - DME Dictionary}

Looking at the elements in Eq.~\eqref{eq:NN_spin_traces}, it is apparent that 
a DME expansion needs to be performed on only three unique OBDM structures:
\beq
  \left\{\rho_1 \; \rho_2 \; , 
  \svec_1 \cdot \svec_2 \; ,
  (\svec_1 \cdot \rhat) (\svec_2 \cdot \rhat)
  \right\}
  \;.
\eeq
Below, we apply the DME parameterization defined in Sec.~\ref{sec:dme_parametrization} 
and keep terms up to second order.
The format for the DME expansions given below has scalar or vector density 
matrices on the left and local densities on the right:
\begin{align*}
  \text{scalar or vector density matrices}
  \xrightarrow{\text{DME}}
  \left\{
  \begin{array}{c}
  \text{local densities}
  \times
  \text{DME expression}
  \end{array}
  \right\}
  \; ,
\end{align*}
where the DME expression on the right hand side has had all 
nonlocal variable integrals done except for the relative distance magnitude $r$.
Below, we restore the spatial dependence in the OBDMs and local densities for clarity.
The expanded structures are given by:
\begin{align*}
  \renewcommand{\arraystretch}{1.3}
  \rho_1 \left(\Rvec - \frac{\rvec}{2}, \Rvec + \frac{\rvec}{2}\right) \;
  \rho_2 \left(\Rvec + \frac{\rvec}{2}, \Rvec - \frac{\rvec}{2}\right)
  \xrightarrow{\text{DME}}
  \left\{
  \begin{array}{lrr}
  & \rho_1 (\Rvec) \rho_2 (\Rvec)  &
  \mathcal{O}_1 
  \\
  + & \rho_1 (\Rvec) \tau_2 (\Rvec)  &
  \mathcal{O}_2
  \\
  + & \tau_1 (\Rvec) \rho_2 (\Rvec)  &
  \mathcal{O}_2
  \\
  + & \rho_1 (\Rvec) \Delta \rho_2 (\Rvec)  &
  \mathcal{O}_3
  \\
  + & \Delta \rho_1 (\Rvec) \rho_2 (\Rvec)  &
  \mathcal{O}_3
  \end{array}
  \right\}
  \;,
  \numberthis
  \label{eq:NN_pp_DME}
\end{align*}
\begin{align*}
  \svec_1 \left(\Rvec - \frac{\rvec}{2}, \Rvec + \frac{\rvec}{2}\right)
  \cdot 
  \svec_2 \left(\Rvec + \frac{\rvec}{2}, \Rvec - \frac{\rvec}{2}\right)
  \xrightarrow{\text{DME}}
  \left\{
  \begin{array}{lr}
  J_{1,ab} (\Rvec) J_{2,ab} (\Rvec)
   &
  \mathcal{O}_4
  \end{array}
  \right\}
  \;,
  \numberthis
  \label{eq:NN_ss_DME}
\end{align*}
\begin{align*}
  \renewcommand{\arraystretch}{1.3}
  \svec_1 \left(\Rvec - \frac{\rvec}{2}, \Rvec + \frac{\rvec}{2} \right)
  \cdot \rhat  
  \
  \svec_2 \left(\Rvec + \frac{\rvec}{2}, \Rvec - \frac{\rvec}{2}\right)
  \cdot \rhat 
  \xrightarrow{\text{DME}}
  \left\{
  \begin{array}{lcr}
  & J_{1,aa} (\Rvec) J_{2,bb} (\Rvec)
   &
  \mathcal{O}_4 / 5
\\
  + & J_{1,ab} (\Rvec) J_{2,ab} (\Rvec)
   &
  \mathcal{O}_4 / 5
\\
  + & J_{1,ab} (\Rvec) J_{2,ba} (\Rvec)
   &
  \mathcal{O}_4 / 5
  \end{array}
  \right\}
  \;,
  \numberthis
  \label{eq:NN_sr_DME}
\end{align*}
where the $\mathcal{O}_i(r, \kf)$ functions, 
which contain the DME $\Pi$ functions and relative distance dependence, are given by,
\bseq
\begin{align*}
  \mathcal{O}_1 &=   
  \left[
  \Pi_0^\rho (\kf r)
  \right]^2
  +
  \frac{r^2 \kf^2}{5} 
  \Pi_0^\rho (\kf r)
  \Pi_2^\rho (\kf r)
  \;,
  \numberthis
\\
  \mathcal{O}_2 &=
  - \frac{r^2}{6}
  \Pi_0^\rho (\kf r)
  \Pi_2^\rho (\kf r)
  \;,
  \numberthis
\\
  \mathcal{O}_3 &=
  \frac{r^2}{24}
  \Pi_0^\rho (\kf r)
  \Pi_2^\rho (\kf r)
  \;,
  \numberthis
\\
  \mathcal{O}_4 &=
  \frac{r^2}{3} 
  \left[
  \Pi_1^s (\kf r)
  \right]^2
  \;.
  \numberthis
\end{align*}
\eseq

\subsection{\texorpdfstring{$NN$}{NN} DME Step 3 - Couplings}

Combining the different local densities in Eqs.~\eqref{eq:NN_pp_DME}, 
\eqref{eq:NN_ss_DME}, and \eqref{eq:NN_sr_DME} with the isospin traces 
in Eq.~\eqref{eq:NN_iso_traces} one can verify the 12 different possible 
local density bilinears seen in Eq.~\eqref{eq:nn_edf}. 
Using Eqs.~\eqref{eq:NN_spin_traces} and \eqref{eq:NN_iso_traces} as input 
into Eq.~\eqref{eq:fock_compact} along with the DME dictionary, symbolic 
software can then perform the algebraic manipulations necessary to isolate 
the equations for each individual DME coupling such that we have an EDF of 
the form of Eq.~\eqref{eq:nn_edf}. 
The coupling expressions are
\bseq
\label{eq:nn_couplings}
  \begin{align*}
  \label{eq:gpp_coupling}
  g^{\rho \rho}_t (\Rvec) &= -\frac{4\pi}{2} \int dr
  \; r^2 \frac{1}{4}  \mathcal{O}_1 \; \Xi^1_t
  \;, \numberthis
\\
  g^{\rho \tau}_t (\Rvec) &= -\frac{4\pi}{2} \int dr
  \; r^2 \frac{1}{2} \mathcal{O}_2 \; \Xi^1_t
  \;, \numberthis
\\
  g^{\rho \Delta \rho}_t (\Rvec) &= -\frac{4\pi}{2} \int dr
  \; r^2 \frac{1}{2} \mathcal{O}_3 \; \Xi^1_t
  \;, \numberthis
\\
  g^{JJ,1}_t (\Rvec) &= -\frac{4\pi}{2} \int dr
  \; r^2 \frac{3}{10} \mathcal{O}_4 \; \Xi^2_t
  \;, \numberthis
\\
  g^{JJ,2}_t (\Rvec) &= -\frac{4\pi}{2} \int dr
  \; r^2 \frac{1}{20} \mathcal{O}_4 \; \Xi^3_t
  \;, \numberthis
\\
  g^{JJ,3}_t (\Rvec) &= -\frac{4\pi}{2} \int dr
  \; r^2 \frac{3}{10} \mathcal{O}_4 \; \Xi^2_t
  \;, \numberthis
\end{align*}
\eseq
where the $\Xi$ functions encapsulate the coupling dependence on the different 
potential form factors from Eq.~\eqref{eq:nn_potential},
\bseq
\begin{align*}
  \Xi^1_t &= 
  \begin{dcases}
  V_C 
  +
  3 W_C
  +
  3 V_S
  + 
  9 W_S
  \quad
  &\enspace t = 0 \;,
\\
  V_C 
  -
  W_C
  +
  3 V_S
  -
  3 W_S
  \quad
  &\enspace t = 1 \;,
  \numberthis
  \end{dcases}
\\
  \Xi^2_t &=
  \begin{dcases}
  V_T + 3 W_T
  \quad
  &\enspace t = 0 \;,
\\
  V_T - W_T
  \quad
  &\enspace t = 1 \;,
  \numberthis
  \end{dcases} 
\\
  \Xi^3_t &=
  \begin{dcases}
  5 V_C + 15 W_C - 5 V_S - 15 W_S - 4 V_T - 12 W_T
  \quad
  &\enspace t = 0 \;,
\\
  5 V_C - 5 W_C - 5 V_S + 5 W_S - 4 V_T + 4 W_T
  \quad
  &\enspace t = 1 \;.
  \numberthis
  \end{dcases}
\end{align*}
\eseq



\section{DME For \texorpdfstring{$3N$}{3N} Forces}
\label{sec:DME_3N}

Utilizing the organization of the $V_C$ operators in Sec.~\ref{sec:3N_forces}, 
the exchange terms are rewritten as: 
\begin{align*}
  \label{eq:fock_3n_se}
  \vnnn_{\text{SE, F}}
  =&\ 
  -
  \sum_{jk} 
  \int d\rvec_1
  d\xvec_2 d\xvec_3 \;
  C_j \;
  D_k \;
  \widetilde{V}_{23,jk} (x_2, x_3)
  \\
  & \null - \frac{1}{2}
  \sum_{jk} 
  \int d\rvec_1
  d\xvec_2 d\xvec_3 \;
  E_j \;
  F_k \;
  \widetilde{V}_{23,jk} (x_2, x_3)
  \;, \numberthis
\\
  \label{eq:fock_3n_de}
  \vnnn_{\text{DE, F}}
  = &\
  \sum_{jk} 
  \int d\rvec_1
  d\xvec_2 d\xvec_3 \;
  G_j \;
  H_k \;
  \widetilde{V}_{23,jk} (x_2, x_3)
  \;, \numberthis
\end{align*}
where the $jk$ sum goes over all the spin-isospin operators in the potential, 
the large Latin letters refer to the result of traces, and 
$\widetilde{V}_{23,jk}(x_2, x_3)$ refers to the corresponding radial 
parts of the potential in the braces in Eqs.~\eqref{eq:vc1}, \eqref{eq:vc2}, and \eqref{eq:vc3} along with the correct chiral prefactor.
The large Latin letters are given by:
\bseq
\label{eq:3n_spin_iso_latin}
\begin{align*}
  C_j &= \frac{1}{8}
  \prod_{i=1}^3   
  \tr_{\sigma_i} 
  \left[
  \bigg(
  \rho_{t,i} (\avec_i)
  + 
  \svec_{t,i} (\avec_i)
  \cdot \sigvec_i
  \bigg)
   \; \mathcal{S}_j P_{12}^{\sigma}
  \right]
  \;, \numberthis
\\
  D_j &= \frac{1}{8}
  \prod_{i=1}^3   
  \tr_{\tau_i}
  \left[
  \bigg(
  \delta^{t,0}_i
  + \delta^{t,1}_i \tau^z_i
  \bigg)
  \mathcal{T}_j P_{12}^{\tau}
  \right]
  \;, \numberthis
\\
  E_j &= \frac{1}{8}
  \prod_{i=1}^3   
  \tr_{\sigma_i} 
  \left[
  \bigg(
  \rho_{t,i} (\avec_i)
  + 
  \svec_{t,i} (\avec_i)
  \cdot \sigvec_i
  \bigg)
   \; \mathcal{S}_j P_{23}^{\sigma}
  \right]
  \;, \numberthis
\\
  F_j &= \frac{1}{8}
  \prod_{i=1}^3   
  \tr_{\tau_i}
  \left[
  \bigg(
  \delta^{t,0}_i
  + \delta^{t,1}_i \tau^z_i
  \bigg)
  \mathcal{T}_j P_{23}^{\tau}
  \right]
  \;, \numberthis
\\
  G_j &= \frac{1}{8}
  \prod_{i=1}^3   
  \tr_{\sigma_i} 
  \left[
  \bigg(
  \rho_{t,i} (\avec_i)
  + 
  \svec_{t,i} (\avec_i)
  \cdot \sigvec_i
  \bigg)
   \; \mathcal{S}_j P_{23}^{\sigma} P_{12}^{\sigma}
  \right]
  \;, \numberthis
\\
  H_j &= \frac{1}{8}
  \prod_{i=1}^3   
  \tr_{\tau_i}
  \left[
  \bigg(
  \delta^{t,0}_i
  + \delta^{t,1}_i \tau^z_i
  \bigg)
  \mathcal{T}_j P_{23}^{\tau} P_{12}^{\tau}
  \right]
  \;, \numberthis
\end{align*}
\eseq
with the $\avec_i$ schematically standing in for the pair of terms appearing in 
the OBDM in Eq.~\eqref{eq:3n_hf}, the $1/8$ prefactor coming from pulling out 
the result of the traces,%
\footnote{For Eqs.~\eqref{eq:fock_3n_se} and \eqref{eq:fock_3n_de}, 
a factor of $1/4^3$ from expanding the OBDMs and $8^2$ from the traces have been 
combined to give unity.} 
and the $\mathcal{S}$, $\T$ terms specified by 
Eqs.~\eqref{eq:nnn_spin_op_1}, 
\eqref{eq:nnn_spin_op_2}, \eqref{eq:nnn_spin_op_3}, and \eqref{eq:nnn_iso_op_1}.
Again to derive coupling expressions, 
we follow the steps outlined in Sec.~\ref{sec:dme_parametrization}.
  


\subsection{\texorpdfstring{$3N$}{3N} DME Step 1 - Traces}

Here we explicitly do the traces for the six spin operator structures 
$\mathcal{S}_{1}$, $\mathcal{S}_{2}$,  $\mathcal{S}_{6}$, $\mathcal{S}_{7}$,  
$\mathcal{S}_{8}$, and $\mathcal{S}_{9}$ along with $\T_1$ and $\T_2$.
Operators $\mathcal{S}_3$, $\mathcal{S}_4$, and $\mathcal{S}_5$, follow from 
the trace of $\mathcal{S}_1$.
After performing the spin traces, we then:
\be
  \item discard all terms with a local spin density, $\svec(\rvec)$, due to time-reversal invariance;
  \item discard terms with three vector densities as we restrict our EDF to second order in derivatives.
\ee  

\subsubsection{Hartree Term}

As previously mentioned, the OBDMs for the Hartree term are diagonal and the term
can thus be evaluated exactly.
However, as all of our $V_{C,i}$ three-body potentials in Eqs.~\eqref{eq:vc1}, 
\eqref{eq:vc2}, and \eqref{eq:vc3} contain at least one Pauli spin matrix, 
the spin traces will yield at least one local spin density for each term.
Therefore, the Hartree term will exactly vanish for time-reversal invariant systems.

Note also that the isospin traces over the operator $\mathcal{T}_2$ 
will identically vanish given isospin symmetry.
As such, even for systems without time-reversal invariance, the $V_{C,3}$ potential 
will not contribute to the Hartree term assuming isospin is a good symmetry.  

\subsubsection{Single Exchange Traces}

The single exchange part has two terms, one corresponding to $P_{12}$ and 
another to $P_{23}$.
Consulting Eq.~\eqref{eq:3n_hf}, it is seen that the density matrix with subscript $3$ 
for the first single exchange term will be diagonal as the $P_{12}$ operator does not act upon it. 
However, all of the three-body potentials under consideration in Eqs.~\eqref{eq:vc1}, 
\eqref{eq:vc2}, and \eqref{eq:vc3} contain a Pauli spin matrix $\sigvec_3$. 
After spin traces, this will yield a local spin density and thus vanish for 
time-reversal invariant systems: 
\beq
  C_i \longrightarrow 0 
  \qquad 
  \text{for all } i 
  \;.
\eeq
Likewise for the second part of the single exchange in Eq.~\eqref{eq:3n_hf}, 
the density matrix with subscript $1$ is diagonal. 
Because all terms of the $V_{C,3}$ potential in Eq.~\eqref{eq:vc3} have a 
$\sigvec_1$ Pauli spin matrix, all $V_{C,3}$ terms will yield a local spin density after traces.
Therefore, this contribution will vanish for time-reversal invariant systems:
\beq
  E_{i} \longrightarrow 0 
  \qquad 
  \text{for } i = 6, 7, 8, 9
  \;.
\eeq
The remaining nonzero single exchange spin and isospin traces,  
$E_i$ and $F_i$ respectively, are given in Appendix~\ref{sec:se_traces}.

\subsubsection{Double Exchange Traces}

The double exchange traces are more involved due to the extra exchange operator 
and the fact that all of the spin operators have a non-zero contribution.
The expressions for the spin and isospin traces, $G_i$ and $H_i$ respectively, 
are given in Appendix~\ref{sec:de_traces}.

\subsection{\texorpdfstring{$3N$}{3N} DME Step 2 - DME Dictionary}

Due to its length, we relegate the DME dictionary for the single exchange and 
double exchange terms to Appendices \ref{sec:se_dict} and \ref{sec:de_dict}.
The format for the DME expansions in these appendices is given schematically by:
\begin{align*}
  \text{density matrices}
  \xrightarrow{\text{DME}}
  \left\{
  \begin{array}{lcr}
  \text{local densities}
  & \times & 
  (\text{LR or IR DME expression})
  \end{array}
  \right\}
  \; ,
\end{align*}
where the LR DME expression has all integrals done 
except for the $x_2$, $x_3$ magnitudes and the relative angle $\theta$ between the two vectors.
For the IR DME expression, only one integral over the magnitude of the nonlocality variable, generically called $x$, remains.

\subsection{\texorpdfstring{$3N$}{3N} DME Step 3 - Couplings}

Again due to length, we relegate the final expressions of the $3N$ couplings 
to Appendix~\ref{sec:nnn_couplings}\footnote{The code used to calculate the $3N$ couplings is available upon request.}.



\section{Results}
\label{sec:results}

\begingroup
\begin{table}[t]
 \caption{\label{tab:param_table}
Summary table for the various LECs and physical parameters appearing in our 
$NN$ and $3N$ potentials up to \NNLO. 
For all potentials, $\mpi = 138\,$MeV, $\fpi = 92.4\,$MeV, and $\ga = 1.29$.
In the table, $\delm$ is given in MeV and $\ha$ is dimensionless. 
The subleading $c_i$s and 
the $b_i$ combination are given in $\text{GeV}^{-1}$, and 
  are taken from the fit to $\pi$--$N$ scattering data in Ref.~\cite{Krebs:2007rh}.}
 \begin{ruledtabular}
 \begin{tabular}{c ||  c | c | c | c | c | c | c}
  &  $\delm$ &  $\ha$ & $c_1$ & $c_2$ & $c_3$ & $c_4$ & $b_3 + b_8$ \\
 \hline
 $\Delta$-less &  --- &  --- &  $-0.57$ & \phantom{$-$}2.84 & $-3.87$ & 2.89 & --- \\
 
 With $\Delta$s  &  293 &  2.74 & $-0.57$ & $-0.25$ & $-0.79$ & 1.33 & 1.40 \\

 \end{tabular}
 \end{ruledtabular}
\end{table}
\endgroup

\begin{figure}[p]
   \includegraphics[width=0.49\textwidth]{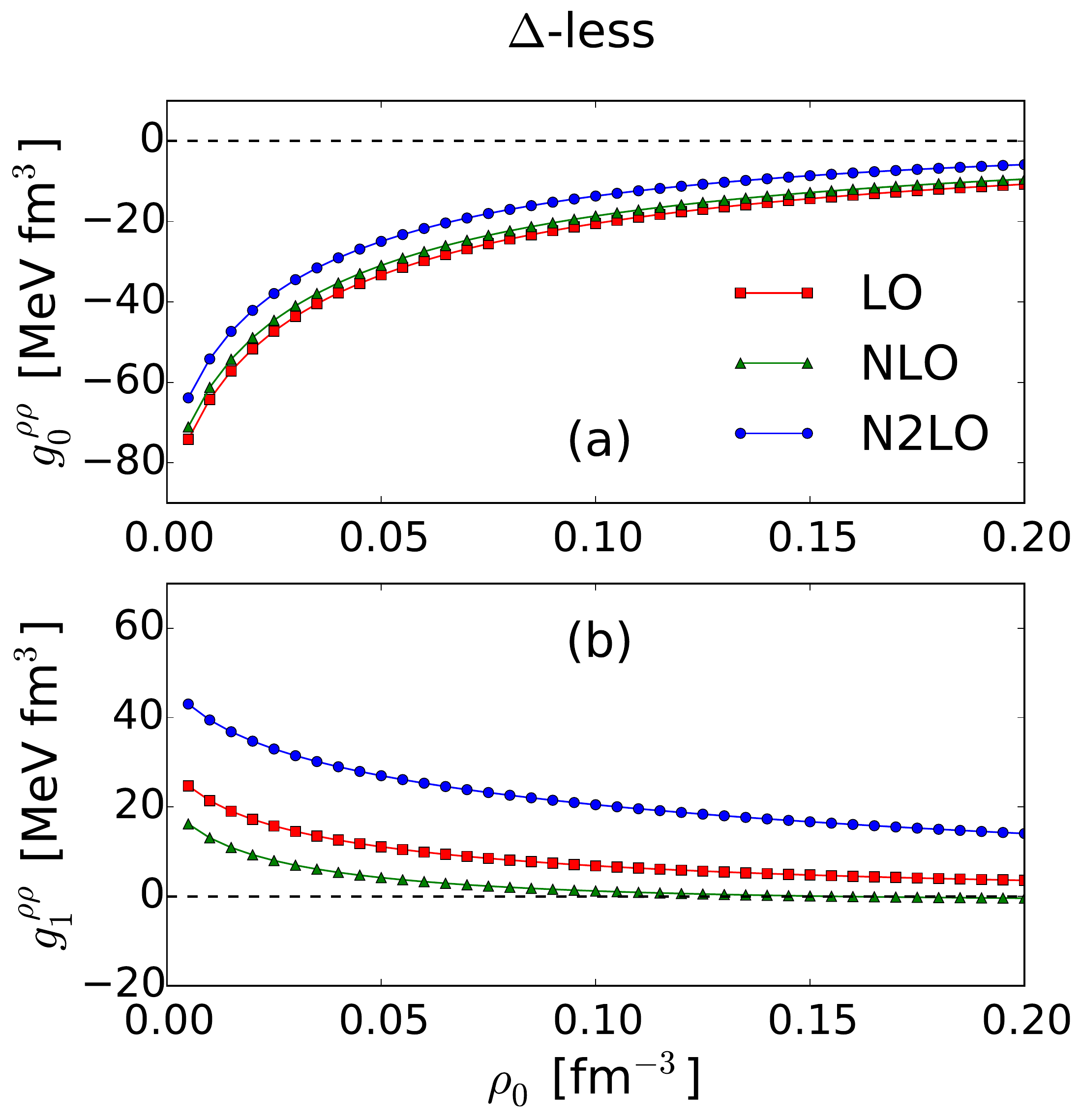}
  \includegraphics[width=0.49\textwidth]{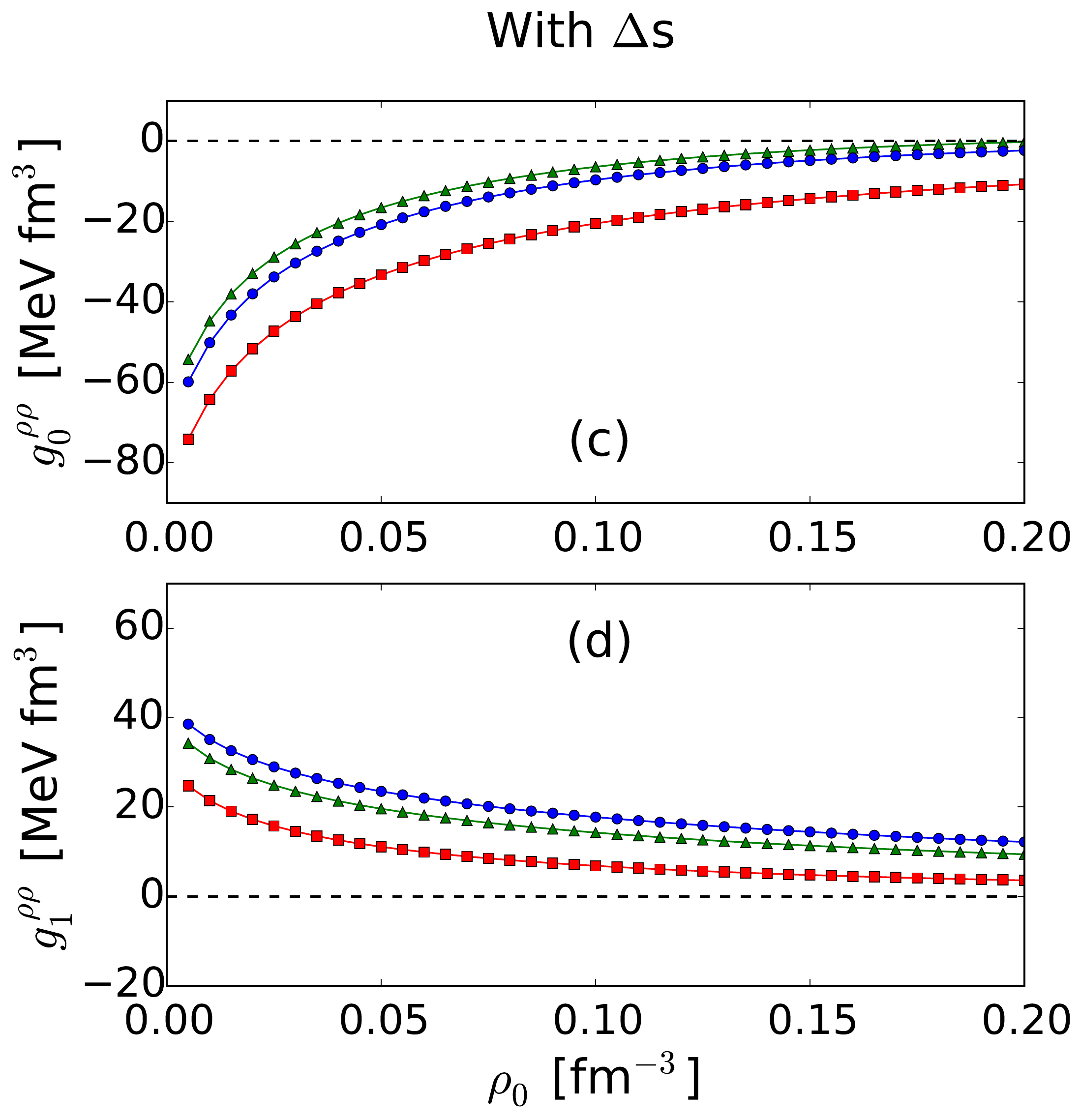}
  \caption{
  The $g^{\rho \rho}_t$ couplings from Eq.~\eqref{eq:gpp_coupling} are plotted as a function of the isoscalar density $\rho_0$ at fixed cutoff $\RNN = 1.2~\fm$ using the regulator in Eq.~\ref{eq:regulator} with $n=6$.
 The values for the couplings are shown at three different chiral orders up to \NNLO.
 The isoscalar coupling $g^{\rho \rho}_0$ is shown without (a) and with (c) $\Delta$ isobars. 
    The isovector coupling $g^{\rho \rho}_1$ is shown without (b) and with (d) $\Delta$ isobars.}
  \label{fig:gpp_plot_deltas_1p2}
  \medskip
   \includegraphics[width=0.49\textwidth]{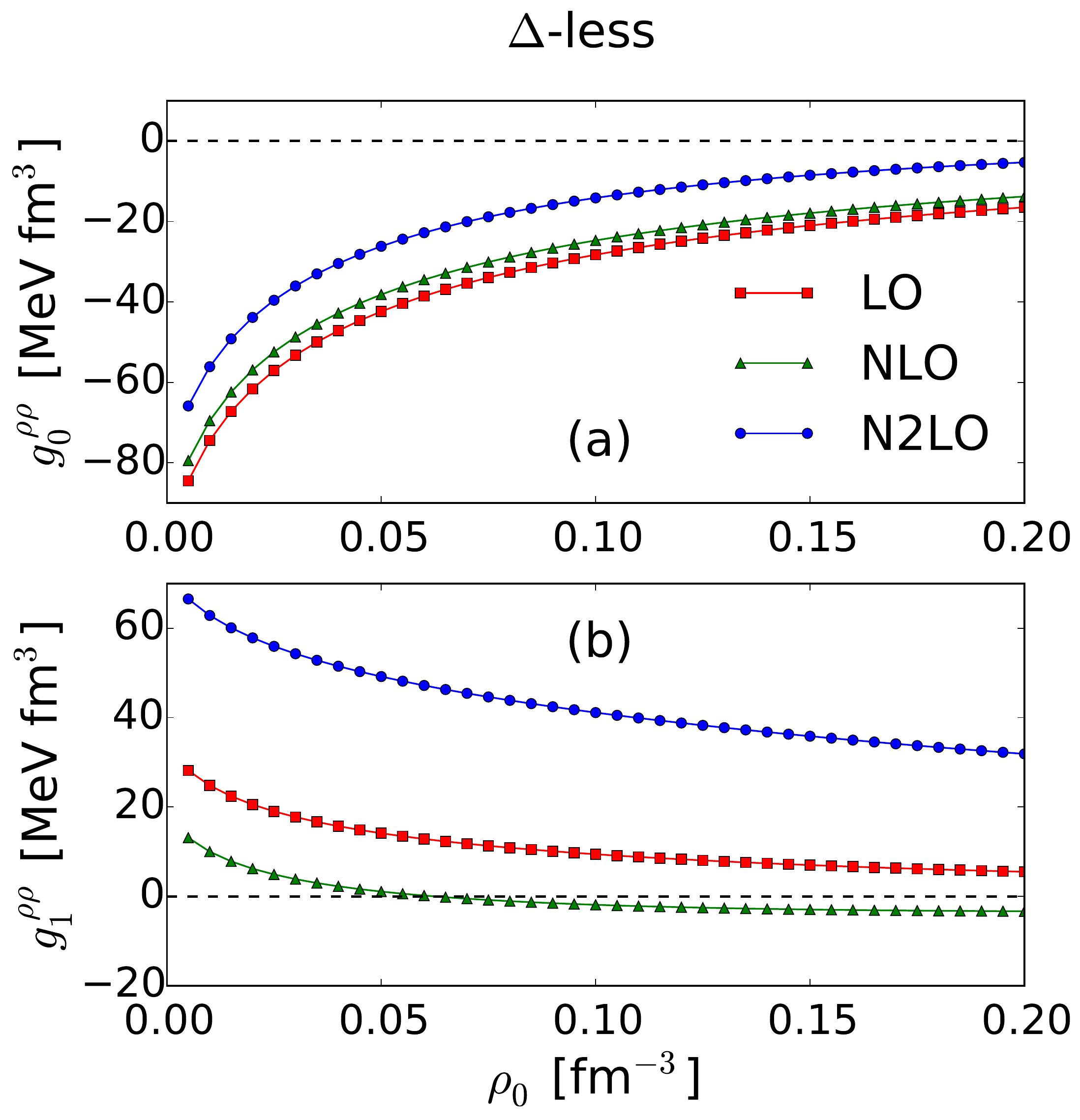}  \includegraphics[width=0.49\textwidth]{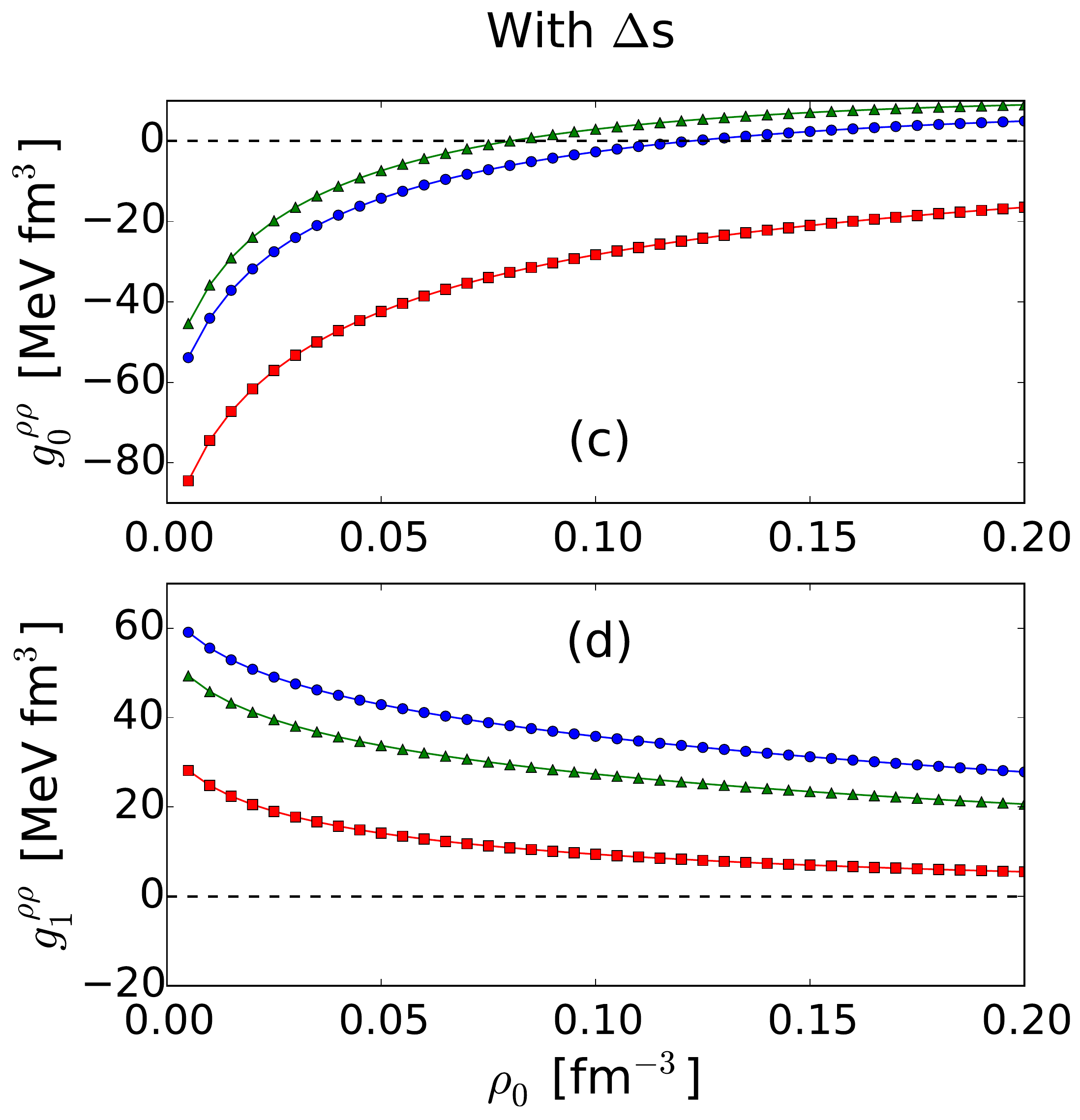}
  \caption{
  The $g^{\rho \rho}_t$ couplings from Eq.~\eqref{eq:gpp_coupling} are plotted as a function of the isoscalar density $\rho_0$ at fixed cutoff $\RNN = 1.0~\fm$ using the regulator in Eq.~\ref{eq:regulator} with $n=6$.
  The values for the couplings are shown at three different chiral orders up to \NNLO.
  The isoscalar coupling $g^{\rho \rho}_0$ is shown without (a) and with (c) $\Delta$ isobars. 
    The isovector coupling $g^{\rho \rho}_1$ is shown without (b) and with (d) $\Delta$ isobars.}
  	\label{fig:gpp_plot_deltas_0p8}
\end{figure}

\begin{figure}[p]
  \includegraphics[width=0.50\textwidth]{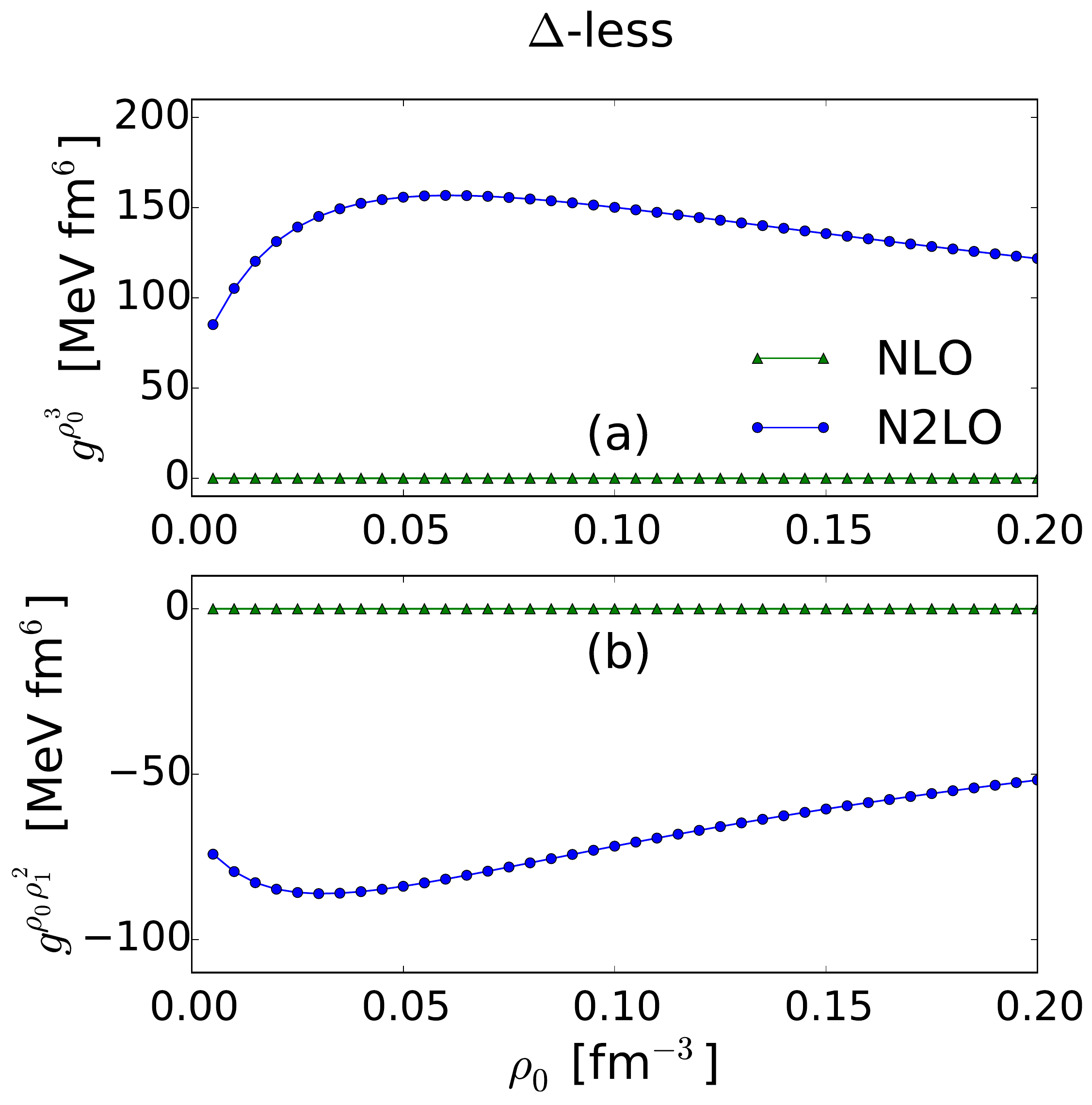}~
  \includegraphics[width=0.50\textwidth]{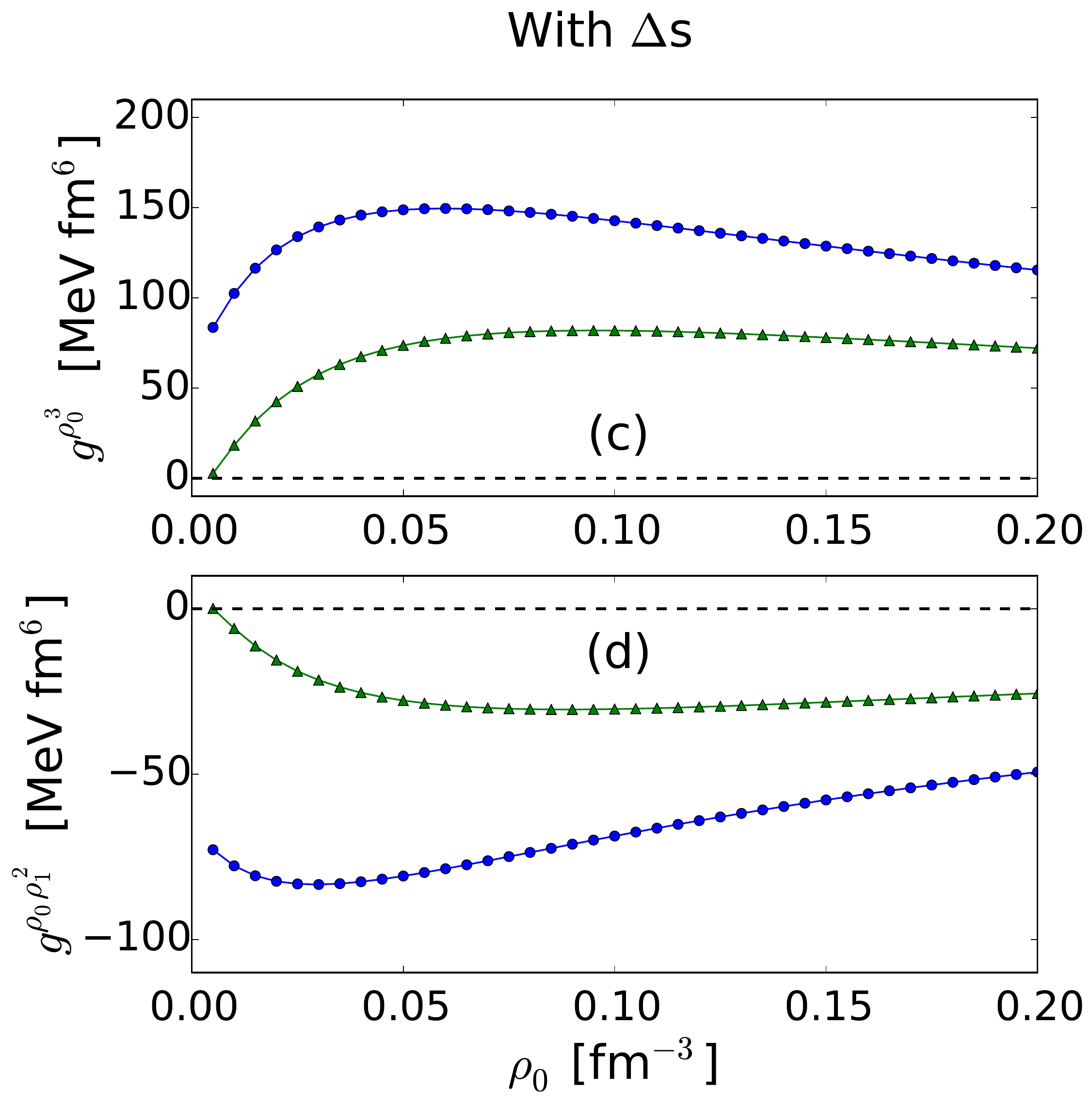}
  \caption{
  The $g^{\rho_0^3}$ and $g^{\rho_0 \rho_1^2}$ couplings are plotted as a function of the isoscalar density $\rho_0$ at fixed cutoff $\RNNN = 1.2~\fm$ using the regulator in Eq.~\ref{eq:regulator} with $n=6$.
  The values for the couplings are shown at two different chiral orders up to \NNLO.
  The coupling $g^{\rho_0^3}$ is shown without (a) and with (c) $\Delta$ isobars. 
    The coupling $g^{\rho_0 \rho_1^2}$ is shown without (b) and with (d) $\Delta$ isobars.}
  \label{fig:gppp_plot_deltas_1p2}
  \medskip
  \includegraphics[width=0.50\textwidth]{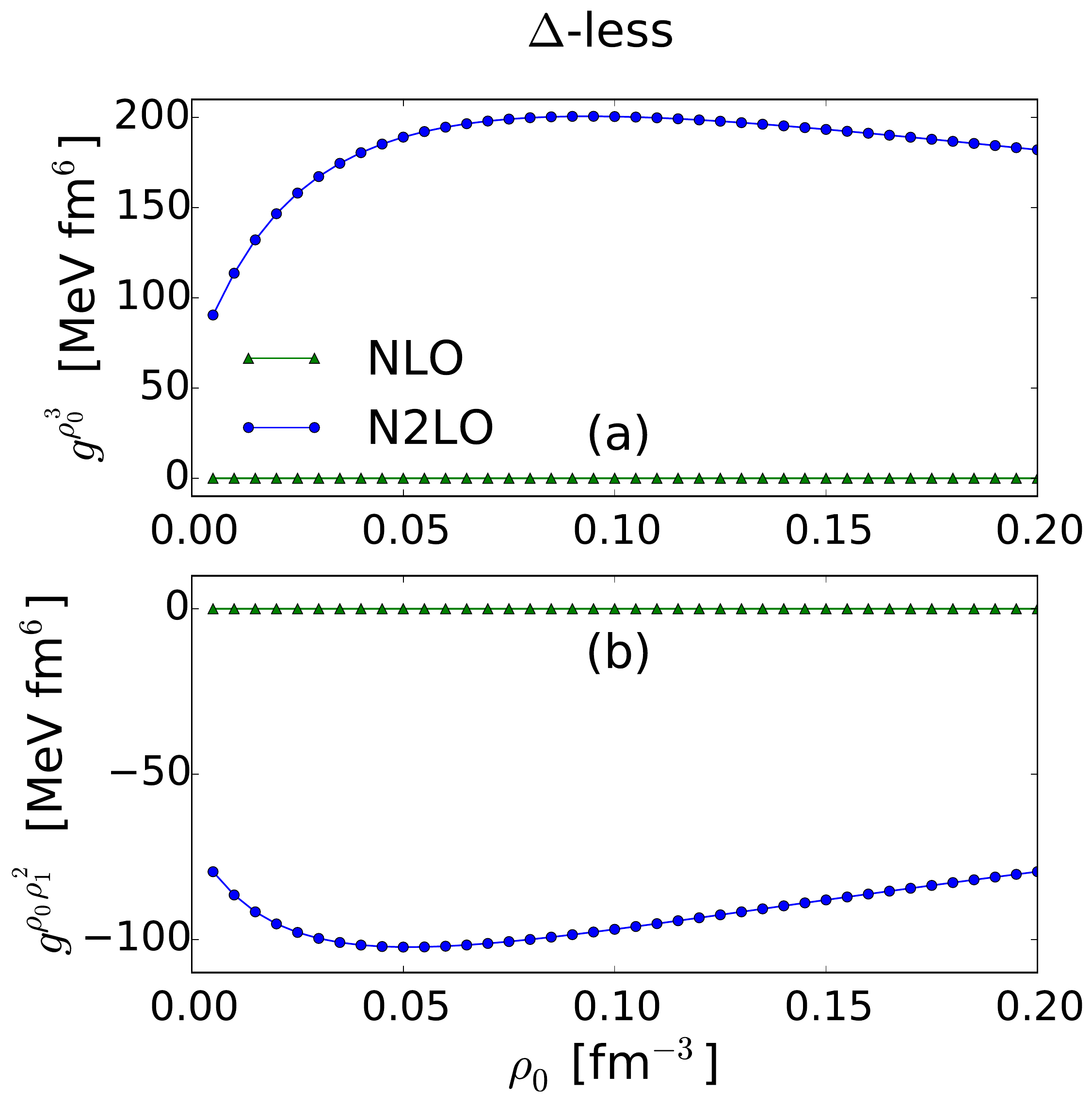}~
  \includegraphics[width=0.50\textwidth]{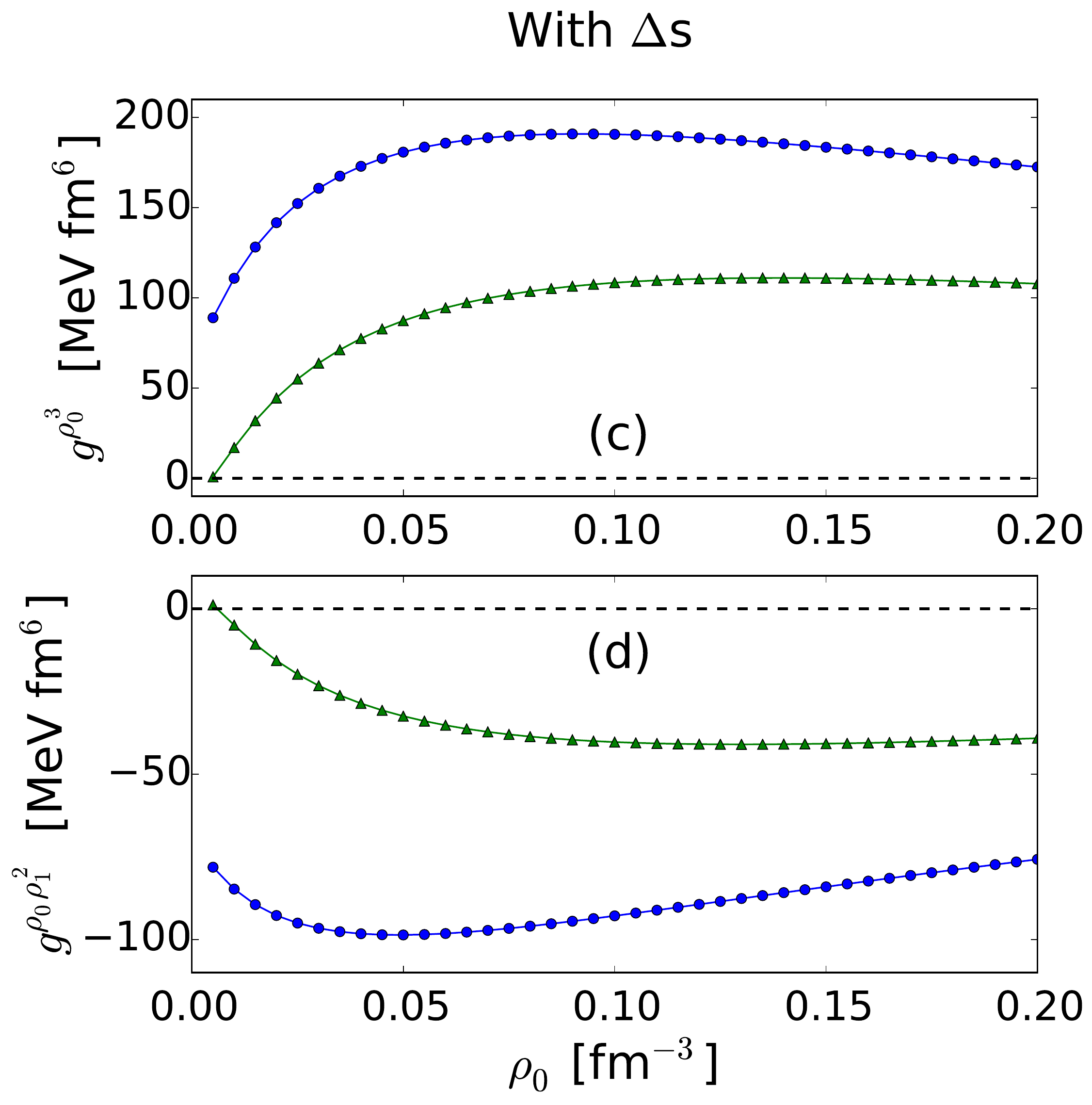}
   \caption{
   The $g^{\rho_0^3}$ and $g^{\rho_0 \rho_1^2}$ couplings are plotted as a function of the isoscalar density $\rho_0$ at fixed cutoff $\RNNN = 1.0~\fm$ using the regulator in Eq.~\ref{eq:regulator} with $n=6$.
  The values for the couplings are shown at two different chiral orders up to \NNLO.
  The coupling $g^{\rho_0^3}$ is shown without (a) and with (c) $\Delta$ isobars. 
    The coupling $g^{\rho_0 \rho_1^2}$ is shown without (b) and with (d) $\Delta$ isobars.}
  \label{fig:gppp_plot_deltas_1p0}
  \end{figure}

  In this section, we show results for some representative examples of the $NN$ 
  and $3N$ couplings $g(\Rvec)$. 
  Values of the various physical parameters and LECs used in these results are 
  given in Table~\ref{tab:param_table}.
  The values of the subleading $c_i$ LECs still have significant uncertainties, but
  we do not consider them in the present discussion.
   For an in-depth discussion on LEC values, see e.g., Ref.~\cite{PhysRevC.94.014620}.
  In the $NN$ sector, we examine the isoscalar and isovector coupling for the density-density term in
  Eq.~\eqref{eq:gpp_coupling}, $g^{\rho \rho}_0$ and $g^{\rho \rho}_1$ respectively.  
  In Fig.~\ref{fig:gpp_plot_deltas_1p2}, we plot these couplings as a function 
  of the isoscalar density at a fixed cutoff $\RNN = 1.2~\fm$ using the regulator defined in Eq.~\eqref{eq:regulator}. 
  The coupling is shown for three different chiral orders up to \NNLO~both with and without explicit $\Delta$ resonances.
  In Fig.~\ref{fig:gpp_plot_deltas_0p8}, we show the same couplings but at a harder cutoff of $\RNN=1.0~\fm$. 
  
  In all cases, the variation of the couplings with density is largest in the low density regime, 
  while at higher densities the couplings tend to asymptote at some finite value.
  This reflects the functional form of the DME $\Pi$ functions 
  used in Sec.~\ref{sec:dme_parametrization}.
  The $\Pi$ functions defined in Eq.~\eqref{eq:dme_pi_functs} are largest at small arguments and have the greatest variation close to zero but die away as the argument increases.
  As the $\Pi$ functions are finite in the limit $\kf\to0$ and $\kf\to\infty$, 
  the couplings are also well-behaved in the same limits.
 Nevertheless, the enhancement of the couplings in the low density limit 
 may necessitate a different fitting procedure.
  As an example, the couplings could be multiplied by the isoscalar density,
  $g^{\rho \rho}_t \to \rho_0 g^{\rho \rho}_t$, 
  such that the very low density behavior is not biasing the functional fit.

  As the cutoff $\RNN$ is lowered from $1.2~\fm$ to $1.0~\fm$, the coupling calculations at 
  different orders tend to move around and spread away from each other.
  This movement is most dramatic at NLO and \NNLO~as these chiral potentials 
  contain increasingly singular terms. 
  However, over the range of the two cutoffs considered here, the \NNLO~couplings vary by less than a factor of 2.
     These differences seen at different chiral orders and cutoffs is expected 
     to be compensated for by a complementary shift in the Skyrme contacts after a global refit is performed.  

  Comparing the $\Delta$-less and the implementation with $\Delta$s, the LO couplings between the 
  two are equivalent as the $\Delta$ does not contribute at this order.
  Going to higher orders, there is a stark difference between the two at NLO which
  is partially restored at \NNLO.
   At the soft cutoff $\RNN = 1.2~\fm$ in
  Fig.~\ref{fig:gpp_plot_deltas_1p2}, this difference is smaller as the soft cutoff excludes a good deal of the chiral potentials.
  Going to the harder cutoff of $\RNN = 1.0~\fm$ in
  Fig.~\ref{fig:gpp_plot_deltas_0p8}, a greater difference between the two formulations becomes evident.  
  
   The convergence pattern of the couplings is also much 
   more systematic when $\Delta$ isobars are included explicitly.
  For the $\Delta$-less case, the LO to NLO difference is 
  quite small and the NLO to \NNLO~difference is rather large.
  This pattern reflects the weakness of the NLO potential in the 
  $\Delta$-less theory as only the lowest order $\pi$-$N$ vertices contribute \cite{Machleidt:20111}.
  In contrast, the couplings with $\Delta$s show a difference 
  between LO and NLO which is larger than the difference between NLO and \NNLO.
  The improved convergence reflects the shift of more physics to the 
  NLO potential when $\Delta$s are included and the more natural $c_i$ coefficients.
  
  In the $3N$ sector, we look at the couplings 
  $g^{\rho_0^3}$ and $g^{\rho_0 \rho_1^2}$.
  In Fig.~\ref{fig:gppp_plot_deltas_1p2}, we plot these couplings as a function of the isoscalar density both with and without explicit $\Delta$s 
  at a soft cutoff of $\RNNN = 1.2~\fm$ using the regulator in Eq.~\eqref{eq:regulator}.
  In Fig.~\ref{fig:gppp_plot_deltas_1p0}, we plot the same couplings only now 
  with a harder cutoff of $\RNNN = 1.0~\fm$.
    Note that the couplings at NLO in the $\Delta$-less theory are exactly zero as the 
  $3N$ diagrams at this order vanish.
  Comparing the couplings at the two different given cutoffs, we do not see especially large variations, i.e., at most a factor of 2 for the \NNLO~couplings. 
  Furthermore at \NNLO, the difference in the couplings between the theory with and without $\Delta$s is small.
  
  Although explicit $\Delta$ resonances do not show drastically different 
  coupling behavior in the few examples here, the larger role of the $\Delta$ remains an open question.
  Furthermore, going to \NNNLO~in the chiral expansion, the ease with which
  the $\Delta$ can be implemented in our formalism will be particularly relevant. 
 At \NNNLO, the $3N$ force in the $\Delta$-less theory is weak, for the same reason the $NN$ potential is weak at NLO (only leading order $\pi$-$N$ vertices contribute).
  Due to resonance saturation of the $c_i$ coefficients, $3N$ diagrams in the $\Delta$-less theory at
  \NNNNLO~are therefore expected to be sizeable while in the theory with $\Delta$s, these diagrams are promoted to \NNNLO~\cite{Machledit:2010}.
  Converging the $3N$ potential may therefore be simpler when working up to \NNNLO~and
  including the $\Delta$ explicitly.



\section{Conclusions}
\label{sec:conclusions}

  In this work, we derived density-dependent couplings from coordinate space chiral potentials 
  working up to
  \NNLO~in the chiral expansion using the DME parameterization outlined in Sec.~\ref{sec:dme_parametrization}. 
    The chiral potentials we used are derived using Weinberg power counting and come in two versions depending on whether $\Delta$ isobars are included as explicit degrees of freedom.
    Our couplings are derived by applying the DME to OBDMs at the Hartree-Fock level in MBPT.
  Working to Hartree-Fock in MBPT is justified here due to both the softness of regulated chiral potentials 
  and the global refit of Skyrme contacts, which are expected to mimic higher-order many-body contributions.
  In Eqs.~\eqref{eq:nn_edf} and \eqref{eq:nnn_edf}, we show the resulting EDF forms in the $NN$ and $3N$ sector respectively.
  We also implemented a new organization scheme which renders tractable and modular the DME algebra associated with three-body interactions.
  The resulting density dependent $NN$ and $3N$ DME couplings then serve as input into a Skyrme-like functional.
  These couplings and their respective local densities can be added to a standard Skyrme functional and used in existing EDF solvers.

  Our work builds upon but ultimately contrasts with the previous DME implementation
  in Refs.~\cite{Gebremariam201117,Stoitsov:2010ha}, which used chiral forces defined
  in momentum space and did not include ultraviolet regulators. 
  We instead work in coordinate space and include regulators for several reasons:
  \be
  \item The DME is naturally formulated as an expansion in coordinate space
  about the OBDM nonlocality, with
  no Fourier integrals to be done for the DME coupling equations.
  \item Including an ultraviolet regulator allows for an adiabatic `turning on' of the the long-range chiral potentials by changing the regularization cutoff.
  \item Coordinate space regulators, which have been shown in certain instances to better control regulator artifacts, can easily be implemented on coordinate space interactions.
  \ee
  Working with coordinate space interactions, numerically performing relevant integrals, and using a new robust organization scheme have also
   rendered the $3N$ DME implementation more transparent than the previous formulation.
  We hope that including regulators will address concerns with functional stability and optimization 
  and ultimately lead to significant improvement over existing functionals,
  or else serve as diagnostics if improvement is not found.

  Going forward, we plan to apply these DME derived couplings in semi-phenomenological calculations of nuclei and address many pertinent questions \cite{Dyhdalo:2016}: 
  How will our new functional calculations compare with existing models?
  Does the re-fit of the Skyrme parameters adequately capture the contributions
  of the omitted $V_D$ term and higher-order MBPT?
  Can we see evidence of pion exchange in medium and heavy nuclei? 
  To what extent are $\Delta$ isobars important as a degree of freedom?
  What are the effects of including $3N$ forces in heavy systems?
  Can we decrease uncertainty in functional calculations to such a degree that outputs will be relevant for astrophysical and standard model experiments?
  Can isovector contributions in the functional be sufficiently constrained so as to make predictions in advance of the Facility for Rare Isotope Beams (FRIB)?
  These questions also inform how one might improve upon the present Weinberg power
   counting for vacuum nucleon interactions. 
  If the addition of pion and delta physics does not provide measurable improvement
  for functional calculations, these degrees of freedom may not relevant at finite density and a 
  different EFT construction should be done for in-medium systems.
  This perspective would translate Skyrme functionals from a 
  successful phenomenology to the lowest rung of a systematic EFT.

  In the future, we aim to eventually construct a truly \textit{ab initio} functional for nuclei. 
   For such a project, reaching convergence is crucial, both in the chiral expansion and the many-body sector.
    The ease with which our formalism deals with including the $\Delta$ isobar
 then becomes particularly relevant especially with respect to $3N$ forces at \NNNLO.
   For renormalization group softened chiral interactions, Hartree-Fock in MBPT becomes a quantitative starting point.
   However, non-trivial challenges remain both in the implementation of nonlocal softened chiral forces as well as including second-order many-body contributions.
  We plan to address these challenges in a future work, with the present 
  semi-phenomenological formulation being a modest step towards first-principle 
  predictions for the full table of nuclides.

\section*{Acknowledgements}

	We would like to thank A. Lovato, R. Navarro Perez, N.~Schunck, I. Tews, and S.~Wesolowski 
  for useful discussions.
  We would like to especially thank T. Coello Pérez and L. Zurek for numerical comparisons and spotting two mistakes in our derivations.
	We are also indebted to the detailed and precise work done by B.
	Gebremariam on the density matrix expansion using chiral interactions. 
	This work was supported in part by the National Science Foundation under 
  Grant Nos. PHY-1306250, PHY-1404159, and PHY-1614460, and
  the NUCLEI SciDAC Collaboration under DOE Grants DE-SC0008533 and DE-SC0008511.

\appendix

\section{\texorpdfstring{$\Delta$}{Delta}-full \texorpdfstring{$NN$}{NN} and 
   \texorpdfstring{$3N$}{3N} Chiral Potentials}
\label{sec:chiral_appendix}

\subsection{\texorpdfstring{$\Delta$}{Delta}-full \texorpdfstring{$NN$}{NN} Potentials}

	Here we list the contributions to the $V_i$ and $W_i$ form factors due to single and double $\Delta$ excitation at NLO and \NNLO. 
	Below, we use the notation that
\beq
	x \equiv r \mpi \; , 
	\qquad y \equiv r \delm \; , 
	\qquad F_{\pi} = 2 \fpi \; .
\eeq
	For the purpose of collating all useful information in a single document, we quote verbatim the results of Appendix A in Ref.~\cite{PhysRevC.91.024003}.
	For single $\Delta$ excitation at NLO, the form factors are given by:
\begin{eqnarray}
V^{(2)}_{C} (r;\Delta)&=&-\frac{1}
 {6 \pi^2  r^5\,y}\frac{g_A^2 h_A^2}{F_{\pi}^4}e^{-2x}\left(6+12x+10x^2+4x^3+x^4\right)\;, 
  \\
W^{(2)}_C (r;\Delta)&=&-\frac{1}
  {216 \pi^3  r^5}\frac{h_A^2}{F_{\pi}^4}\Bigg[\int_{0}^{\infty}d\mu \frac{\mu^2}
  {\sqrt{\mu^2+4x^2}}e^{-\sqrt{\mu^2+4x^2}}(12x^2+5\mu^2+12y^2) \nonumber\\
  && \quad\null -12y\int_{0}^{\infty}d\mu \frac{\mu}
  {\sqrt{\mu^2+4x^2}}e^{-\sqrt{\mu^2+4x^2}}(2x^2+\mu^2+2y^2)\arctan{\frac{\mu}{2y}}\Bigg]
     \nonumber\\
  && \quad\null -\frac{1}{216 \pi^3  r^5}\frac{g_A^2 h_A^2}{F_{\pi}^4}
    \Bigg[-\int_{0}^{\infty}d\mu \frac{\mu^2}{\sqrt{\mu^2+4x^2}}e^{-\sqrt{\mu^2+4x^2}}(24x^2+11\mu^2+12y^2)\nonumber\\
  && \quad\null +\frac{6}{y}\int_{0}^{\infty}d\mu \frac{\mu}
   {\sqrt{\mu^2+4x^2}}e^{-\sqrt{\mu^2+4x^2}}(2x^2+\mu^2+2y^2)^2\arctan{\frac{\mu}{2y}}\Bigg]
    \;,\\
V^{(2)}_S (r;\Delta)&=&-\frac{1}
{72 \pi^3  r^5}\frac{g_A^2 h_A^2}{F_{\pi}^4}\Bigg[2\,\int_{0}^{\infty}d\mu \frac{\mu^2}
{\sqrt{\mu^2+4x^2}}e^{-\sqrt{\mu^2+4x^2}}(\mu^2+4x^2)\nonumber \\
  && \quad\null -\frac{1}{y}\int_{0}^{\infty}d\mu \frac{\mu}
  {\sqrt{\mu^2+4x^2}}e^{-\sqrt{\mu^2+4x^2}}(\mu^2+4x^2)(\mu^2+4y^2)\arctan{\frac{\mu}{2y}}\Bigg]\;,\\
W^{(2)}_S (r;\Delta)&=&\frac{1}
{54 \pi^2  r^5\,y}\frac{g_A^2 h_A^2}{F_{\pi}^4}e^{-2x}\left(1+x \right)\left(3+3x+x^2\right)
   \;, \\
V^{(2)}_T (r;\Delta)&=&\frac{1}
   {144 \pi^3  r^5}\frac{g_A^2 h_A^2}{F_{\pi}^4}\Bigg[2\,\int_{0}^{\infty}d\mu \frac{\mu^2}
   {\sqrt{\mu^2+4x^2}}e^{-\sqrt{\mu^2+4x^2}}(3+3\sqrt{\mu^2+4x^2}+\mu^2+4x^2)\nonumber \\
  && \quad\null -\frac{1}{y}\int_{0}^{\infty}d\mu\frac{\mu}
     {\sqrt{\mu^2+4x^2}}e^{-\sqrt{\mu^2+4x^2}}(\mu^2+4y^2) \nonumber \\
  && \quad\null \times (3+3\sqrt{\mu^2+4x^2}
     +\mu^2+4x^2)\arctan{\frac{\mu}{2y}}\Bigg]\;, \\
W^{(2)}_T (r;\Delta)&=&-\frac{1}
   {54 \pi^2  r^5\,y}\frac{g_A^2 h_A^2}{F_{\pi}^4}e^{-2x}\left(1+x \right)\left(3+3x+2x^2\right) \;.
\end{eqnarray}	
	The double $\Delta$ excitations at NLO are given by:
\begin{eqnarray}
 V^{(2)}_C (r;2\Delta) &=& {-\frac{1}{108 \pi^3  r^5}} \frac{h_A^4}{F_{\pi}^4}
   \Bigg[\int_{0}^{\infty}d\mu\frac{\mu^2}
   {\sqrt{\mu^2+4x^2}}e^{-\sqrt{\mu^2+4x^2}}
   \left[4y^2+2\frac{(2x^2+\mu^2+2y^2)^2}{(\mu^2+4y^2)}\right]\nonumber\\
  && \quad\null+\frac{1}{y}\int_{0}^{\infty}d\mu\frac{\mu}
   {\sqrt{\mu^2+4x^2}}e^{-\sqrt{\mu^2+4x^2}}\nonumber\\
  && \quad\null\times(2x^2+\mu^2+2y^2)(2x^2+\mu^2-6y^2)
     \arctan{\frac{\mu}{2y}}\Bigg]\;,\\
W^{(2)}_C (r;2\Delta) &=&-\frac{1}
  {1944 \pi^3  r^5}\frac{h_A^4}{F_{\pi}^4}\Bigg[\int_{0}^{\infty}d\mu\frac{\mu^2}
  {\sqrt{\mu^2+4x^2}}e^{-\sqrt{\mu^2+4x^2}}
  \nonumber\\
 && \quad\null\times \biggl[(24x^2+11\mu^2+24y^2)
   +6\frac{(2x^2+\mu^2+2y^2)^2}{(\mu^2+4y^2)}\biggr]\nonumber\\
 && \quad\null -\frac{3}{y}\int_{0}^{\infty}d\mu\frac{\mu}
   {\sqrt{\mu^2+4x^2}}e^{-\sqrt{\mu^2+4x^2}}\nonumber\\
 && \quad\null\times(2x^2+\mu^2+2y^2)(2x^2+\mu^2+10y^2)\arctan{\frac{\mu}{2y}}\Bigg]\;,\\
V^{(2)}_S (r;2\Delta) &=&-\frac{1}
  {1296 \pi^3  r^5}\frac{h_A^4}{F_{\pi}^4}\Bigg[{-6} \int_{0}^{\infty}d\mu\frac{\mu^2}
  {\sqrt{\mu^2+4x^2}}e^{-\sqrt{\mu^2+4x^2}}
  (\mu^2+4x^2)\nonumber\\
&& \quad\null +\frac{1}{y}\int_{0}^{\infty}d\mu\frac{\mu}
  {\sqrt{\mu^2+4x^2}}e^{-\sqrt{\mu^2+4x^2}}(\mu^2+4x^2)
  (\mu^2+12y^2)\arctan{\frac{\mu}{2y}}\Bigg]\;,\\
W^{(2)}_S (r;2\Delta) &=&-\frac{1}
  {7776 \pi^3  r^5}\frac{h_A^4}{F_{\pi}^4}\Bigg[{-2} \int_{0}^{\infty}d\mu\frac{\mu^2}
  {\sqrt{\mu^2+4x^2}}e^{-\sqrt{\mu^2+4x^2}}
  (\mu^2+4x^2)\nonumber\\
 && \quad\null +\frac{1}{y}\int_{0}^{\infty}d\mu\frac{\mu}
  {\sqrt{\mu^2+4x^2}}e^{-\sqrt{\mu^2+4x^2}}(\mu^2+4x^2)
  (-\mu^2+4y^2)\arctan{\frac{\mu}{2y}}\Bigg]\;,\\
V^{(2)}_T (r;2\Delta) &=&\frac{1}
  {2592 \pi^3  r^5}\frac{h_A^4}{F_{\pi}^4}\Bigg[{-6} \int_{0}^{\infty}d\mu\frac{\mu^2}
  {\sqrt{\mu^2+4x^2}}e^{-\sqrt{\mu^2+4x^2}}
  (3+3\sqrt{\mu^2+4x^2}+\mu^2+4x^2)\nonumber\\
 && \quad\null +\frac{1}{y}\int_{0}^{\infty}d\mu\frac{\mu}
  {\sqrt{\mu^2+4x^2}}e^{-\sqrt{\mu^2+4x^2}}\nonumber\\
 && \quad\null \times(3+3\sqrt{\mu^2+4x^2}+\mu^2+4x^2)
(\mu^2+12y^2)\arctan{\frac{\mu}{2y}}\Bigg]\;,\\
W^{(2)}_T (r;2\Delta) &=&\frac{1}
  {15552 \pi^3  r^5}\frac{h_A^4}{F_{\pi}^4}\Bigg[{-2} \int_{0}^{\infty}d\mu\frac{\mu^2}
  {\sqrt{\mu^2+4x^2}}e^{-\sqrt{\mu^2+4x^2}}
  (3+3\sqrt{\mu^2+4x^2}+\mu^2+4x^2)\nonumber\\
 && \quad\null +\frac{1}{y}\int_{0}^{\infty}d\mu\frac{\mu}
  {\sqrt{\mu^2+4x^2}}e^{-\sqrt{\mu^2+4x^2}}\nonumber\\
 && \quad\null \times(3+3\sqrt{\mu^2+4x^2}+\mu^2+4x^2)
  (-\mu^2+4y^2)\arctan{\frac{\mu}{2y}}\Bigg]\;.
\end{eqnarray}
	The single $\Delta$ excitation form factors at \NNLO~are given by:
\begin{eqnarray}
V^{(3)}_C (r;\Delta)&=&\frac{1}
  {18 \pi^3  r^6}\frac{h_A^2\,y}{F_{\pi}^4}\Bigg[\int_{0}^{\infty}d\mu\frac{\mu^2}
  {\sqrt{\mu^2+4x^2}}e^{-\sqrt{\mu^2+4x^2}}\nonumber\\
 && \qquad\qquad\qquad\quad\null\times\left[-24c_1x^2+c_2(5\mu^2+12x^2+12y^2)-6c_3(\mu^2+2x^2) \right] 
   \nonumber\\
 && \quad\null +\frac{6}{y}\int_{0}^{\infty}\!\!d\mu\frac{\mu}
  {\sqrt{\mu^2+4x^2}}e^{-\sqrt{\mu^2+4x^2}} (\mu^2\!+\!2x^2\!+\!2y^2)\nonumber \\
 && \qquad\qquad\quad\null \times 
   [4c_1x^2-2c_2y^2+c_3(\mu^2+2x^2)]\,\arctan{\frac{\mu}{2y}}\Bigg]\;,\\
W^{(3)}_C (r;\Delta)&=&-\frac{1}
  {54 \pi^3  r^6}\frac{(b_3+b_8)\,h_A\,y}{F_{\pi}^4}\Bigg[+\int_{0}^{\infty}d\mu\frac{\mu^2}
  {\sqrt{\mu^2+4x^2}}e^{-\sqrt{\mu^2+4x^2}}(5\mu^2+12x^2+12y^2)\nonumber\\
 && \quad\null -12\,y\int_{0}^{\infty}d\mu\frac{\mu}
  {\sqrt{\mu^2+4x^2}}e^{-\sqrt{\mu^2+4x^2}}(\mu^2+2x^2+2y^2)\,\arctan{\frac{\mu}{2y}}\Bigg]\nonumber\\
 && \quad\null -\frac{1}{54 \pi^3  r^6}\frac{(b_3+b_8)\,h_A\,g_A^2\,y}{F_{\pi}^4}\Bigg[-\int_{0}^{\infty}d\mu\frac{\mu^2}
  {\sqrt{\mu^2+4x^2}}e^{-\sqrt{\mu^2+4x^2}} \nonumber\\
 && \qquad\qquad\qquad\qquad\qquad\qquad\quad\null\times (11\mu^2+24x^2+12y^2)\nonumber\\
 && \quad\null +\frac{6}{y}\int_{0}^{\infty}d\mu\frac{\mu}{\sqrt{\mu^2+4x^2}}e^{-\sqrt{\mu^2+4x^2}}
  \left(\mu^2+2x^2+2y^2\right)^2\arctan{\frac{\mu}{2y}}\Bigg]\;,
\end{eqnarray}
\begin{eqnarray}
V^{(3)}_S (r;\Delta)&=&-\frac{1}
  {18 \pi^3  r^6}\frac{(b_3+b_8)\,h_A\,g_A^2\,y}{F_{\pi}^4}\Bigg[2\int_{0}^{\infty}d\mu\frac{\mu^2}
  {\sqrt{\mu^2+4x^2}}e^{-\sqrt{\mu^2+4x^2}}(\mu^2+4x^2)\nonumber\\
 && \quad\null -\frac{1}{y}\int_{0}^{\infty}d\mu\frac{\mu}{\sqrt{\mu^2+4x^2}}e^{-\sqrt{\mu^2+4x^2}}
  (\mu^2+4x^2) \nonumber\\
 && \qquad\qquad\qquad\qquad\qquad\qquad\quad\null\times  (\mu^2+4y^2)\arctan{\frac{\mu}{2y}}\Bigg]\;,\\
W^{(3)}_S (r;\Delta)&=&-\frac{1}
  {108 \pi^3  r^6}\frac{c_4\,h_A^2\,y}{F_{\pi}^4}\Bigg[2\,\int_{0}^{\infty}d\mu\frac{\mu^2}
  {\sqrt{\mu^2+4x^2}}e^{-\sqrt{\mu^2+4x^2}}(\mu^2+4x^2)\nonumber\\
 && \quad\null -\frac{1}{y}\int_{0}^{\infty}d\mu\frac{\mu}
  {\sqrt{\mu^2+4x^2}}e^{-\sqrt{\mu^2+4x^2}}(\mu^2+4x^2) \nonumber\\
 && \qquad\qquad\qquad\qquad\qquad\qquad\quad\null\times  (\mu^2+4y^2)\,\arctan{\frac{\mu}{2y}}\Bigg]\;,
\end{eqnarray}
\begin{eqnarray}
V^{(3)}_T (r;\Delta)&=&\frac{1}
  {36 \pi^3  r^6}\frac{(b_3+b_8)\,h_A\,g_A^2\,y}{F_{\pi}^4}\Bigg[2\int_{0}^{\infty}d\mu\frac{\mu^2}
  {\sqrt{\mu^2+4x^2}}e^{-\sqrt{\mu^2+4x^2}} \nonumber\\
 && \qquad\qquad\qquad\qquad\qquad\qquad\quad\null \times
  (3+3\sqrt{\mu^2+4x^2}+\mu^2+4x^2)\nonumber\\
  && \quad\null -\frac{1}{y}\int_{0}^{\infty}d\mu\frac{\mu}{\sqrt{\mu^2+4x^2}}e^{-\sqrt{\mu^2+4x^2}}
\nonumber \\
  && \qquad\qquad\quad\null \times(3+3\sqrt{\mu^2+4x^2}+\mu^2+4x^2)(\mu^2+4y^2)\arctan{\frac{\mu}{2y}}\Bigg]\;,\\
W^{(3)}_T (r;\Delta)&=&\frac{1}
{216 \pi^3  r^6}\frac{c_4\,h_A^2\,y}{F_{\pi}^4}\Bigg[2\,\int_{0}^{\infty}d\mu\frac{\mu^2}
{\sqrt{\mu^2+4x^2}}e^{-\sqrt{\mu^2+4x^2}}(3+3\sqrt{\mu^2+4x^2}+\mu^2+4x^2)\nonumber\\
&& \quad\null -\frac{1}{y}\int_{0}^{\infty}d\mu\frac{\mu}
{\sqrt{\mu^2+4x^2}}e^{-\sqrt{\mu^2+4x^2}}
\nonumber\\
&& \qquad\qquad\quad\null \times(3+3\sqrt{\mu^2+4x^2}+\mu^2+4x^2)
(\mu^2+4y^2)\,\arctan{\frac{\mu}{2y}}\Bigg]\  .
\end{eqnarray}
	The double $\Delta$ excitation form factors at \NNLO~are given by:
\begin{eqnarray}
V^{(3)}_C (r;2\Delta)&=&-\frac{2}
  {81 \pi^3  r^6}\frac{(b_3+b_8)\,h_A^3\,y}{F_{\pi}^4}\Bigg[
  \int_{0}^{\infty}d\mu\frac{\mu^2}{\sqrt{\mu^2+4x^2}}e^{-\sqrt{\mu^2+4x^2}}
  \nonumber\\
 &&\times \left[ 6\frac{(\mu^2+2x^2+2y^2)^2}{\mu^2+4y^2}+11\mu^2+24x^2+12y^2 \right]
  \nonumber\\
 &&-\frac{3}{y}\int_{0}^{\infty}d\mu\frac{\mu}
  {\sqrt{\mu^2+4x^2}}e^{-\sqrt{\mu^2+4x^2}}(\mu^2+2x^2+10y^2) \nonumber\\
 && \qquad\qquad\qquad\qquad\qquad\qquad\quad\null\times (\mu^2+2x^2+2y^2)\arctan{\frac{\mu}{2y}}\Bigg]\;,  \\
 W^{(3)}_C (r;2\Delta)&=&-\frac{1}
  {243 \pi^3  r^6}\frac{(b_3+b_8)\,h_A^3\,y}{F_{\pi}^4}\Bigg[
  \int_{0}^{\infty}d\mu\frac{\mu^2}{\sqrt{\mu^2+4x^2}}e^{-\sqrt{\mu^2+4x^2}}
  \nonumber\\
 &&\times \left[ 6\frac{(\mu^2+2x^2+2y^2)^2}{\mu^2+4y^2}+11\mu^2+24x^2+12y^2 \right]
  \nonumber\\
 &&-\frac{3}{y}\int_{0}^{\infty}d\mu\frac{\mu}
  {\sqrt{\mu^2+4x^2}}e^{-\sqrt{\mu^2+4x^2}}(\mu^2+2x^2+10y^2) \nonumber\\
 && \qquad\qquad\qquad\qquad\qquad\qquad\quad\null\times(\mu^2+2x^2+2y^2)
  \arctan{\frac{\mu}{2y}}\Bigg]\;, 
\end{eqnarray}
\begin{eqnarray}
V^{(3)}_S (r;2\Delta)&=&-\frac{1}
  {162 \pi^3  r^6}\frac{(b_3+b_8)\,h_A^3\,y}{F_{\pi}^4}\Bigg[
  {-}6\int_{0}^{\infty}d\mu\frac{\mu^2}{\sqrt{\mu^2+4x^2}}e^{-\sqrt{\mu^2+4x^2}}
  (\mu^2+4x^2)\nonumber\\
 && \quad\null +\frac{1}{y}\int_{0}^{\infty}d\mu\frac{\mu}
 {\sqrt{\mu^2+4x^2}}e^{-\sqrt{\mu^2+4x^2}}(\mu^2+4x^2) \nonumber\\
 && \qquad\qquad\qquad\qquad\qquad\qquad\quad\null\times(\mu^2+12y^2)
  \arctan{\frac{\mu}{2y}}\Bigg]\;,\\
W^{(3)}_S (r;2\Delta)&=&-\frac{1}
  {972 \pi^3  r^6}\frac{(b_3+b_8)\,h_A^3\,y}{F_{\pi}^4}\Bigg[
  {-}6\int_{0}^{\infty}d\mu\frac{\mu^2}{\sqrt{\mu^2+4x^2}}e^{-\sqrt{\mu^2+4x^2}}
  (\mu^2+4x^2)\nonumber\\
 && \quad\null +\frac{1}{y}\int_{0}^{\infty}d\mu\frac{\mu}
  {\sqrt{\mu^2+4x^2}}e^{-\sqrt{\mu^2+4x^2}}(\mu^2+4x^2) \nonumber\\
 && \qquad\qquad\qquad\qquad\qquad\qquad\quad\null\times(\mu^2+12y^2)
  \arctan{\frac{\mu}{2y}}\Bigg]\;,
\end{eqnarray}
\begin{eqnarray}
V^{(3)}_T (r;2\Delta)&=&\frac{1}
  {324 \pi^3  r^6}\frac{(b_3+b_8)\,h_A^3\,y}{F_{\pi}^4}\Bigg[
  {-}6\int_{0}^{\infty}d\mu\frac{\mu^2}{\sqrt{\mu^2+4x^2}}e^{-\sqrt{\mu^2+4x^2}}
  \nonumber\\
 && \qquad\qquad\qquad\qquad\qquad\qquad\quad\null\times  (3+3\sqrt{\mu^2+4x^2}+\mu^2+4x^2)\nonumber\\
 && \quad\null+\frac{1}{y}\int_{0}^{\infty}d\mu\frac{\mu}
  {\sqrt{\mu^2+4x^2}}e^{-\sqrt{\mu^2+4x^2}}(3+3\sqrt{\mu^2+4x^2}+\mu^2+4x^2)
  \nonumber \\
  && \qquad\qquad\qquad\qquad\null\times(\mu^2+12y^2)
\arctan{\frac{\mu}{2y}}\Bigg]\;, \\
W^{(3)}_T (r;2\Delta)&=&\frac{1}
  {1944 \pi^3  r^6}\frac{(b_3+b_8)\,h_A^3\,y}{F_{\pi}^4}\Bigg[
  {-}6\int_{0}^{\infty}d\mu\frac{\mu^2}{\sqrt{\mu^2+4x^2}}e^{-\sqrt{\mu^2+4x^2}}
  \nonumber \\
 && \qquad\qquad\qquad\qquad\null\times  (3+3\sqrt{\mu^2+4x^2}+\mu^2+4x^2)\nonumber\\
 && \quad\null +\frac{1}{y}\int_{0}^{\infty}d\mu\frac{\mu}
   {\sqrt{\mu^2+4x^2}}e^{-\sqrt{\mu^2+4x^2}}(3+3\sqrt{\mu^2+4x^2}+\mu^2+4x^2)
   \nonumber \\
 && \qquad\qquad\qquad\qquad\null\times (\mu^2+12y^2)
  \arctan{\frac{\mu}{2y}}\Bigg]\;.
\end{eqnarray}
	
\subsection{\texorpdfstring{$3N$}{3N} Potentials}

	Here, we give the complete form of the $V_C$ $3N$ potentials at NLO and \NNLO~for the potential term 
	$V_{23} \left( \rvec_{21}, \rvec_{31}, \left\{\sigma \tau \right\} \right)$ 
	including all short-range contact terms. The first $V_C$ term is given by \cite{Tews:2015,PhysRevC.85.024003}:
\begin{align*}
  V_{C,1} &= 
  \; \tauvec_2 \cdot \tauvec_3 \;
  (\sigvec_2 \cdot \rhat_{21})
  (\sigvec_3 \cdot \rhat_{31})
  \;
  U(r_{21}) Y(r_{21})
  U(r_{31}) Y(r_{31})
  \;,
  \numberthis
  \label{eq:3n_vc1}
\end{align*}
where again we have only written down the $V_{23} (\rvec_{21}, \rvec_{31}, 
\{\sigma \tau\})$ part of the potential.
 The second $V_C$ term is more complicated~\cite{Tews:2015,PhysRevC.85.024003}:
\begin{align*}
  V_{C,2} &=  \
  \tauvec_2 \cdot \tauvec_3 
  \Bigl\{
  \frac{16\pi^2}{\mpi^6} 
  \sigvec_2 \cdot \sigvec_3 \;
  \delta^3(\rvec_{21}) \;
  \delta^3(\rvec_{31})
\\
  & \quad\null
  - 
  \frac{4\pi}{\mpi^3}
  \Bigl[
  S_{23} (\rhat_{21}) T(r_{21}) 
  + \sigvec_2 \cdot \sigvec_3
  \Bigl] Y (r_{21}) \;
  \delta^3(\rvec_{31})
\\
  & \quad\null
  -
  \frac{4\pi}{\mpi^3}
  \Bigl[
  S_{23} (\rhat_{31}) T(r_{31}) 
  + \sigvec_2 \cdot \sigvec_3
  \Bigl] Y (r_{31}) \;
  \delta^3(\rvec_{21})
\\
  & \quad\null +
  \Bigl[
  9 (\sigvec_2 \cdot \rhat_{21})
  (\sigvec_3 \cdot \rhat_{31})
  (\rhat_{21} \cdot \rhat_{31})
  -
  3 (\sigvec_2 \cdot \rhat_{21})
  (\sigvec_3 \cdot \rhat_{21})
\\
  & \quad\null -
  3 (\sigvec_2 \cdot \rhat_{31})
  (\sigvec_3 \cdot \rhat_{31})
  +
  \sigvec_2 \cdot \sigvec_3
  \Bigr] T(r_{21}) Y(r_{21}) 
  T(r_{31}) Y(r_{31})
  +
  (\sigvec_2 \cdot \sigvec_3) 
  Y(r_{21}) Y(r_{31})
\\
  & \quad\null +
  S_{23}(\rhat_{21}) T(r_{21})Y(r_{21}) Y(r_{31})
  +
  S_{23}(\rhat_{31}) T(r_{31})Y(r_{31}) Y(r_{21})
  \Bigr\}
  \;.
  \numberthis
  \label{eq:3n_vc2}
\end{align*}
The third $V_C$ term is the most involved \cite{Tews:2015,PhysRevC.85.024003}:
\begin{align*}
  V_{C,3} & =  \
  \tauvec_2 \cdot (\tauvec_3 \times \tauvec_1)
  \Bigl\{
  \frac{16\pi^2}{\mpi^6} 
  \delta^3(\rvec_{21}) 
  \delta^3(\rvec_{31})
  \sigvec_2 \cdot (\sigvec_3 \times \sigvec_1)
\\
  & \quad\null -
  \frac{12\pi}{\mpi^3}
  (\sigvec_2 \cdot \rhat_{21})
  \rhat_{21} \cdot (\sigvec_3 \times \sigvec_1)
  T(r_{21}) Y(r_{21}) \delta^3(\rvec_{31})
\\
  & \quad\null  -
  \frac{4\pi}{\mpi^3} 
  \sigvec_2 \cdot (\sigvec_3 \times \sigvec_1)
  (1 - T(r_{21})) Y(r_{21}) \delta^3(\rvec_{31})
\\
  & \quad\null -
  \frac{12\pi}{\mpi^3}
  (\sigvec_3 \cdot \rhat_{31})
  \rhat_{31} \cdot (\sigvec_1 \times \sigvec_2)
  T(r_{31}) Y(r_{31}) \delta^3(\rvec_{21})
\\
  & \quad\null  -
  \frac{4\pi}{\mpi^3} 
  \sigvec_2 \cdot (\sigvec_3 \times \sigvec_1)
  (1 - T(r_{31})) Y(r_{31}) \delta^3(\rvec_{21})
\\
  & \quad\null + 9 (\sigvec_2 \cdot \rhat_{21})
  (\sigvec_3 \cdot \rhat_{31})
  \sigvec_1 \cdot (\rhat_{21} \times \rhat_{31})
  T(r_{21})Y(r_{21}) T(r_{31}) Y(r_{31})
\\
  & \quad\null + 3 (\sigvec_2 \cdot \rhat_{21})
  \rhat_{21} \cdot (\sigvec_3 \times \sigvec_1)
  T(r_{21}) (1 - T(r_{31})) Y(r_{21}) Y(r_{31})
\\
  & \quad\null + 3 (\sigvec_3 \cdot \rhat_{31})
  \rhat_{31} \cdot (\sigvec_1 \times \sigvec_2)
  T(r_{31}) (1 - T(r_{21})) Y(r_{21}) Y(r_{31})
\\
  & \quad\null +
  \sigvec_1 \cdot (\sigvec_2 \times \sigvec_3)
  (1 - T(r_{21}))(1 - T(r_{31})) Y(r_{21}) Y(r_{31})
  \Bigr\}
  \;.
  \numberthis
  \label{eq:3n_vc3}
\end{align*}

\section{\texorpdfstring{$3N$}{3N} DME Expansion Example}
\label{sec:dme_expansion_example}

In this appendix, we show an example derivation of the simplified PSA-DME expansion
equations for the three-body system.
In the following, all isospin labels are suppressed.
We derive the vector part of the generalized PSA-DME with two nonlocality 
variables $\xvec_2, \xvec_3$, where
first the vector density matrix is exactly rewritten in a derivative expansion,
\beq
	s_b (\rvec_1 + \xvec_2, \rvec_1 + \xvec_3) = 
	e^{\xvec_2 \cdot \nabla_2}
	e^{\xvec_3 \cdot \nabla_3}
	s_b (\rvec_2, \rvec_3) 
	|_{\rvec_1 = \rvec_2 = \rvec_3} \;.
\eeq
Introducing an arbitrary momentum vector $\kvec$ and 
expanding the exponentials with derivatives to linear order, the above becomes,
\begin{align*}
	s_b (\rvec_1 + \xvec_2, \rvec_1 + \xvec_3) 
	&= 
	e^{i \kvec \cdot \left(
	\xvec_2 - \xvec_3
	\right)}
	e^{\xvec_2 \cdot \left(
	\nabla_2
	- i \kvec
	\right)}
	e^{\xvec_3 \cdot \left(
	\nabla_3
	+ i \kvec
	\right)}
	s_b (\rvec_2, \rvec_3) 
	|_{\rvec_1 = \rvec_2 = \rvec_3}
\\
	&\approx
	e^{i \kvec \cdot \left(
	\xvec_2 - \xvec_3
	\right)}
	\left[
	1 + 
	\xvec_2 \cdot
	\left(
	\nabla_2 - i \kvec
	\right)
	+
	\xvec_3 \cdot
	\left(
	\nabla_3 + i \kvec
	\right)
	\right]
	s_b (\rvec_2, \rvec_3) 
	|_{\rvec_1 = \rvec_2 = \rvec_3} \;.
	\numberthis
\end{align*}
The leading term vanishes as it will give a local spin density $\svec(\rvec)$. 
The linear terms $\xvec_i \cdot i \kvec$ also give a local spin density and so vanish as well.
Therefore, the only terms that contribute are:
\beq
	\label{eq:vector_deriv_1}
	s_b (\rvec_1 + \xvec_2, \rvec_1 + \xvec_3)
	\approx
	e^{i \kvec \cdot \left(
	\xvec_2 - \xvec_3
	\right)}
	\left[
	\xvec_2 \cdot \nabla_2
	+
	\xvec_3 \cdot \nabla_3
	\right]
	s_b (\rvec_2, \rvec_3) 
	|_{\rvec_1 = \rvec_2 = \rvec_3} \;.
\eeq 
Now an identity can be used to transform the derivatives acting on
the vector density matrix into two local densities,
\beq
	\nabla_{2,a}
	s_b (\rvec_2, \rvec_3) 
	|_{\rvec_1 = \rvec_2 = \rvec_3}
	=
	\frac{1}{2} \nabla_a
	s_b (\rvec_1)
	+
	i J_{ab} (\rvec_1) \;.
\eeq
Starting from the right-hand side and inserting the local densities from Eq.~\eqref{eq:local_densities}, this identity can easily be verified:
\begin{align*}
	\frac{1}{2} \nabla_a
	s_b (\rvec)
	+
	i J_{ab} (\rvec)
	&=
	\frac{1}{2}
	\sum_i 
	\Big[
	\phi_i^* (\rvec) 
	\sigma_b
	\nabla_a \phi_i (\rvec)
	+
	\nabla_a \phi_i^* (\rvec)
	\sigma_b
	\phi_i (\rvec)
	\Big]
\\
	& \quad\null +
	\frac{1}{2}
	\sum_i
	\Big[
	\phi_i^* (\rvec)
	\sigma_b
	\nabla_a \phi_i (\rvec)
	-
	\nabla_a \phi_i^* (\rvec)
	\sigma_b
	\phi_i (\rvec)
	\Big]
\\
	&=	
	\sum_i
	\phi_i^* (\rvec)
	\sigma_b
	\nabla_a \phi_i (\rvec)
\\
	&= \nabla_{2,a}
	s_b (\rvec_2, \rvec_3) 
	|_{\rvec = \rvec_2 = \rvec_3} \;.
	\numberthis
\end{align*}
An identical derivation follows for the
other derivative term,
\beq
	\nabla_{3,a} s_b (\rvec_2, \rvec_3) |_{\rvec_1 = \rvec_2 = \rvec_3}
	= 
	\frac{1}{2} \nabla_a
	s_b (\rvec_1)
	-
	i J_{ab} (\rvec_1) \;.
\eeq
Using these identities, Eq.~\eqref{eq:vector_deriv_1} can be rewritten as,
\beq
	s_b (\rvec_1 + \xvec_2, \rvec_1 + \xvec_3)
	\approx
	i 
	e^{i \kvec \cdot \lvec}
	\sum_{a=x}^z
	l_a J_{ab} (\rvec_1) \;,
\eeq
where again we dropped the local spin density term and used $\lvec = \xvec_2 - \xvec_3$.
The arbitrary momentum scale $k = |\kvec|$ is identified with the local density at a given point,
\beq
	k \equiv \kf (\rvec_1)
	= 
	\left(	
	\frac{3 \pi^2}{2}
	\rho_0 (\rvec_1)
	\right)^{1/3} \;.
\eeq
We average over this momentum scale in the phase factor following the simplified 
PSA-DME initially used in Ref.~\cite{Gebremariam201117}.
In this approach, the averaging is done over the phase space of infinite nuclear matter,
equivalent to setting the phase space quadrupole anistropy to zero for the 
full PSA-DME \cite{PhysRevC.82.014305}. 
As described in Ref.~\cite{Gebremariam201117}, this is an intermediate choice that
avoids the complications of the full PSA-DME,
while still being an improvement for the vector density matrix over the 
original Negele-Vautherin choice.
Performing the relevant integrals, the DME $\Pi$ function is derived,
\beq
	\frac{3}{4 \pi \kf^3}
	\int d^3\kvec \
	\theta \left(
	\kf - |\kvec| 	
	\right)	
	e^{i \kvec \cdot \lvec}
	=
	3 \frac{j_1 (\kf l)}
	{\kf l} 
	= \Pi_1^s (\kf l) \;,
\eeq
finally giving the DME expansion equation, 
\beq
	s_b (\rvec_1 + \xvec_2, \rvec_1 + \xvec_3)
	\approx
	i \Pi_1^s (\kf l)
	\sum_{a=x}^z
	l_a J_{ab} (\rvec_1) \;.	
\eeq

\section{Single Exchange Traces}
\label{sec:se_traces}

  In this appendix, we give the single exchange spin and isospin traces for the
  $3N$ operators which do not contain a local spin density $\svec(\rvec)$.
    All traces in this section have been checked explicitly with an included mathematica notebook.
  For the long-range (LR) terms with two nonlocality magnitudes, OBDMs will have arguments of the form,
\beq
\zeta_1 (\rvec_1) \
  \zeta_2 (\rvec_1 + \xibv, \rvec_1 + \xiav) \
  \zeta_3 (\rvec_1 + \xiav, \rvec_1 + \xibv) \;,
\eeq
where $\zeta$ can be either a scalar or vector part of the OBDM. 
  For notational convenience, we omit the arguments of the density matrices and drop the isoscalar-isovector $t$ subscript.
  The nine spin operators appearing in the local $3N$ potentials are, 
\bseq
\label{eq:nnn_spin}
 \begin{align*}
  \mathcal{S}_{1} &\equiv (\sigvec_2 \cdot \xia)
  (\sigvec_3 \cdot \xib) \; ,
  \numberthis
\\
  \mathcal{S}_{2} &\equiv \sigvec_2 \cdot \sigvec_3 \; ,
  \numberthis
\\
  \mathcal{S}_{3} &\equiv (\sigvec_2 \cdot \xia)
  (\sigvec_3 \cdot \xia) \; ,
  \numberthis
\\
  \mathcal{S}_{4} &\equiv (\sigvec_2 \cdot \xib)
  (\sigvec_3 \cdot \xib) \; ,
  \numberthis
\\
  \mathcal{S}_{5} &\equiv (\sigvec_2 \cdot \xia)
  (\sigvec_3 \cdot \xib) (\xia \cdot \xib) \; ,
  \numberthis
\\
  \mathcal{S}_{6} &\equiv \sigvec_1 \cdot (\sigvec_2 \times \sigvec_3) \; ,
  \numberthis
\\
  \mathcal{S}_{7} &\equiv (\sigvec_2 \cdot \xia) 
  \; \xia \cdot (\sigvec_3 \times \sigvec_1) \; ,
  \numberthis
\\
  \mathcal{S}_{8} &\equiv(\sigvec_3 \cdot \xib) 
  \; \xib \cdot (\sigvec_1 \times \sigvec_2) \; ,
  \numberthis
\\
  \mathcal{S}_{9} &\equiv (\sigvec_2 \cdot \xia) (\sigvec_3 \cdot \xib) 
  \; 
  \sigvec_1 \cdot (\xia \times \xib) 
  \; ,
  \numberthis
\end{align*}
\eseq
while the two isospin operators are given by,
\bseq
\label{eq:nnn_iso}
\begin{align*}
  \mathcal{T}_1 &\equiv \tauvec_2 \cdot \tauvec_3 \; ,
  \numberthis
\\
  \mathcal{T}_2 &\equiv \tauvec_1 \cdot (\tauvec_2 \times \tauvec_3) \; .
  \numberthis
\end{align*}
\eseq

	The single exchange spin and isospin traces are evaluated by,
\bseq	
\begin{align*}
	 E_j &= \frac{1}{8}
  \prod_{i=1}^3   
  \tr_{\sigma_i} 
  \left[
  \bigg(
  \rho_{t,i} (\ldots)
  + 
  \svec_{t,i} (\ldots)
  \cdot \sigvec_i
  \bigg)
   \; \mathcal{S}_j P_{23}^{\sigma}
  \right]
  \;, \numberthis
\\
  F_j &= \frac{1}{8}
  \prod_{i=1}^3   
  \tr_{\tau_i}
  \left[
  \bigg(
  \delta^{t,0}_i
  + \delta^{t,1}_i \tau^z_i
  \bigg)
  \mathcal{T}_j P_{23}^{\tau}
  \right]
  \;, \numberthis
\end{align*}
\eseq
where the $\ldots$ stand in for the arguments of the different OBDMs.
  The traces associated with the spin operators in Eqs.~\eqref{eq:nnn_spin} are,
\bseq
\label{eq:se_spin_traces}
\begin{align*}
  E_{1} &= \frac{1}{2}
  \Big[
  \rho_1
  (\svec_2 \cdot \xia)
  (\svec_3 \cdot \xib)
  +
  \rho_1 \rho_2 \rho_3
  (\xia \cdot \xib)
  + i \xia \cdot 
  (\xib \times \svec_2) \rho_1 \rho_3 
\\
  & \quad\null + i \xib \cdot (\xia \times \svec_3)
  \rho_1 \rho_2 
  -
  \rho_1 (\svec_2 \cdot \svec_3)
  (\xia \cdot \xib)
  + 
  \rho_1
  (\svec_3 \cdot \xia)
  (\svec_2 \cdot \xib) 
  \Big] \;,
  \numberthis
\\
  E_{2} &= \frac{1}{2}
  \left(
  3 \rho_1 \rho_2 \rho_3
  - 
  \rho_1 \svec_2 \cdot \svec_3
  \right) \;,
  \numberthis
\\
  E_{3} &= \frac{1}{2}
  \Big[
  2 \rho_1
  (\svec_2 \cdot \xia)
  (\svec_3 \cdot \xia)
  +
  \rho_1 \rho_2 \rho_3
  -
  \rho_1 (\svec_2 \cdot \svec_3) 
  \Big] \;,
  \numberthis
\\
  E_{4} &= E_{3} (\xia \to \xib) \;,
  \numberthis
\\
  E_{5} &= E_{1} \times 
  (\xia \cdot \xib) \; .
  \numberthis
\end{align*}
\eseq
  Note that the traces for the spin operators $\mathcal{S}_6 - \mathcal{S}_9$ are not considered as each result will have a local spin density.
  The traces over the isospin operators in Eq.~\eqref{eq:nnn_iso} are,
\bseq
\label{eq:se_iso_traces}
\begin{align*}
  F_1 &= \frac{1}{2}
  \left(
  3 \delta_1^0 \delta_2^0 \delta_3^0
  -
  \delta_1^0 \delta_2^1 \delta_3^1
  \right) \;,
  \numberthis
\\
  F_2 &=
  i \left(
  \delta_1^1 \delta_2^1 \delta_3^0
  -
  \delta_1^1 \delta_2^0 \delta_3^1
  \right) \;.
  \numberthis
\end{align*}
\eseq

	For intermediate-range (IR) terms with one nonlocality magnitude, the OBDMs will have the form, 
\beq
  \zeta_1 (\rvec_1) \
  \zeta_2 (\rvec_1 + \xibv, \rvec_1) \
  \zeta_3 (\rvec_1, \rvec_1 + \xibv) \;,
  \qquad
  \zeta_1 (\rvec_1) \
  \zeta_2 (\rvec_1, \rvec_1 + \xiav) \
  \zeta_3 (\rvec_1 + \xiav, \rvec_1) \;,
\eeq
for terms with $\delta^3(\xvec_2)$ and $\delta^3(\xvec_3)$ respectively.
	The Dirac $\delta$ functions lead to simplified trace structures as the terms $E_1$ and $E_5$ now vanish. 
	The remaining terms are unmodified but are given below as $E_i'$ for distinctive clarity,
\bseq
\begin{align*}
  E_{2}' &= \frac{1}{2}
  \left(
  3 \rho_1 \rho_2 \rho_3
  - 
  \rho_1 \svec_2 \cdot \svec_3
  \right) \;,
  \numberthis
\\
  E_{3}' &= \frac{1}{2}
  \Big[
  2 \rho_1
  (\svec_2 \cdot \xia)
  (\svec_3 \cdot \xia)
  +
  \rho_1 \rho_2 \rho_3
  -
  \rho_1 (\svec_2 \cdot \svec_3) 
  \Big] \;,
  \numberthis
\\
  E_{4}' &= E_{3} (\xia \to \xib) \;,
  \numberthis
\end{align*}
\eseq

\section{Double Exchange Traces}  
\label{sec:de_traces}

  In this appendix, we give the double exchange spin and isospin traces for the
  $3N$ operators.
  All traces in this section have been checked explicitly with an included mathematica notebook.
   For the LR terms with two nonlocality magnitudes, OBDMs will have arguments of the form,
\beq
\zeta_1 (\rvec_1 + \xibv, \rvec_1) \
  \zeta_2 (\rvec_1, \rvec_1 + \xiav) \
  \zeta_3 (\rvec_1 + \xiav, \rvec_1 + \xibv) \;,
\eeq
where $\zeta$ can be either a scalar or vector part of the OBDM. 
  For notational convenience, we omit the arguments of the density matrices and drop the isoscalar-isovector $t$ subscript.
  The double exchange traces are found by,
\bseq
\begin{align*}
  G_j &= \frac{1}{8}
  \prod_{i=1}^3   
  \tr_{\sigma_i} 
  \left[
  \bigg(
  \rho_{t,i} (\ldots)
  + 
  \svec_{t,i} (\ldots)
  \cdot \sigvec_i
  \bigg)
   \; \mathcal{S}_j P_{23}^{\sigma} P_{12}^{\sigma}
  \right]
  \;, \numberthis
\\
  H_j &= \frac{1}{8}
  \prod_{i=1}^3   
  \tr_{\tau_i}
  \left[
  \bigg(
  \delta^{t,0}_i
  + \delta^{t,1}_i \tau^z_i
  \bigg)
  \mathcal{T}_j P_{23}^{\tau} P_{12}^{\tau}
  \right]
  \;, \numberthis
\end{align*}
\eseq
where the $\ldots$ stand in for the arguments of the different OBDMs.
  The traces associated with the spin operators in Eq.~\eqref{eq:nnn_spin} are,
\bseq
\label{eq:de_spin_traces}
\begin{align*}
  G_{1} &= \frac{1}{4} 
  \Bigl[
  \rho_1 \rho_2 \rho_3
  (\xia \cdot \xib)
\\
  & \quad\null +
  \rho_1 
  (\svec_2 \cdot \xia)
  (\svec_3 \cdot \xib)
  +
  \rho_1
  (\svec_2 \cdot \xib) 
  (\svec_3 \cdot \xia)
  -
  \rho_1 
  (\svec_2 \cdot \svec_3)
  (\xia \cdot \xib)
\\
  & \quad\null + 
  \rho_2 
  (\svec_1 \cdot \xia) 
  (\svec_3 \cdot \xib) 
  +
  \rho_2 
  (\svec_1 \cdot \xib)
  (\svec_3 \cdot \xia)
  -
  \rho_2 
  (\svec_1 \cdot \svec_3)
  (\xia \cdot \xib)
\\
  & \quad\null -
  \rho_3 (\svec_1 \cdot \xia)
  (\svec_2 \cdot \xib)
  +
  \rho_3 (\svec_1 \cdot \xib)
  (\svec_2 \cdot \xia)
  +
  \rho_3 (\svec_1 \cdot \svec_2)
  (\xia \cdot \xib) 
\\
  & \quad\null + 
  i \xia \cdot (\xib \times \svec_1)
  \rho_2 \rho_3 
  + 
  i \xia \cdot 
  (\xib \times \svec_2) \rho_1 \rho_3
  - 
  i \xia \cdot (\xib \times \svec_3)
  \rho_1 \rho_2 
  \Bigr] \;,
  \numberthis
\\
  G_{2} &= \frac{1}{4}
  \left(
  3 \rho_1 \rho_2 \rho_3
  + 3 \svec_1 \cdot \svec_2 \rho_3
  - \rho_1 \svec_2 \cdot \svec_3
  - \rho_2 \svec_1 \cdot \svec_3
  \right) \;,
  \numberthis
\end{align*}

\begin{align*}
  G_{3} &= \frac{1}{4} 
  \Bigl[
  \rho_1 \rho_2 \rho_3
  -
  \rho_1 (\svec_2 \cdot \svec_3)
  -
  \rho_2 (\svec_1 \cdot \svec_3)
  +
  \rho_3 (\svec_1 \cdot \svec_2)
\\
  & \quad\null +
  2 \rho_1 
  (\svec_2 \cdot \xia)
  (\svec_3 \cdot \xia)
  + 
  2 \rho_2 (\svec_1 \cdot \xia) 
  (\svec_3 \cdot \xia) 
  \Bigr] \;,
  \numberthis
\\
  G_{4} &= G_{3} (\xia \to \xib) \;,
  \numberthis
\\
  G_{5} &= G_{1} \times (\xia \cdot \xib) \;,
  \numberthis
\end{align*}

\begin{align*}
  G_{6} &= \frac{i}{2}
  \Bigl[
  3 \rho_1 \rho_2 \rho_3
  -
  \rho_3 \svec_1 \cdot \svec_2
  -
  \rho_1 \svec_2 \cdot \svec_3
  - 
  \rho_2 \svec_1 \cdot \svec_3
  \Bigr] \  ,
  \numberthis
\\
  G_{7} &= \frac{i}{2} 
  \Bigl[
  \rho_1 \rho_2 \rho_3
  - 
  \rho_1 (\svec_2 \cdot \svec_3)
  +
  2 \rho_1 (\svec_2 \cdot \xia) (\svec_3 \cdot \xia)
\\
   & \quad\null -
   \rho_3 (\svec_1 \cdot \xia) (\svec_2 \cdot \xia)
  -
   \rho_2 (\svec_1 \cdot \xia) (\svec_3 \cdot \xia)
  \Bigr] \;,
  \numberthis
\\
  G_{8} &= \frac{i}{2}
  \Bigl[
  \rho_1 \rho_2 \rho_3 
  - 
  \rho_2 (\svec_1 \cdot \svec_3)
  +
  2 \rho_2 (\svec_1 \cdot \xib) (\svec_3 \cdot \xib)
\\
  & \quad\null -
  \rho_1 (\svec_2 \cdot \xib) (\svec_3 \cdot \xib)
  -
  \rho_3 (\svec_1 \cdot \xib) (\svec_2 \cdot \xib)
  \Bigr] \;,
  \numberthis
\end{align*}

\begin{align*}
  G_{9} &= \frac{i}{4}
  \Bigl[
  \rho_1 \rho_2 \rho_3 
  -
  \rho_1 \rho_2 \rho_3 (\xia \cdot \xib) (\xia \cdot \xib)
  - i
  \rho_1 \rho_3 \svec_2 \cdot (\xia \times \xib) (\xia \cdot \xib)
\\
  & \qquad\quad\null 
  + i
  \rho_1 \rho_2 \svec_3 \cdot (\xia \times \xib) (\xia \cdot \xib)
  - i
  \rho_2 \rho_3 \svec_1 \cdot (\xia \times \xib) (\xia \cdot \xib)
\\
  & \qquad\quad\null + 
  \rho_3 \svec_1 \cdot (\xia \times \xib) \svec_2 \cdot (\xia \times \xib)
  - 
  \rho_2 \svec_1 \cdot (\xia \times \xib) \svec_3 \cdot (\xia \times \xib)
\\
  & \qquad\quad\null 
  +
  \rho_2 (\svec_1 \cdot \xib) (\svec_3 \cdot \xib)
  - 
  \rho_1 (\svec_2 \cdot \xib) (\svec_3 \cdot \xib)
  +
  2 \rho_3 (\svec_2 \cdot \xia) (\svec_1 \cdot \xib) (\xia \cdot \xib)
\\
  & \qquad\quad\null 
  -
  \rho_3 (\svec_1 \cdot \xia) (\svec_2 \cdot \xia)
  +
  \rho_1 (\svec_2 \cdot \xia) (\svec_3 \cdot \xia)
  - 
  \rho_3 (\svec_1 \cdot \xib) (\svec_2 \cdot \xib)
\\
  & \qquad\quad\null 
  -
  \rho_2 (\svec_3 \cdot \xia) (\svec_1 \cdot \xia)
  -
  \rho_1 \svec_3 \cdot (\xia \times \xib) \svec_2 \cdot (\xia \times \xib)
  \Bigr] \;.
  \numberthis
\end{align*}
\eseq

  The traces associated with the isospin operators in Eq.~\ref{eq:nnn_iso} are given by,
\bseq
\label{eq:de_iso_traces}
\begin{align*}
  H_1 &= \frac{1}{4} \left(
  3 \delta_1^0 \delta_2^0 \delta_3^0
  +
  3 \delta_1^1 \delta_2^1 \delta_3^0
  -
  \delta_1^0 \delta_2^1 \delta_3^1
  -
  \delta_1^1 \delta_2^0 \delta_3^1
  \right) \;,
  \numberthis
\\
  H_2 &= \frac{i}{2}
  \left(
  3 \delta_1^0 \delta_2^0 \delta_3^0 
  -
  \delta_1^1 \delta_2^1 \delta_3^0
  -
  \delta_1^1 \delta_2^0 \delta_3^1
  -
  \delta_1^0 \delta_2^1 \delta_3^1
  \right) \;.
  \numberthis
\end{align*}
\eseq

	For IR terms with one nonlocality magnitude, the OBDMs will have the form, 
\beq
  \zeta_1 (\rvec_1 + \xibv, \rvec_1) \
  \zeta_2 (\rvec_1) \
  \zeta_3 (\rvec_1, \rvec_1 + \xibv) \;,
  \qquad
  \zeta_1 (\rvec_1) \
  \zeta_2 (\rvec_1, \rvec_1 + \xiav) \
  \zeta_3 (\rvec_1 + \xiav, \rvec_1) \;,
\eeq
for terms with $\delta^3(\xvec_2)$ and $\delta^3(\xvec_3)$ respectively.
	The Dirac $\delta$ functions render some of the OBDMs diagonal leading to simplified trace structures due to local spin densities vanishing.
	These are easily derived but the modified trace structures are given below as $G_i'$,
\bseq
\label{eq:de_con_traces}
\begin{align*}
	G'_{2,\delta^3(\xvec_2)} &= \frac{1}{4} 
	\bigl[
	3 \rho_1 \rho_2 \rho_3 - \rho_2 \svec_1 \cdot \svec_3
	\bigl] \; ,
	\numberthis
\\
	G'_{2,\delta^3(\xvec_3)} &= \frac{1}{4}
	\bigl[
	3 \rho_1 \rho_2 \rho_3 - \rho_1 \svec_2 \cdot \svec_3
	\bigl] \; ,
	\numberthis
\\
	G'_3 &= \frac{1}{4} 
	\bigl[
	\rho_1 \rho_2 \rho_3 - \rho_1 \svec_2 \cdot \svec_3
	+ 2 \rho_1 (\svec_2 \cdot \xia) (\svec_3 \cdot \xia)
	\bigl] \; ,
	\numberthis
\\
	G'_4 &= \frac{1}{4} 
	\bigl[
	\rho_1 \rho_2 \rho_3 - \rho_2 \svec_1 \cdot \svec_3
	+ 2 \rho_2 (\svec_1 \cdot \xib) (\svec_3 \cdot \xib)
	\bigl] \; ,
	\numberthis
\\
	G'_{6,\delta^3(\xvec_2)} &= 2 i G'_{2,\delta^3(\xvec_2)} \; , 
	\numberthis
\\
	G'_{6,\delta^3(\xvec_3)} &= 2 i G'_{2,\delta^3(\xvec_3)} \; ,
	\numberthis
\\
	G'_{7} &= 2i G'_3 \; ,
	\numberthis
\\
	G'_{8} &= 2i G'_4 \; .
	\numberthis
\end{align*}
\eseq

\section{Single Exchange DME Dictionary}
\label{sec:se_dict}

In this appendix, we give the DME expansion equations for the single exchange Fock term corresponding to $P_{23}$.
The format for the expansion equations is given by,
\begin{align*}
  \text{density matrices}
  \xrightarrow{\text{DME}}
  \left\{
  \begin{array}{lcr}
  \text{local densities}
  & \times & 
  (\text{LR or IR DME expression})
  \end{array}
  \right\}
  \; ,
\end{align*}
where the LR and IR DME terms are the expansions for the different arguments in the OBDMs. 
The LR DME expression on the right hand side has had all nonlocality variable integrals done except for the magnitudes of the two variables, 
$x_2$ and $x_3$, and their relative angle (i.e., $\xhat_2 \cdot \xhat_3 = \ct$).
The IR DME expression has had all nonlocality variable integrals done except for a generic magnitude $x$.
The LR OBDMs have arguments of the form,
\beq
\zeta_1 (\rvec_1) \
  \zeta_2 (\rvec_1 + \xibv, \rvec_1 + \xiav) \
  \zeta_3 (\rvec_1 + \xiav, \rvec_1 + \xibv) \;,
\eeq
where $\zeta$ can be either a scalar or vector part of the OBDM. 
The IR OBDMs have arguments of the form,
\beq
  \zeta_1 (\rvec_1) \
  \zeta_2 (\rvec_1 + \xibv, \rvec_1) \
  \zeta_3 (\rvec_1, \rvec_1 + \xibv) \;,
  \qquad
  \zeta_1 (\rvec_1) \
  \zeta_2 (\rvec_1, \rvec_1 + \xiav) \
  \zeta_3 (\rvec_1 + \xiav, \rvec_1) \;,
\eeq
for terms with $\delta^3(\xvec_2)$ and $\delta^3(\xvec_3)$ respectively.
All local densities in this section depend purely on $\rvec_1$. 
Spatial arguments for OBDMs and local densities have been suppressed below for brevity.

The $\W$ functions have the DME expansions for the LR OBDMs while the $\Wp$ functions have the DME expansions for the IR OBDMs.
The $\W$ functions are listed in Table~\ref{tab:se_ang} and the $\Wp$ functions are listed in Table~\ref{tab:se_dme_functions_con}.
For convenience, we employ the following shorthand for the DME $\Pi$ functions in the $\W$ functions,
\beq
	\Sub_{ij} 
	= \Pi_i (\kf l) \Pi_j (\kf l) \; .
\eeq
As a reminder, the vectors $\Nvec$ and $\lvec$ are defined as,
\beq
 \lvec = \xvec_2 - \xvec_3 \; ,
 \qquad
 \Nvec = (1-a) \xvec_2 + a \xvec_3 \; .
\eeq
where $a$ describes our freedom in performing the DME expansion with respect to the $23$ subsystem, see Sec.~\ref{sec:dme_parametrization}.

\subsection{Three Scalars}

The LR three scalar term is given by:

\begin{align*}
  \renewcommand{\arraystretch}{1.2}
  \rho_1 \; \rho_2 \; \rho_3  
  &\xrightarrow{\text{DME}}
  \rho_1
  \left\{\begin{array}{lcr}
  & \rho_2 \rho_3  &  \Se_1
\\
  + & \rho_2 \tau_3  & (-1) \Se_2
 \\
  + & \tau_2 \rho_3  & (-1) \Se_2
\\
  + & \rho_3 \Delta \rho_2  & \Se_3
\\
  + & \rho_2 \Delta \rho_3  & \Se_3
\\
  + & \nabla \rho_2 \cdot \nabla \rho_3  & \Se_4
  \end{array}\right\} \;,
  \numberthis
\end{align*}
while the IR term is given by:

\begin{align*}
  \renewcommand{\arraystretch}{1.2}
  \rho_1 \; \rho_2 \; \rho_3  
   &\xrightarrow{\text{DME}}
  \rho_1
  \left\{\begin{array}{lcr}
  & \rho_2 \rho_3  &  
  \Wp_1
\\
  + & \rho_2 \tau_3  & 
  \Wp_2
\\
  + & \tau_2 \rho_3  & 
  \Wp_2
\\
  + & \rho_3 \Delta \rho_2  &
  \left(- \frac{1}{2} \Wp_2\right)
\\
  + & \rho_2 \Delta \rho_3  &
  \left(- \frac{1}{2} \Wp_2\right)
  \end{array}\right\} \;,
  \numberthis
\end{align*}

\subsection{Two Scalars, One Vector}

Both of these are LR terms.
\bseq
\begin{align*}
  i \xia \cdot 
  (\xib \times \svec_2) \rho_1 \rho_3
  &\xrightarrow{\text{DME}}  \rho_1
  \left\{\begin{array}{lr}
  \epsilon_{ijk} 
  \left(
  \nabla_i \; \rho_3
  \right) 
  J_{2,jk}  
   &
   \Se_5
  \end{array}
  \right\} \;,
  \numberthis
\\
  i \xib \cdot 
  (\xia \times \svec_3) \rho_1 \rho_2
&\xrightarrow{\text{DME}}
  \rho_1 
  \left\{\begin{array}{lr}
  \epsilon_{ijk} 
  \left(
  \nabla_i \; \rho_2
  \right) 
  J_{3,jk}  &  
  \Se_5
  \end{array}
  \right\} \;,
  \numberthis
\end{align*}
\eseq

\subsection{One Scalar, Two Vectors}

The LR terms in this section are given by:

\bseq
\begin{align*}
  \renewcommand{\arraystretch}{1.2}
  \rho_1 
  (\svec_2 \cdot \svec_3)
&\xrightarrow{\text{DME}}
  \rho_1
  \left\{\begin{array}{lr}
  J_{2,ab} \; J_{3,ab}  &
  \Se_6
  \end{array}
  \right\} \;,
  \numberthis
\end{align*}

\begin{align*}
  \renewcommand{\arraystretch}{1.2}
  \rho_1  \; 
  (\svec_2 \cdot \xia)
  (\svec_3 \cdot \xia)
&\xrightarrow{\text{DME}}  
  \rho_1
  \left\{\begin{array}{lcr}
  & J_{2,ab} J_{3,ab}  &
  \Se_7
\\
  + & J_{2,aa} J_{3,bb}  &
  \Se_8
\\
  + & J_{2,ab} J_{3,ba}  &
  \Se_8
  \end{array}
  \right\} \;,
  \numberthis
\end{align*}

\begin{align*}
  \renewcommand{\arraystretch}{1.2}
  \rho_1  \; 
  (\svec_2 \cdot \xib)
  (\svec_3 \cdot \xib)
&\xrightarrow{\text{DME}}
  \rho_1
  \left\{\begin{array}{lcr}
  & J_{2,ab} J_{3,ab}  &
  \Se_9
\\
  + & J_{2,aa} J_{3,bb}  &
  \Se_{10}
\\
  + & J_{2,ab} J_{3,ba}  &
  \Se_{10}
  \end{array}
  \right\} \;,
  \numberthis
\end{align*}

\begin{align*}
  \renewcommand{\arraystretch}{1.2}
  \rho_1  \; 
  (\svec_2 \cdot \xia)
  (\svec_3 \cdot \xib)
 &\xrightarrow{\text{DME}}
  \rho_1
  \left\{\begin{array}{lcr}
    & J_{2,ab} J_{3,ab}  &
    \Se_{11}
  \\
  + & J_{2,aa} J_{3,bb}  &
  \Se_{12}
  \\
  + & J_{2,ab} J_{3,ba}  &
  \Se_{12}
  \end{array}
  \right\} \;,
  \numberthis
\end{align*}

\begin{align*}
  \renewcommand{\arraystretch}{1.2}
  \rho_1  \; 
  (\svec_3 \cdot \xia)
  (\svec_2 \cdot \xib)
&\xrightarrow{\text{DME}}
  \rho_1
  \left\{\begin{array}{lcr}
    & J_{2,ab} J_{3,ab}  &
    \Se_{11}
  \\
  + & J_{2,aa} J_{3,bb}  &
  \Se_{12}
  \\
  + & J_{2,ab} J_{3,ba}  &
  \Se_{12}
  \end{array}
  \right\} \;,
  \numberthis
\end{align*}
\eseq
while the IR terms are given by:
\bseq
\begin{align*}
  \renewcommand{\arraystretch}{1.2}
  \rho_1 
  (\svec_2 \cdot \svec_3)
&\xrightarrow{\text{DME}}
  \rho_1
  \left\{\begin{array}{lr}
  J_{2,ab} \; J_{3,ab}  &
  \Wp_3
  \end{array}
  \right\} \;,
  \numberthis
\end{align*}
\begin{align*}
  \renewcommand{\arraystretch}{1.2}
  \rho_1  \; 
  (\svec_2 \cdot \xia)
  (\svec_3 \cdot \xia)
&\xrightarrow{\text{DME}}  \rho_1 
  \left\{\begin{array}{lcr}
  & J_{2,ab} J_{3,ab}  &
  \frac{1}{5} \Wp_3
\\
  + & J_{2,aa} J_{3,bb}  &
  \frac{1}{5} \Wp_3
\\
  + & J_{2,ab} J_{3,ba}  &
  \frac{1}{5} \Wp_3
  \end{array}
  \right\} \;,
  \numberthis
\end{align*}

\begin{align*}
  \renewcommand{\arraystretch}{1.2}
  \rho_1  \; 
  (\svec_2 \cdot \xib)
  (\svec_3 \cdot \xib)
&\xrightarrow{\text{DME}}
  \rho_1 
  \left\{\begin{array}{lcr}
  & J_{2,ab} J_{3,ab}  &
   \frac{1}{5} \Wp_3
\\
  + & J_{2,aa} J_{3,bb}  &
   \frac{1}{5} \Wp_3
\\
  + & J_{2,ab} J_{3,ba}  &
   \frac{1}{5} \Wp_3
  \end{array}
  \right\} \; .
  \numberthis
\end{align*}
\eseq

\begin{table}[th]
  \renewcommand{\arraystretch}{1.4}
  \caption{\label{tab:se_ang} Table of $\Se$ functions}
  \begin{tabular}{l|l}
  \hline\hline
  $\Se_1$ & 
  $\Sub_{00} + \Sub_{02} 
  \frac{\kf^2 l^2}{5}$
 \\ \hline
  $\Se_2$ & 
  $\Sub_{02} \frac{l^2}{6}$
\\ \hline
  $\Se_3$ & 
  $\Sub_{00} \frac{1}{6} N^2
  + \Sub_{02} \frac{l^2 \gamma}{6}$
  \\ \hline
  $\Se_4$ & 
  $\Sub_{00}
  \frac{1}{3} 
  N^2$
  \\ \hline
  $\Se_5$ &
  $\Sub_{01} 
  \frac{-1}{6}
  x_2 x_3
  \sin(\theta)^2$
  \\ \hline
  $\Se_6$ &
  $\Sub_{11} \frac{1}{3}
  l^2$
  \\ \hline
  $\Se_7$ &
  $\Sub_{11} 
  \frac{1}{15}
  \left(
  2 l^2
  - 
  (\lvec \cdot \xia)^2
  \right)$
  \\ \hline
  $\Se_8$ &
  $\Sub_{11} 
  \frac{1}{30}
  \left(
  3
  (\lvec \cdot \xia)^2
  - 
  l^2
  \right)$
  \\ \hline
  $\Se_9$ &
  $\Sub_{11} 
  \frac{1}{15}
  \left(
  2 l^2
  - 
  (\lvec \cdot \xib)^2
  \right)$
  \\ \hline
  $\Se_{10}$ &
  $\Sub_{11} 
  \frac{1}{30}
  \left(
  3
  (\lvec \cdot \xib)^2
  - 
  l^2
  \right)$
  \\ \hline
  $\Se_{11}$ &
  $\Sub_{11} 
   \frac{1}{15}
  \left(
  2 l^2
  \ct
  - 
  (\lvec \cdot \xia) 
  (\lvec \cdot \xib)
  \right)$
    \\ \hline
  $\Se_{12}$ &
  $\Sub_{11} 
  \frac{1}{30}
  \left(
  3
  (\lvec \cdot \xia)
  (\lvec \cdot \xib)
  - 
  l^2 \ct
  \right)$
  \\ \hline \hline
  \end{tabular}
\end{table}  
\begin{table}[th]
  \renewcommand{\arraystretch}{1.4}
  \caption{\label{tab:se_dme_functions_con} Table of $\Wp$ functions}
  \begin{tabular}{l|l}
  \hline\hline
  $\Wp_1$ &   
  $\left[\Pi_0^\rho (\kf x) \right]^2 
   +
   \Pi_0^\rho (\kf x) \Pi_2^\rho (\kf x)
   \frac{x^2 \kf^2}{5}$
 \\ \hline
  $\Wp_2$ & 
  ${-\frac{x^2}{6}} \Pi_0^\rho (\kf x)
  \Pi_2^\rho (\kf x) $
\\ \hline
  $\Wp_3$ &
  $\frac{1}{3}
  x^2 
  \left[\Pi_1^s(\kf x) \right]^2$
  \\ \hline\hline
  \end{tabular}
\end{table}  

\section{Double Exchange DME Dictionary}
\label{sec:de_dict}

In this appendix, we give the DME expansion equations for the double exchange Fock term corresponding to $P_{23} P_{12}$.
The format for the expansion equations is given by,
\begin{align*}
  \text{density matrices}
  \xrightarrow{\text{DME}}
  \left\{
  \begin{array}{lcccr}
  \text{local densities}
  & \times & 
  (\text{LR or IR DME expression})
  \end{array}
  \right\}
  \; ,
\end{align*}
where the LR and IR DME terms are the expansions for the different arguments in the OBDMs.
The LR DME expression on the right hand side has had all nonlocality variable integrals done except for the magnitudes
$x_2$ and $x_3$ and their relative angle (i.e., $\xhat_2 \cdot \xhat_3 = \ct$).
The IR DME expression has had all nonlocality variable integrals done except for a generic magnitude $x$.
The LR OBDMs have arguments of the form,
\beq
\zeta_1 (\rvec_1 + \xibv, \rvec_1) \
  \zeta_2 (\rvec_1, \rvec_1 + \xiav) \
  \zeta_3 (\rvec_1 + \xiav, \rvec_1 + \xibv) \;,
\eeq
where $\zeta$ can be either a scalar or vector part of the OBDM.
The IR OBDMs have arguments of the form,
\beq
  \zeta_1 (\rvec_1 + \xibv, \rvec_1) \
  \zeta_2 (\rvec_1) \
  \zeta_3 (\rvec_1, \rvec_1 + \xibv) \;,
  \qquad
  \zeta_1 (\rvec_1) \
  \zeta_2 (\rvec_1, \rvec_1 + \xiav) \
  \zeta_3 (\rvec_1 + \xiav, \rvec_1) \;,
\eeq
for terms with $\delta^3(\xvec_2)$ and $\delta^3(\xvec_3)$ respectively.
Note that because these two OBDMs are diagonal with respect to different particle indices, the DME expansion equations for the two will correspond to local densities with different particle indices.
However, the isospin traces in Eq.~\eqref{eq:de_iso_traces} are symmetric under the interchange of index 1 and 2.
Hence, in the following, we show \textit{only} the DME expansion equations for the IR OBDMs corresponding to the $\delta^3(\xvec_2)$ term.
All local densities in this section depend purely on $\rvec_1$. 
Spatial arguments for OBDMs and local densities have been suppressed below for brevity.

The $\Z$ functions have the DME expansions for the LR OBDMs while the $\Zp$ functions have the DME expansions for the IR OBDMs.
The $\Z$ functions are listed in Table~\ref{tab:de_dme_functions} 
and the $\Zp$ functions are listed in Table~\ref{tab:de_dme_functions_con}.
For convenience, we employ the following shorthand for the DME $\Pi$ functions in the $\Z$ functions,
\beq
	\Dub_{ijk} = 
	\Pi_i (\kf x_2)
	\Pi_j (\kf x_3)
	\Pi_k (\kf l) \; .
\eeq
As a reminder, the vectors $\Nvec$ and $\lvec$ are defined as,
\beq
 \lvec = \xvec_2 - \xvec_3 \; ,
 \qquad
 \Nvec = (1-a) \xvec_2 + a \xvec_3 \; .
\eeq
where $a$ describes our freedom in performing the DME expansion with respect to the $23$ subsystem, see Sec.~\ref{sec:dme_parametrization}.

\subsection{Three Scalars}

The three scalar LR term is given by:

\beq
  \rho_1 \rho_2 \rho_3
  \xrightarrow{\text{DME}}
  \left\{\begin{array}{lcr}
  & \rho_1 \rho_2 \rho_3  &
	\De_1
\\
  + & \rho_1 \rho_2 \Delta \rho_3  &
  \De_2
\\
  + & \Delta \rho_1 \rho_2 \rho_3  &
  \De_3
\\
  + & \rho_1 \Delta \rho_2 \rho_3  &
  \De_4
\\
  + & \rho_1 \rho_2 \tau_3  &
  \left(-1
  \right)
  \De_5
\\
  + &\tau_1 \rho_2 \rho_3  &
  \left(
  - 2
  \right)
  \De_3
\\
  + & \rho_1 \tau_2 \rho_3  &
  \left(
  - 2
  \right)
  \De_4
  \end{array}
  \right\} \;,
\eeq

The three scalar IR term is given by:

\beq
  \rho_1 \rho_2 \rho_3
  \xrightarrow{\text{DME}}
  \left\{\begin{array}{lcr}
  & \rho_1 \rho_2 \rho_3  &
  \Zp_1
\\
  + & \rho_1 \rho_2 \Delta \rho_3  &
  \Zp_2
\\
  + & \Delta \rho_1 \rho_2 \rho_3  &
 \Zp_2
\\
  + & \rho_1 \rho_2 \tau_3  &
  (- 2 \Zp_2)
\\
  + &\tau_1 \rho_2 \rho_3  &
  (- 2 \Zp_2)
  \end{array}
  \right\} \;,
\eeq

\subsection{Two Scalars, One Vector}

All of the contributions here come from LR terms:

\bseq
\begin{align*}
  i \xia \cdot 
  (\xib \times \svec_2) \rho_1 \rho_3
  &\xrightarrow{\text{DME}}
  \left\{\begin{array}{lr}
  \rho_1 \epsilon_{ijk} 
  (\nabla_i \rho_3)
  J_{2,jk}  &
  \De_6
  \end{array}
  \right\} \;,
  \numberthis
\\
  i \xia \cdot 
  (\xib \times \svec_1) \rho_2 \rho_3
   &\xrightarrow{\text{DME}}
  \left\{\begin{array}{lr}
  \rho_2 \epsilon_{ijk}
  (\nabla_i \rho_3)
  J_{1,jk}  &
  \De_7 \; (-1)
  \end{array}
  \right\} \;,
  \numberthis
\\
  i \xia \cdot
  (\xib \times \svec_3) \rho_1 \rho_2
  &\xrightarrow{\text{DME}}
  \left\{\begin{array}{lcr}
  0
  \end{array}
  \right\} \;,
  \numberthis
\end{align*}
\eseq

\subsection{One Scalar, Two Vectors - Spin Dots}

The LR terms are given by:

\bseq
\begin{align*}
  \rho_1 (\svec_2 \cdot \svec_3) 
  &\xrightarrow{\text{DME}}
  \left\{
  \begin{array}{lr}
  \rho_1 J_{2,ab} J_{3,ab}  &
  \De_8
  \end{array}
  \right\} \;,
  \numberthis
\\
  \rho_2 (\svec_1 \cdot \svec_3) 
  &\xrightarrow{\text{DME}}
  \left\{
  \begin{array}{lr}
  \rho_2 J_{1,ab} J_{3,ab}  &
  \De_9 \; (-1)
  \end{array}
  \right\} \;,
  \numberthis
\\
  \rho_3 (\svec_1 \cdot \svec_2) 
   &\xrightarrow{\text{DME}}
  \left\{
  \begin{array}{lr}
  \rho_3 J_{1,ab} J_{2,ab} &
  \De_{10}
  \end{array}
  \right\} \;,
  \numberthis
\end{align*}
\eseq

The IR terms are given by:

\beq
  \rho_2 (\svec_1 \cdot \svec_3) 
  \xrightarrow{\text{DME}}
  \left\{
  \begin{array}{lr}
  \rho_2 J_{1,ab} J_{3,ab}  &
  \Zp_3
  \end{array}
  \right\} \;,
\eeq

\subsection{One Scalar, Two Vectors - Tensor Terms}

The LR terms are given by:

\bseq
\begin{align*}
  \rho_1 (\svec_2 \cdot \xia) (\svec_3 \cdot \xia)
  &\xrightarrow{\text{DME}}
  \rho_1 
	\frac{\De_8}{5}
	\left\{
  \begin{array}{lcr} 
    & J_{2,ab} J_{3,ab} &
\\
  + & J_{2,aa} J_{3,bb} &
\\
  + & J_{2,ab} J_{3,ba} &
  \end{array}
  \right\} \;,
  \numberthis
\\
  \rho_1 (\svec_2 \cdot \xib) (\svec_3 \cdot \xib)
  &\xrightarrow{\text{DME}}
  \rho_1  
  \left\{
  \begin{array}{lcr}
  & J_{2,ab} J_{3,ab}  &
    \De_{11}
\\
  + & J_{2,aa} J_{3,bb} &
    \De_{12}
\\
  + & J_{2,ab} J_{3,ba} &
    \De_{12}
  \end{array}
  \right\} \;,
  \numberthis
\\
  \rho_2 (\svec_1 \cdot \xia) (\svec_3 \cdot \xia)
  &\xrightarrow{\text{DME}}
  \rho_2
  \left\{
  \begin{array}{lcr}
  & J_{1,ab} J_{3,ab}  &
  \De_{13}
\\
  + & J_{1,aa} J_{3,bb} &
    \De_{14}
\\
  + & J_{1,ab} J_{3,ba} &
    \De_{14}
  \end{array}
  \right\} (-1) \;,
  \numberthis
\\
  \rho_2 (\svec_1 \cdot \xib) (\svec_3 \cdot \xib)
  &\xrightarrow{\text{DME}}
  \rho_2
	\frac{\De_9}{5}
	\left\{
  \begin{array}{lcr}
    & J_{1,ab} J_{3,ab}  & 
\\
  + & J_{1,aa} J_{3,bb}  &
\\
  + & J_{1,ab} J_{3,ba}  &
  \end{array}
  \right\} (-1) \;,
  \numberthis
\\
  \rho_3 (\svec_1 \cdot \xia) (\svec_2 \cdot \xia)
  &\xrightarrow{\text{DME}}
  \rho_3
  \frac{\De_{10}}{5}
  \left\{
  \begin{array}{lcr}
    & J_{1,ab} J_{2,ab} &
\\
  + & J_{1,aa} J_{2,bb} &
\\
  + & J_{1,ab} J_{2,ba} &
  \end{array}
  \right\} \;,
  \numberthis
\\
  \rho_3 (\svec_1 \cdot \xib) (\svec_2 \cdot \xib)
  &\xrightarrow{\text{DME}}
  \rho_3
  \frac{\De_{10}}{5}
  \left\{
  \begin{array}{lcr}
    & J_{1,ab} J_{2,ab} &
\\
  + & J_{1,aa} J_{2,bb} &
\\
  + & J_{1,ab} J_{2,ba} &
  \end{array}
  \right\} \;,
  \numberthis
\end{align*}
\eseq

The IR terms are given by:

\beq
  \rho_2 (\svec_1 \cdot \xib) (\svec_3 \cdot \xib)
  \xrightarrow{\text{DME}}
  \rho_2
  \left\{
  \begin{array}{lcr}
    & J_{1,ab} J_{3,ab}  & 
  \frac{1}{5} \Zp_3
\\
  + & J_{1,aa} J_{3,bb}  &
  \frac{1}{5} \Zp_3
\\
  + & J_{1,ab} J_{3,ba}  &
  \frac{1}{5} \Zp_3
  \end{array}
  \right\} \;,
\eeq

\subsection{One Scalar, Two Vectors - Tensor-Like Terms}

All of the contributions here come from LR terms:

\bseq
\begin{align*}
  \rho_1 (\svec_2 \cdot \xia) (\svec_3 \cdot \xib)
  \xrightarrow{\text{DME}}
  \rho_1
  \left\{
  \begin{array}{lcr}
    & J_{2,ab} J_{3,ab}  &
    \De_{15}
\\
  + & J_{2,aa} J_{3,bb}  &
    \De_{16}
\\
  + & J_{2,ab} J_{3,ba}  &
    \De_{15}
  \end{array}
  \right\} \;,
  \numberthis
\end{align*}

\begin{align*}
  \rho_1 (\svec_2 \cdot \xib) (\svec_3 \cdot \xia)
  \xrightarrow{\text{DME}}
  \rho_1
  \left\{
  \begin{array}{lcr}
  & J_{2,ab} J_{3,ab}  &
    \De_{15}
\\
  + & J_{2,aa} J_{3,bb}  &
    \De_{15}
\\
  + & J_{2,ab} J_{3,ba}  &
    \De_{16}
   \end{array}
  \right\} \;,
  \numberthis
\end{align*}

\begin{align*}
  \rho_2 (\svec_1 \cdot \xia) (\svec_3 \cdot \xib)
  \xrightarrow{\text{DME}}
  \rho_2
  \left\{
  \begin{array}{lcr}
  & J_{1,ab} J_{3,ab}  &
    \De_{17}
\\
  + & J_{1,aa} J_{3,bb}  &
    \De_{17}
\\
  + & J_{1,ab} J_{3,ba}  &
    \De_{18} 
  \end{array}
  \right\} (-1) \;,
  \numberthis
\end{align*}

\begin{align*}
  \rho_2 (\svec_1 \cdot \xib) (\svec_3 \cdot \xia)
  \xrightarrow{\text{DME}}
  \rho_2
  \left\{
  \begin{array}{lcr}
    & J_{1,ab} J_{3,ab}  &
    \De_{17}
\\
  + & J_{1,aa} J_{3,bb}  &
    \De_{18}
\\
  + & J_{1,ab} J_{3,ba}  &
    \De_{17}
  \end{array}
  \right\} (-1)\;,
  \numberthis
\end{align*}

\begin{align*}
  \rho_3 (\svec_1 \cdot \xia) (\svec_2 \cdot \xib)
  \xrightarrow{\text{DME}}
  \rho_3 
  \left\{
  \begin{array}{lcr}
  & J_{1,ab} J_{2,ab}  &
  \De_{19}
\\
  + & J_{1,aa} J_{2,bb}  &
  \De_{19}
\\
  + & J_{1,ab} J_{2,ba}  &
  \De_{20}
  \end{array}
  \right\} \;,
  \numberthis
\end{align*}

\begin{align*}
  \rho_3 (\svec_1 \cdot \xib) (\svec_2 \cdot \xia)
  \xrightarrow{\text{DME}}
  \rho_3
  \left\{
  \begin{array}{lcr}
    & J_{1,ab} J_{2,ab}  &
    \De_{19}
\\
  + & J_{1,aa} J_{2,bb}  &
    \De_{20}
\\
  + & J_{1,ab} J_{2,ba}  &
    \De_{19}
  \end{array}
  \right\} \;,
  \numberthis
\end{align*}
\eseq
\subsection{One Scalar, Two Vectors - Spin Crosses}

All of the contributions here come from LR terms:

\bseq
\begin{align*}
  \rho_1 \svec_2 \cdot (\xia \times \xib)
  \svec_3 \cdot (\xia \times \xib)
  &\xrightarrow{\text{DME}}
  \rho_1
  \De_{21}
  \left\{
  \begin{array}{lcr}
    & J_{2,ab} J_{3,ab}  &
  \left(
  1 
  \right)
\\
  + & J_{2,aa} J_{3,bb}  & 
  \left(
  - \frac{1}{4}
  \right)
\\
  + & J_{2,aa} J_{3,ba}  &
  \left(
  - \frac{1}{4}
  \right)
  \end{array}
  \right\} \;,
  \numberthis
\\
  \rho_2 \svec_1 \cdot (\xia \times \xib)
  \svec_3 \cdot (\xia \times \xib)
  &\xrightarrow{\text{DME}}
  \rho_2
  \De_{22}
  \left\{
  \begin{array}{lcr}
    & J_{1,ab} J_{3,ab}  &
  \left(
  1
  \right)
\\
  + & J_{1,aa} J_{3,bb}  &
  \left(
  - \frac{1}{4}
  \right)
\\
  + & J_{1,ab} J_{3,ba}  & 
  \left(
  - \frac{1}{4}
  \right)
  \end{array}
  \right\} (-1) \;,
  \numberthis
\\
  \rho_3 \svec_1 \cdot (\xia \times \xib)
  \svec_2 \cdot (\xia \times \xib)
  &\xrightarrow{\text{DME}}
  \rho_3
  \De_{23}
  \left\{
  \begin{array}{lcr}
    & J_{1,ab} J_{2,ab}  &
    \left( 1 \right)
\\
  + & J_{1,aa} J_{2,bb}  & 
    \left( - \frac{1}{4} \right)
\\
  + & J_{1,ab} J_{2,ba}  & 
    \left( - \frac{1}{4} \right)
  \end{array}
  \right\} \; .
  \numberthis
\end{align*}
\eseq

\begin{table}[th]
  \renewcommand{\arraystretch}{1.4}
  \caption{Table of $\De$ functions \label{tab:de_dme_functions}}
  \begin{tabular}{l|l|l|l}
  \hline\hline
  $\De_1$ & 
  $\Dub_{000} 
  + \Dub_{200} \frac{x_2^2 \kf^2}{10}
  + \Dub_{020} \frac{x_3^2 \kf^2}{10}
  +
  \Dub_{002} \frac{l^2 \kf^2}{10}$ & 
  $\De_{13}$ & 
  $\Dub_{011}
  \frac{1}{15}
  \left(
  2 \lvec \cdot \xvec_3
  - 
  (\lvec \cdot \xia)
  x_3 \ct
  \right) $
\\ \hline
  $\De_2$ &   
  $\Dub_{000}
  \frac{1}{6}
  N^2 +
  \Dub_{002} \;
  \frac{l^2 \gamma}{6}$ &
    $\De_{14}$ &
  $\Dub_{011}
  \frac{1}{30}
  \left(
  3 (\lvec \cdot \xia)
  x_3 \ct
  -
  \lvec \cdot \xvec_3
  \right) $
\\ \hline
  $\De_3$ &   
  $\Dub_{020} \;
  \frac{x_3^2}{12}$ &
    $\De_{15}$ &
  $\Dub_{101}
  \frac{1}{30}
  \left(
  3 (\lvec \cdot \xvec_2) \ct
  -
  x_2 (\lvec \cdot \xib)
  \right)$
  \\ \hline  
  $\De_4$ & 
  $  \Dub_{200}  \;
  \frac{x_2^2}{12}$ &
    $\De_{16}$ &
  $\Dub_{101}
  \frac{1}{15}
  \left(
  2 x_2 (\lvec \cdot \xib)
  -
  (\lvec \cdot \xvec_2) \ct
  \right)$
\\ \hline
  $\De_5$ & $\Dub_{002} \;
  \frac{l^2}{6}$ &
    $\De_{17}$ &
  $\Dub_{011}
  \frac{1}{30}
  \left(  
  3 (\lvec \cdot \xvec_3) \ct
  -
  x_3 (\lvec \cdot \xia)
  \right)$
\\ \hline
  $\De_6$ & 
  $  \Dub_{100}
  \frac{-1}{6}
  a x_2 x_3
  \sin(\theta)^2$ &
    $\De_{18}$ &
  $\Dub_{011}
  \frac{1}{15}
  \left(  
  2 x_3 (\lvec \cdot \xia)
  -  
  (\lvec \cdot \xvec_3) \ct
  \right)$
\\ \hline
  $\De_7$ & 
  $\Sub_{010} 
  \frac{1}{6}
  (1-a) x_2 x_3
  \sin(\theta)^2$ &
    $\De_{19}$ &
  $\Dub_{110}
  \frac{1}{30}
  \left(
  3 x_2 x_3 \cos^2 \theta
  -
  x_2 x_3
  \right)$
\\ \hline
  $\De_8$ & 
  $\Dub_{101}
 \frac{1}{3}
 \lvec \cdot \xvec_2$ &
    $\De_{20}$ &
  $\Dub_{110}
  \frac{1}{15}
  \left(
  2 x_2 x_3
  -
  x_2 x_3 \cos^2 \theta
  \right)$
\\ \hline
  $\De_9$ & 
  $\Dub_{011}
 \frac{1}{3} 
 \lvec \cdot \xvec_3$ &
    $\De_{21}$ &
  $\Dub_{101}
  \frac{2}{15} (\lvec \cdot \xvec_2)
  \sin^2\theta$
\\ \hline
  $\De_{10}$ & 
  $\Dub_{110}
  \frac{1}{3}
  \xvec_2 \cdot \xvec_3$ &
  $\De_{22}$ &
  $\Dub_{011}
  \frac{2}{15} (\lvec \cdot \xvec_3)
  \sin^2\theta$
\\ \hline
  $\De_{11}$ &
  $\Dub_{101}
  \frac{1}{15}
  \left(
  2 \lvec \cdot \xvec_2
  - 
  (\lvec \cdot \xib)
  x_2 \ct
  \right)$ &
    $\De_{23}$ &
  $\Dub_{110}
  \frac{2}{15} (\xvec_2 \cdot \xvec_3)
  \sin^2\theta$
\\ \hline
  $\De_{12}$ &
  $\Dub_{101} 
  \frac{1}{30}
  \left(
  3 (\lvec \cdot \xib)
  x_2 \ct
  -
  \lvec \cdot \xvec_2
  \right)$
\\ \hline\hline
  \end{tabular}
\end{table} 
\begin{table}[th]
  \renewcommand{\arraystretch}{1.4}
  \caption{Table of $\Zp$ functions \label{tab:de_dme_functions_con}}
  \begin{tabular}{l|l}
  \hline\hline
  $\Zp_1$ &   $\left[ \Pi_0 (\kf x) \right]^2
  +   
  \frac{1}{5} x^2 \kf^2 
  \Pi_0 (\kf x) \Pi_2 (\kf x)$
\\ \hline
  $\Zp_2$ & $\frac{1}{12} x^2 \Pi_0(\kf x) \Pi_2 (\kf x)$
\\ \hline
  $\Zp_3$ & 
  $\frac{1}{3} x^2
  \left[ \Pi_1^s(\kf x) \right]^2$
  \\ \hline\hline
  \end{tabular}
\end{table}

\section{\texorpdfstring{$3N$}{3N} DME Couplings}
\label{sec:nnn_couplings}

	For our $3N$ potentials, the resulting EDF for the Fock term will consist of 23
trilinears of local densities with the form:
\begin{align*}
  V_{\text{F}} \approx& \ \int d\Rvec \;
  g^{\rho_0^3} \rho_0^3 
  + 
  g^{\rho_0^2 \tau_0} \rho_0^2 \tau_0 
  +
  g^{\rho_0^2 \Delta \rho_0}\rho_0^2 \Delta \rho_0 
  +
  g^{\rho_0 (\nabla \rho_0)^2}
  \rho_0 \nabla \rho_0 \cdot \nabla \rho_0 
  +
  g^{\rho_0 \rho_1^2}\rho_0 \rho_1^2
\\
  & \quad\null + 
  g^{\rho_1^2 \tau_0} \rho_1^2 \tau_0
  +
  g^{\rho_1^2 \Delta \rho_0}
  \rho_1^2 \Delta \rho_0
  +
  g^{\rho_0 \rho_1 \tau_1}
  \rho_0 \rho_1 \tau_1
  +
  g^{\rho_0 \rho_1 \Delta \rho_1}
  \rho_0 \rho_1 \Delta \rho_1
  +
  g^{\rho_0 (\nabla \rho_1)^2}
  \rho_0 \nabla \rho_1 \cdot \nabla \rho_1
\\
  & \quad\null + 
  \rho_0 \epsilon_{ijk}
  \left[
  g^{\rho_0 \nabla \rho_0 J_0}
  \nabla_i \rho_0 J_{0,jk}
  +
  g^{\rho_0 \nabla \rho_1 J_1}
  \nabla_i \rho_1 J_{1,jk}
  \right]
\\
  & \quad\null + 
  \rho_1 \epsilon_{ijk}
  \left[
  g^{\rho_1 \nabla \rho_1 J_0}
  \nabla_i \rho_1 J_{0,jk}
  +
  g^{\rho_1 \nabla \rho_0 J_1}
  \nabla_i \rho_0 J_{1,jk}
  \right]
\\
  & \quad\null + 
  \rho_0 \left[
  g^{\rho_0 J_0^2, 1}
  J_{0,aa} J_{0,bb}
  +
  g^{\rho_0 J_0^2, 2}
  J_{0,ab} J_{0,ab} 
  +
  g^{\rho_0 J_0^2, 3}
  J_{0,ab} J_{0,ba}
  \right]
\\
  & \quad\null + 
  \rho_0 \left[
  g^{\rho_0 J_1^2, 1}
  J_{1,aa} J_{1,bb}
  +
  g^{\rho_0 J_1^2, 2}
  J_{1,ab} J_{1,ab} 
  +
  g^{\rho_0 J_1^2, 3}
  J_{1,ab} J_{1,ba}
  \right]
\\
  & \quad\null + 
  \rho_1 \left[
  g^{\rho_1 J_0 J_1, 1}
  J_{1,aa} J_{0,bb}
  +
  g^{\rho_1 J_0 J_1, 2}
  J_{1,ab} J_{0,ab} 
  +
  g^{\rho_1 J_0 J_1, 3}
  J_{1,ab} J_{0,ba}
  \right]
  \;.
  \numberthis
  \label{eq:nnn_edf}
\end{align*}
  Combining the single exchange trace results in Eqs.~\eqref{eq:se_spin_traces} and \eqref{eq:se_iso_traces} along with the DME expansion equations in Appendix~\ref{sec:se_dict}, one can verify 16 of the 23 local density trilinears in Eq.~\eqref{eq:nnn_edf}. 
  Combining the remaining double exchange trace results in Eqs.~\eqref{eq:de_spin_traces}, \eqref{eq:de_iso_traces}, and
  \eqref{eq:de_con_traces} along with the DME expansion equations in Appendix~\ref{sec:de_dict} gives the remaining 7 structures in Eq.~\eqref{eq:nnn_edf}.
  Writing the $3N$ DME couplings $g^{i}$ in terms of the relevant integrals we find,
\beq
  \label{eq:g_3n_couplings}
  g^{i} = 
  8 \pi^2 
  \int_0^{\infty} x_2^2 \; dx_2 
  \int_0^{\infty} x_3^2 \; dx_3
  \int_0^{\pi} \sin\theta \; d\theta  \;
  h^{i} \;
  + 4 \pi \int_0^{\infty} dx \; x^2 j^{i} ,
\eeq
where $h^{i}$ encodes the relevant information about the LR DME expansion terms and $\B$ radial functions.
The $j^i$ term encodes the relevant information about the IR DME expansion terms and the $\B_{c}$ radial functions.
The coupling kernels $h^i$ and $j^i$ are derived using the DME dictionaries and trace results in the appendices and combined using an included Mathematica notebook.

  Below, the $\W$ variables refer to single exchange LR DME functions from Table~\ref{tab:se_ang} while the $\Z$ variables refer to the double exchange LR DME functions from Table~\ref{tab:de_dme_functions}.
  Their primed counterparts refer to the IR DME functions in Tables~\ref{tab:se_dme_functions_con} and \ref{tab:de_dme_functions_con}.
 To condense our output, we also define the new variable,
\beq
	\R \equiv \alpha_3 / \alpha_2 \; ,
\eeq
where $\alpha_i$ are the chiral prefactors in,
\beq
  V_{3N}^{(i)} = 
  \alpha_1^{(i)} V_{C,1}
  +
  \alpha_2^{(i)} V_{C,2}
  +
  \alpha_3^{(i)} V_{C,3}
  \;,
\eeq
	This is done such that the radial functions $\B_{6-9}$ can be represented by $\B_{2-5}$ with an extra factor of $\R$.

  To manage notation, we introduce auxiliary variables $\F$ and $\Fp$ to encapsulate dependence on the different radial parts of the potential, $\B$ and $\B_{c}$ respectively (LR and IR). 
  These are given in Tables~\ref{tab:f_functions_1} and \ref{tab:f_functions_2}.
  For the $\F_i$ functions, if the subscript $i$ is a Latin letter, then the function exclusively appears in coupling equations corresponding to two spin-current densities.
  For the first 16 DME couplings, single exchange contributions are given by the first term while double exchange contributions, where applicable, are given by the second term. The LR contributions to these terms are given by:
\bseq
\label{eq:nnn_couplings}
\begin{align*}
  h^{\rho_0^3} &= 
  - \frac{3}{8}
  \W_1 \F_2
  +
  \frac{3}{16} \Z_1 \F_4 \;,
  \numberthis
\\
  h^{\rho_0^2 \tau_0} &=
  \frac{3}{4}
  \W_2 \F_2
  -
  \frac{3}{16} \left(
  2 \Z_3 + 2 \Z_4 + \Z_5
  \right) \F_4 \;,
  \numberthis
\\
  h^{\rho_0^2 \Delta \rho_0} &=
  - \frac{3}{4} 
  \W_3 \F_2
  +
  \frac{3}{16} \left(
  \Z_2 + \Z_3 + \Z_4
  \right) \F_4 \;,
  \numberthis
\\
  h^{\rho_0 (\nabla \rho_0)^2} &=
  - \frac{3}{8}
  \W_4 \F_2
  +
  0 \;,
  \numberthis
\end{align*}
\begin{align*}
  h^{\rho_0 \rho_1^2} &=
  \frac{1}{8}
  \W_1 \F_2
  +
  \frac{1}{16} \Z_1 \F_5 \;,
  \numberthis 
\\
  h^{\rho_0 \rho_1 \tau_1} &=
  - \frac{1}{4}  
  \W_2 \F_2
  + \frac{1}{8}
  \left(
  \F_4 \Z_5 - 2 \F_6 (\Z_3 + \Z_4)
  \right) \;,
  \numberthis
\\
  h^{\rho_0 \rho_1 \Delta \rho_1} &=
  \frac{1}{4}
  \W_3 \F_2
  + \frac{1}{8}
  \left(
  \F_6 (\Z_3 + \Z_4) - \F_4 \Z_2
  \right) \;,
  \numberthis
\\
  h^{\rho_0 (\nabla \rho_1)^2} &=
  \frac{1}{8}
  \W_4 \F_2
  + 0 \;,
  \numberthis
\\
  h^{\rho_0 \nabla \rho_0 J_0} &=
  - \frac{3}{4} \W_5
  \F_1
  +  
  \frac{3}{16} 
  (\F_1 + \F_7) 
  \left(\Z_6 - \Z_7 \right) \;,
  \numberthis
\\
  h^{\rho_0 \nabla \rho_1 J_1} &=
  \frac{1}{4} 
  \W_5 \F_1
  +
  \frac{1}{16} 
  (\F_1 + \F_7) 
  \left( \Z_7 - \Z_6 \right) \;,
  \numberthis
\end{align*}
\begin{align*}
  h^{\rho_0 J_0^2, 1} &=   
  - \frac{3}{4}
  \F_a
  + 
  \left(
  \F_d + \F_c \R
  \right) \;,
  \numberthis
\\
  h^{\rho_0 J_0^2, 2} &=
  \frac{3}{8}
  \F_b
  +  \left(
  \F_f + \F_e \R
  \right) \;,
  \numberthis
\\
  h^{\rho_0 J_0^2, 3} &=
  - \frac{3}{4}
  \F_a
  + 
  \left(
  \F_d + \F_i + \F_g \R
  \right) \;,
  \numberthis
\\
  h^{\rho_0 J_1^2, 1} &=
  \frac{1}{4}
  \F_a
  +
  \frac{1}{3}
  \left(
  - \F_d -2 \F_i - \F_c \R
  \right)\;,
  \numberthis
\\
  h^{\rho_0 J_1^2, 2} &=
  - \frac{1}{8}
  \F_b
  + \frac{1}{3}
  \left(
  3 \F_h - \F_e \R
  \right) \;,
  \numberthis
\\
  h^{\rho_0 J_1^2, 3} &=
  \frac{1}{4}
  \F_a
  +  \frac{1}{3}
  \left(
  - \F_d + \F_i - \F_g \R
  \right) \;.
  \numberthis
\end{align*}
\eseq

The IR contributions to these 16 couplings are given by:

\bseq
\label{eq:nnn_couplings_con}
\begin{align*}
  j^{\rho_0^3} &= 
  - \frac{3}{4} \Wp_1 \Fp_1
  + \frac{3}{8}
  (\Fp_1 - 4 \Fp_4) \Zp_1
  \;,
  \numberthis
\\
  j^{\rho_0^2 \tau_0} &= 
  - \frac{3}{2}  
  \Wp_2 \Fp_1
  -
  \frac{3}{2}
  (\Fp_1 - 4 \Fp_4) \Zp_2
  \;,
  \numberthis
\\
  j^{\rho_0^2 \Delta \rho_0} &= 
  \frac{3}{4} \Wp_2 \Fp_1
  +
  \frac{3}{4}
  (\Fp_1 - 4 \Fp_4) \Zp_2
  \;,
  \numberthis
\\
  j^{\rho_0 (\nabla \rho_0)^2} &= 0 + 0
  \;,
  \numberthis
\end{align*}
\begin{align*}
  j^{\rho_0 \rho_1^2} &=
  \frac{1}{4}
  \Wp_1 \Fp_1
  +
  \frac{1}{8} (\Fp_1 + 12 \Fp_4) \Zp_1
  \;,
  \numberthis 
\\
  j^{\rho_0 \rho_1 \tau_1} &=
  \frac{1}{2}
  \Wp_2 \Fp_1
  -
  4 \Fp_4 \Zp_2
  \;,
  \numberthis
\\
  j^{\rho_0 \rho_1 \Delta \rho_1} &=
  - \frac{1}{4}
  \Wp_2 \Fp_1
  +
  2 \Fp_4 \Zp_2
  \;,
  \numberthis
\\
  j^{\rho_0 (\nabla \rho_1)^2} &= 0 + 0
  \;,
  \numberthis
\\
  j^{\rho_0 \nabla \rho_0 J_0} &= 0 + 0
  \;,
  \numberthis
\\
  j^{\rho_0 \nabla \rho_1 J_1} &= 0 + 0
  \;,
  \numberthis
\end{align*}
\begin{align*}
  j^{\rho_0 J_0^2, 1} &=   
  - \frac{3}{10} \Wp_3 \Fp_2
  +
  \frac{3}{20} \Fp_5 \Zp_3
  \;,
  \numberthis
\\
  j^{\rho_0 J_0^2, 2} &=
  \frac{3}{20} \Wp_3 \Fp_3
   - \frac{3}{40}
   (\Fp_3 - 4 \Fp_6) \Zp_3
  \;,
  \numberthis
\\
  j^{\rho_0 J_0^2, 3} &=
  - \frac{3}{10} \Wp_3 \Fp_2
  +
  \frac{3}{20} \Fp_5 \Zp_3
  \;,
  \numberthis
\\
  j^{\rho_0 J_1^2, 1} &=
  \frac{1}{10}
  \Wp_3 \Fp_2
  - \frac{1}{20} 
  \Fp_5 \Zp_3
  \;,
  \numberthis
\\
  j^{\rho_0 J_1^2, 2} &=
  - \frac{1}{20}
  \Wp_3 \Fp_3
  +
  \frac{1}{40}
  (\Fp_3 - 4 \Fp_6) \Zp_3
  \;,
  \numberthis
\\
  j^{\rho_0 J_1^2, 3} &=
  \frac{1}{10}
  \Wp_3 \Fp_2
  - \frac{1}{20} 
  \Fp_5 \Zp_3
  \;.
  \numberthis
\end{align*}
\eseq
  The 7 couplings which only have a contribution from the double exchange term are given by, for the LR terms,
\bseq
\begin{align*}
  h^{\rho_1^2 \tau_0} &=
  \frac{1}{16}
  \left[
  2 \F_4 (\Z_3 + \Z_4) - \F_8 \Z_5
  \right] \;,
  \numberthis  
\\
  h^{\rho_1^2 \Delta \rho_0} &=
  \frac{1}{16}
  \left[
  \F_8 \Z_2 - \F_4 (\Z_3 + \Z_4)
  \right] \;,
  \numberthis
\\
  h^{\rho_1 \nabla \rho_1 J_0} &=
  \frac{1}{16}
  \left[
  (\F_1 + \F_7)
  (\Z_7 - \Z_6)
  \right] \;,
  \numberthis
\\
  h^{\rho_1 \nabla \rho_0 J_1} &=
  \frac{1}{16}
  \left[
  (3 \F_1 - \F_7)
  (\Z_6 - \Z_7)
  \right] \;,
  \numberthis
\\
  h^{\rho_1 J_0 J_1, 1} &=
  \frac{2}{3}
  \left[
  \F_d + \F_i
  - \F_c \R
  \right] \;,
  \numberthis
\\
  h^{\rho_1 J_0 J_1, 2} &=
  \frac{1}{3}
  \left[
  \F_f - 3 \F_h - 2 \F_e \R
  \right] \;,
  \numberthis
\\
  h^{\rho_1 J_0 J_1, 3} &=
  \frac{2}{3}
  \left[
  \F_d - \F_g \R
  \right] \;.
  \numberthis
\end{align*}
\eseq
The IR contributions are given by,
\bseq
\begin{align*}
  j^{\rho_1^2 \tau_0} &= - \frac{1}{2} (\Fp_1 + 4 \Fp_4) \Zp_2
  \;,
  \numberthis  
\\
  j^{\rho_1^2 \Delta \rho_0} &= 
  \frac{1}{4} (\Fp_1 + 4 \Fp_4) \Zp_2
  \;,
  \numberthis
\\
  j^{\rho_1 \nabla \rho_1 J_0} &= 0
  \;,
  \numberthis
\\
  j^{\rho_1 \nabla \rho_0 J_1} &= 0
  \;,
  \numberthis
\\
  j^{\rho_1 J_0 J_1, 1} &= 
  \frac{1}{10}
  \Fp_7 \Zp_3
  \;,
  \numberthis
\\
  j^{\rho_1 J_0 J_1, 2} &=
  - \frac{1}{20} (\Fp_3 + 4 \Fp_6) 
  \Zp_3
  \;,
  \numberthis
\\
  j^{\rho_1 J_0 J_1, 3} &=
  \frac{1}{10}
  \Fp_7 \Zp_3
  \;.
  \numberthis
\end{align*}
\eseq

\begin{table}[th]
  \renewcommand{\arraystretch}{1.2}
  \caption{\label{tab:f_functions_1} Table of $\F$ functions}
  \begin{tabular}{l|l}
   \hline\hline
    $\F_1$ &   
    $\B_1 + \B_5 \ct$ 
    \\ \hline
    $\F_2$ &   
    $3 \B_2 + \B_3  + \B_4 + \F_1 \ct$ 
    \\ \hline
    $\F_3$ &
    $6 \B_2 + 2 \B_3 + 2 \B_4 + \B_5 - \B_5 \ct^2$
    \\ \hline
    $\F_4$ &
    $\F_2 - 2 \F_3 \R$
    \\ \hline
    $\F_5$ &
    $\F_2 + 6 \F_3 \R$
    \\ \hline
    $\F_6$ &
    $\F_2 + 2 \F_3 \R$
    \\ \hline
    $\F_7$ &
    $2 \R \B_5 \ct$
    \\ \hline
    $\F_8$ &
    $3 \F_2 + 2 \F_3 \R$
  \\ \hline \hline  
  $\F_a$ & 
  $ \B_3 \W_8 
  + \B_4 \W_{10}
  + \F_1 \W_{12}$
  \\ \hline
  $\F_b$ &  
  $\W_6 \left(
   \B_2 + \B_3 + \B_4 + \F_1 \ct
  \right)
  - 2 \left( 
  \B_3 \W_7 + \B_4 \W_9
  + \F_1 \W_{11}
  \right)$
  \\ \hline
  $\F_c$ &
  $\frac{3}{160} \Big[
  \B_5 \left(8 \Z_{10}
  + 20 \Z_{12} - 20 \Z_{14}
  - 40 \ct \Z_{20}
  - 5 \Z_{21} + 5 \Z_{22} + 5 \Z_{23}
  - 4 \Z_8 + 4 \Z_9 \right)$
  \\
  &
  $+ 8 \B_3 \left(\Z_{10}
  - 5 \Z_{14} - 2 \Z_8 \right) 
  + 8 \B_4 \left(
  \Z_{10} + 5 \Z_{12} + 2 \Z_9
  \right)
  \Big]$  
  \\ \hline
  $\F_d$ &
  $\frac{3}{80}
  \Big[
  5 \F_1 \left(
  \Z_{15}+\Z_{16}-\Z_{17}
  -Z_{18}-Z_{19}+Z_{20}
  \right)
  +2\B_3 (\Z_8 - 5 Z_{14})
  +2\B_4 (5\Z_{12}-Z_9) 
  \Big]$  
  \\ \hline   
  $\F_e$ &
  $\frac{3}{40}
  \Big[
  10 \B_2 (
  \Z_{10} + \Z_8 - \Z_9)  
  + 2 \B_3 (\Z_{10} - 5 \Z_{13} + 3 \Z_8)
  +
  \B_4 
  (
  2 \Z_{10} + 10 \Z_{11} - 6 \Z_9
  )$
  \\
  &
  $+ \B_5 (
  2 \Z_{10} + 5 \Z_{11} - 5 \Z_{13}
  -10 \ct \Z_{19} + 5 \Z_{21} - 5 \Z_{22} - 5 \Z_{23} - \Z_8 + \Z_9
  )
  \Big]$
  \\ \hline
  $\F_f$ &
  $\frac{3}{80} \Big[
  \F_1 \left(
  5 \ct \Z_{10} + 10 \Z_{15} -
  10 \Z_{17} - 5 \ct \Z_8 + 5
  \ct \Z_9
  \right)
  + 5 \B_2 (
  3 \Z_{10} - \Z_8 + \Z_9
  )$
  \\
  &
  $ + \B_3 (
  5 \Z_{10} - 10 \Z_{13} - 3 \Z_8 + 5 \Z_9
  )
  + \B_4 (
  5 \Z_{10} + 10 \Z_{11} - 5 \Z_8 + 3 \Z_9
  )
  \Big]$
  \\ \hline
  $\F_g$ &
  $\F_c (\Z_{20} \to
  \Z_{19})$
  \\ \hline
  $\F_h$ &
  $\frac{1}{80}
  \Big[5 \B_2 (9 \Z_{10} + \Z_8 - Z_9)
  + \B_3 (15 \Z_{10}
  +10 \Z_{13}
  +3 \Z_8 -5 \Z_9) + 
  \B_4 (
  15 \Z_{10} - 10 \Z_{11}
  +5 \Z_8 -3 \Z_9
  )$
  \\
  &
  $ 
  + \F_1 (
  15 \ct \Z_{10} 
  -10 \Z_{15}
  +10 \Z_{17} 
  +5 \ct \Z_8 
  -5 \ct \Z_9
  )
  \Big]$
  \\
  \hline
  $\F_i$ & $\frac{3}{8}
  \F_1 (\Z_{19} - \Z_{20})$
  \\ \hline\hline
  \end{tabular}
\end{table}  
\begin{table}[th]
  \renewcommand{\arraystretch}{1.2}
  \caption{\label{tab:f_functions_2} Table of $\Fp$ functions}
  \begin{tabular}{l|l}
   \hline\hline
    $\Fp_1$ &   $3 \B_{c,2} + \B_{c,3}$ 
     \\ \hline 
    $\Fp_2$ &   $\B_{c,3}$ 
     \\ \hline 
    $\Fp_3$ &   $5 \B_{c,2} + 3 \B_{c,3}$
     \\ \hline 
   $\Fp_4$ & 
    $3 \B_{c,6} + \B_{c,7}$
     \\ \hline 
    $\Fp_5$ & $\B_{c,3} - 4 \B_{c,7}$
  \\ \hline 
    $\Fp _6$ &  $5 \B_{c,6} + 3 \B_{c,7}$
  \\ \hline 
    $\Fp _7$ &  
	$\B_{c,3} + 4 \B_{c,7}$
  \\ \hline\hline
  \end{tabular}
\end{table}

\bibliography{dme_refs}

\end{document}